\documentclass[preprint,3p,times,sort&compress]{elsarticle}

\usepackage{multirow,setspace,times,amssymb,amsmath,graphicx,color,rotating,subfigure,url}
\usepackage{lineno}
\usepackage{natbib}
\usepackage{booktabs}%
\usepackage{longtable}%
\graphicspath{{Figures/}}
\usepackage[table]{xcolor}
\usepackage{tabularx}
\usepackage{graphicx} 
\usepackage{mathrsfs}
\usepackage{makecell}
\usepackage[bookmarks=true,colorlinks,linkcolor=blue,anchorcolor=blue,citecolor=blue,unicode]{hyperref}
\usepackage{bookmark}

\journal{}

\begin{document}

\begin{frontmatter}
\newpage
\title{Impact of the Russia-Ukraine conflict on the international staple agrifood trade networks}

\author[SB]{Yin-Ting Zhang}
\author[SB,RCE]{Mu-Yao Li}
\author[SB,RCE,DM]{Wei-Xing Zhou\corref{WXZhou}}
\ead{wxzhou@ecust.edu.cn}
\cortext[WXZhou]{Corresponding author.} 

\address[SB]{School of Business, East China University of Science and Technology, Shanghai 200237, China}
\address[RCE]{Research Center for Econophysics, East China University of Science and Technology, Shanghai 200237, China}
\address[DM]{Department of Mathematics, East China University of Science and Technology, Shanghai 200237, China}

\begin{abstract}
The Russia-Ukraine conflict is a growing concern worldwide and poses serious threats to regional and global food security. Using monthly trade data for maize, rice, and wheat from 2016/1 to 2022/12, this paper constructs three international crop trade networks (iCTNs) and an aggregate international food trade network (iFTN). We aim to examine the structural changes following the occurrence of the Russia-Ukraine conflict. We find significant shifts in the number of edges, average degree, density, efficiency, and natural connectivity in the third quarter of 2022, particularly in the international wheat trade network. Additionally, we have shown that political reasons have caused more pronounced changes in the trade connections between the economies of the North Atlantic Treaty Organization and Russia than with Ukraine. This paper could provide insights into the negative impact of geopolitical conflicts on the global food system and encourage a series of effective strategies to mitigate the negative impact of the conflict on global food trade.
\end{abstract}


\end{frontmatter}


\section{Introduction}
\label{S1:Introduction}

The Russia-Ukraine conflict is a growing concern worldwide \cite{Shumilova-Tockner-Sukhodolov-Khilchevskyi-DeMeester-Stepanenko-Trokhymenko-HernandezAgueero-Gleick-2023-NatSustain,Zhang-Hu-Su-Xing-Liu-2023-SciTotalEnviron}. It has had significant ramifications not only on political and security fronts but also on various socio-economic aspects, including economic development \cite{Liadze-Macchiarelli-MortimerLee-Juanino-2023-WorldEcon,Kumari-Kumar-Pandey-2023-JBehavExpFinanc}, regional environment \cite{Rawtani-Gupta-Khatri-Rao-Hussain-2022-SciTotalEnviron,Pereira-Basic-Bogunovic-Barcelo-2022-SciTotalEnviron}, international trade \cite{Braun-Braun-Gyimesi-Iloskics-Sebestyen-2023-WorldEcon,Korovkin-Makarin-2023-AmEconRev}and food security 
 \cite{LiverpoolTasie-Reardon-Parkhi-Dolislager-2023-NatFood}. Since Russia and Ukraine have emerged as key players in the global agricultural and energy markets over the past few years \cite{Abay-Breisinger-Glauber-Kurdi-Laborde-Siddig-2023-GlobFoodSecur-AgricPolicy}, the conflict poses serious global and regional food and energy security challenges \cite{Stulberg-2015-ProblPost-Communism,Zhou-Lu-Xu-Yan-Khu-Yang-Zhao-2023-ResourConservRecycl}.

According to UN Comtrade\footnote{Accessed on 2023/7/15, see \url{https://comtradeplus.un.org/}} and FAO\footnote{Accessed on 2023/7/15, see \url{https://www.fao.org/faostat/en/}}, Russia and Ukraine together accounted for significant shares of global wheat, maize, and sunflower oil exports, representing approximately 34 percent, 17 percent, and 73 percent, respectively. Additionally, they held substantial market shares in the global barley and maize trade, contributing around 27 percent and 17 percent, respectively \cite{2022-Laborde-MamunNature-IFPRI}. These exports play a crucial role in global consumption and diets, providing approximately 12 percent of the total calories traded worldwide. Furthermore, Russia holds a significant position as a major exporter of nitrogen and potash fertilizers, contributing to approximately 15 percent of global trade in nitrogenous fertilizers. Collectively, Russia and Belarus account for approximately 33 percent of global potash fertilizer exports \cite{Dillon-Barrett-2016-AmJAgrEcon}.

Food trade is crucial to global food systems \cite{Wu-Guclu-2013-RiskAnal,Laborde-Martin-Swinnen-Vos-2020-Science}. Economies and trade relationships create global food trade networks \cite{D'Odorico-Carr-Laio-Ridolfi-Vandoni-2014-EarthFuture,MacDonald-Brauman-Sun-Carlson-Cassidy-Gerber-West-2015-Bioscience}. Trade provides a way to bridge the gap between regions with surplus food production and those facing deficits, thereby contributing to global food security goals. However, climate change \cite{Zhao-Liu-Piao-Wang-Lobell-Huang-Huang-Yao-Bassu-Ciais-Durand-Elliott-Ewert-Janssens-Li-Lin-Liu-Martre-Muller-Peng-Penuelas-Ruane-Wallach-Wang-Wu-Liu-Zhu-Zhu-Asseng-2017-ProcNatlAcadSciUSA,Zurek-Hebinck-Selomane-2022-Science}, geopolitical conflict \cite{Olsen-Fensholt-Olofsson-Bonifacio-Butsic-Druce-Ray-Prishchepov-2021-NatFood,Li-Li-Fan-Mi-Kandakji-Song-Li-Song-2022-NatFood}, pandemics \cite{Farrell-Thow-Wate-Nonga-Vatucawaqa-Brewer-Sharp-Farmery-Trevena-Reeve-Eriksson-Gonzalez-Mulcahy-Eurich-Andrew-2020-FoodSecur,Laborde-Martin-Swinnen-Vos-2020-Science}, and the financial crisis \cite{Headey-Fan-2008-AgricEcon}will all conspire to put unprecedented stress on the global food trade system. When economies that are main food producers or exporters experience a reduction in food production \cite{Chamorro-Barriga-2023-RiskAnal} due to extreme events or implement trade restrictions, it will result in shortages, price spikes, and increased vulnerability to hunger and malnutrition. Therefore, the strong presence of Russia and Ukraine in agricultural sectors underscores their importance in global food trade and food security.

Recent studies have dedicated attention to examining the potential impacts of the Russia-Ukraine conflict on food security from various perspectives. One branch of this literature analyzes the effects of the conflict on food yields \cite{Lin-Li-Jia-Feng-Huang-Huang-Fan-Ciais-Song-2023-GlobFoodSecur-AgricPolicy}. The scarcity of labor, suspension of transportation, and disturbances to the supply of chemical fertilizers, as well as pest and disease controls, could have a profound impact on wheat cultivation in Ukraine. Additionally, several studies have explored the influence of the conflict on global food prices \cite{Arndt-Diao-Dorosh-Pauw-Thurlow-2023-GlobFoodSecur-AgricPolicy} and trade \cite{Neik-Siddique-Mayes-Edwards-Batley-Mabhaudhi-Song-Massawe-2023-FrontSustainFoodSyst}. These studies have found that the conflict would cause sharp increases in food prices, a decline in trade, and acute food insecurity, particularly for nations that rely significantly on grain imports from Russia and Ukraine \cite{Feng-Jia-Lin-2023-ChinaAgricEconRev,Laborde-Pineiro-2023-NatFood}. Furthermore, researchers have investigated how shocks generated by the conflict propagate through the global food trade network \cite{Feng-Laber-Klimek-Bruckner-Yang-Thurner-2023-NatFood}.

Some of these studies have utilized historical data prior to 2022 \cite{Korovkin-Makarin-2023-AmEconRev} or employed algorithms and models to simulate the impact of the Russia-Ukraine conflict on the food system \cite{Feng-Jia-Lin-2023-ChinaAgricEconRev}. However, there is a lack of research that has examined the effects of the conflict using post-conflict food trade data. In this paper, we leverage monthly trade data for maize, rice, and wheat from January 2016 to December 2022, obtained from the UN Comtrade Database, to construct three distinct international crop trade networks (iCTNs) and an aggregated international food trade network (iFTN). By comparing the changes in network structure and properties before and after the Russia-Ukraine conflict, we aim to analyze its influence on the global food trade network.

Due to the inherent delays in data collection and updates, most economies, including Russia and Ukraine, have not provided food trade data beyond 2022. Only a subset of European economies and a small number of economies in North and South America have reported timely and reliable trade data. To address this limitation, we fill in the missing data using bilateral trade data provided by the reporting economies. This approach enables us to capture the trade relationships between Russia, Ukraine, and select economies. Although we cannot construct a comprehensive global food trade network, our study still provides valuable insights into parts of global food trade networks, thereby filling a research gap in analyzing the actual impact of the Russia-Ukraine conflict using up-to-date data.

The subsequent sections of this paper are organized as follows: In Section~\ref{sec2:DataMethod}, we show an overview of the data source and methodology used, which includes network construction and formulation of topological metrics. Section~\ref{sec3:Results} presents the empirical findings, highlighting the effects of the Russia-Ukraine conflict on the three iCTNs, the iFTN, and economies. Finally, we summarize the key findings, discuss our contributions, and present future prospects in Section~\ref{sec4:Conclusions}.

\section{Data and method}
\label{sec2:DataMethod}

\subsection{Data description}
\label{subsec2:Data}

\begin{table}[!ht]    
\caption{Economies that report crop trade data sets during the period from 2016/1 to 2022/12. Economies with superscript * belong to the North Atlantic Treaty Organization (NATO). We also show the regions to which the reporting economies belong, where AS means Asia, AF means Africa, NA means North America, SA means South America, EU means Europe, and OC means Oceania. }
    \smallskip
    \setlength\tabcolsep{15.5mm}
    \renewcommand{\arraystretch}{0.8}
    \centering
\begin{tabular}{llll}
    \toprule
       Maize & Rice & Wheat & Region \\ 
    \midrule
        AUS & AUS & AUS & OC \\ 
        ARM & ARM & ARM & AS \\  
        BRB & BRB & ~ & NA \\  
        BEL* & BEL & ~ & EU \\  
        BOL & BOL & ~ & SA \\  
        BIH & BIH & BIH & EU \\  
        BRA & BRA & BRA & SA \\  
        BLZ & BLZ & ~ & NA \\  
        CAN* & CAN* & CAN* & NA \\  
        CHL & CHL & CHL & SA \\  
        HRV* & HRV* & ~ & EU \\  
        DNK* & DNK* & DNK* & EU \\  
        SLV & SLV & SLV & NA \\  
        ~ & FJI & ~ & OC \\  
        FIN & FIN & FIN & EU \\  
        GEO & GEO & GEO & AS \\  
        DEU* & DEU* & DEU* & EU \\  
        GRC* & GRC* & GRC* & EU \\  
        GRD & GRD & ~ & NA \\  
        HUN* & HUN* & ~ & EU \\  
        ISL* & ~ & ~ & EU \\  
        IRL & IRL & IRL & EU \\  
        ITA* & ITA* & ~ & EU \\  
        JPN & JPN & JPN & AS \\  
        LVA* & LVA* & LVA* & EU \\  
        LTU* & LTU* & LTU* & EU \\  
        ~ & MUS & ~ & AF \\  
        MEX & MEX & ~ & NA \\  
        NLD & NLD & NLD & EU \\  
        NZL & NZL & NZL & OC \\  
        NOR* & NOR* & NOR* & EU \\  
        PHL & PHL & ~ & AS \\  
        POL* & POL* & POL* & EU \\  
        ~ & PRT* & ~ & EU \\  
        ROU* & ROU* & ~ & EU \\  
        RWA & RWA & ~ & AF \\  
        ~ & STP & ~ & AF \\  
        SVK* & ~ & ~ & EU \\  
        SVN* & SVN* & ~ & EU \\  
        SWE & SWE & ~ & EU \\  
        CHE & CHE & CHE & EU \\  
        MKD* & MKD* & MKD* & EU \\  
        EGY & EGY & ~ & AF \\  
        GBR* & ~ & GBR* & EU \\  
        USA* & USA* & USA* & NA \\
    \bottomrule
\end{tabular}
  \label{Table:iCTN:reporters}
\end{table}

We consider three crops (maize, rice, and wheat) which meet the main calorie needs of the world's population. Our research incorporates monthly import and export data for these crops, spanning from January 2016 to December 2022, sourced from the UN Comtrade database. Due to delayed updates in data reporting, numerous economies have not yet provided monthly trade data for these crops since 2022. To mitigate this data gap, we have adopted a strategy that leverages information from trade partners to supplement the missing data. When a particular economy, such as China, fails to disclose its crop import data for a given month (e.g., April 2022), we rely on data reported by other countries, such as Brazil, who have shared their crop export data with China during the same period. By extrapolating from Brazil's data, we estimate China's crop import volume for that specific month. Likewise, in instances where Russia has not reported any crop trade data since April 2022, we utilize import and export data provided by its trade partners to fill in the gaps and obtain a comprehensive picture of Russia's crop trade activity during that period.

We establish trade relationships between reporting economies and their trading partners. Because of big monthly swings, we perform quarterly analysis of the international crop trade networks rather than monthly analysis, thereby maximizing coverage of the economies included in our study. Following data clean-up, we focus on those economies that constantly provided data during the period from January 2016 to December 2022, which include 30 reporting economies for maize, 32 for rice, and 25 for wheat (see Table~\ref{Table:iCTN:reporters}). To compare and aggregate different crop trade data, we ultimately use three types of crop trade data provide by 24 reporting economies (see the reporting economies for crops shown in Table~\ref{Table:iCTN:reporters}). Our data set contains crop trade flows for the selected economies only. As a result, we are unable to investigate the Russia-Ukraine conflict's impact on some economies, which would underestimate the conflict's impact on the global crop trade system. Nevertheless, we could still establish international food trade networks based on the available data, allowing us to assess the conflict's effects on some economies' crop trade.

\subsection{Network construction}
\label{S2:NetworkConstruction}

\begin{figure}[h!]
    \centering
    \includegraphics[width=0.4733\linewidth]{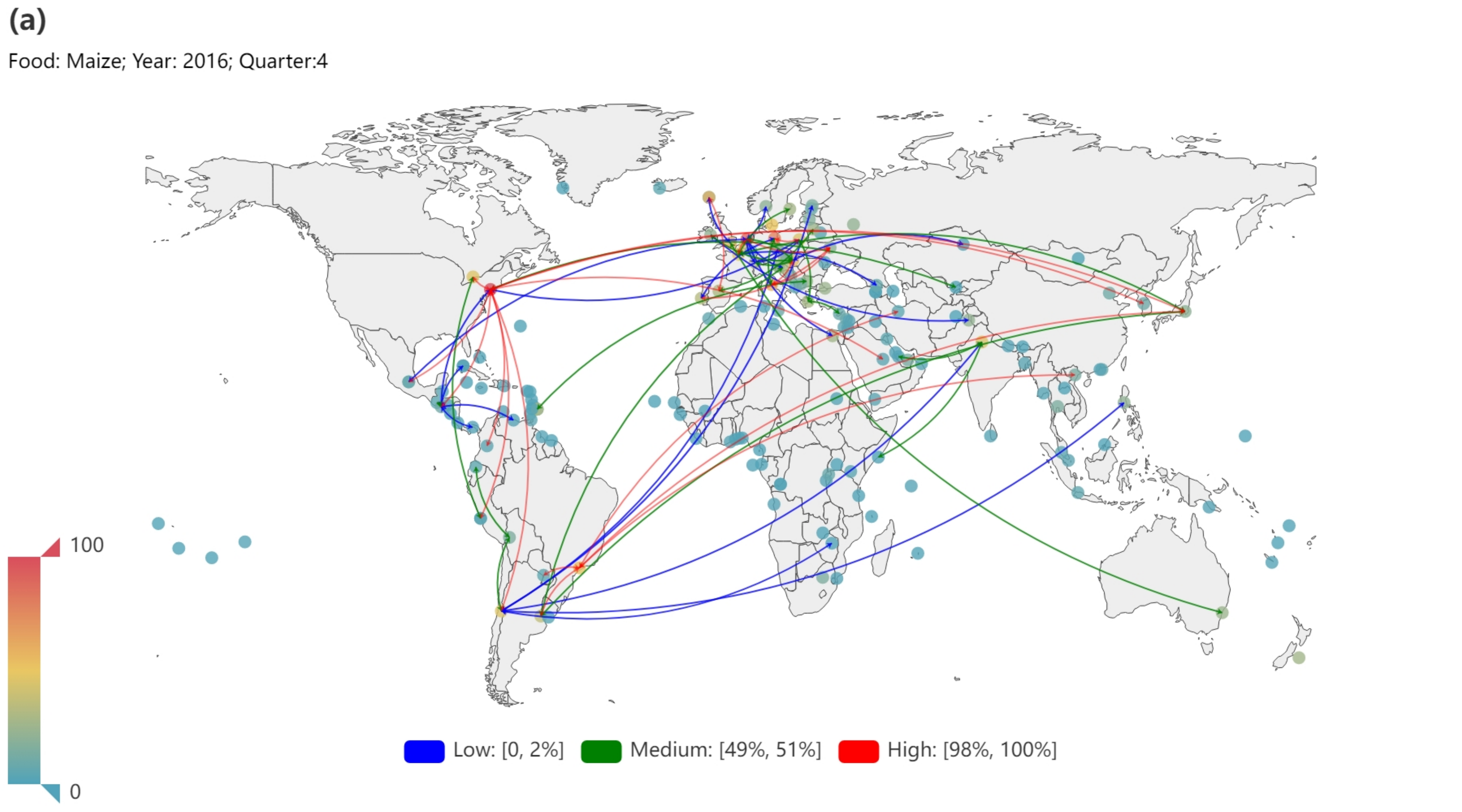}    
    \includegraphics[width=0.4733\linewidth]{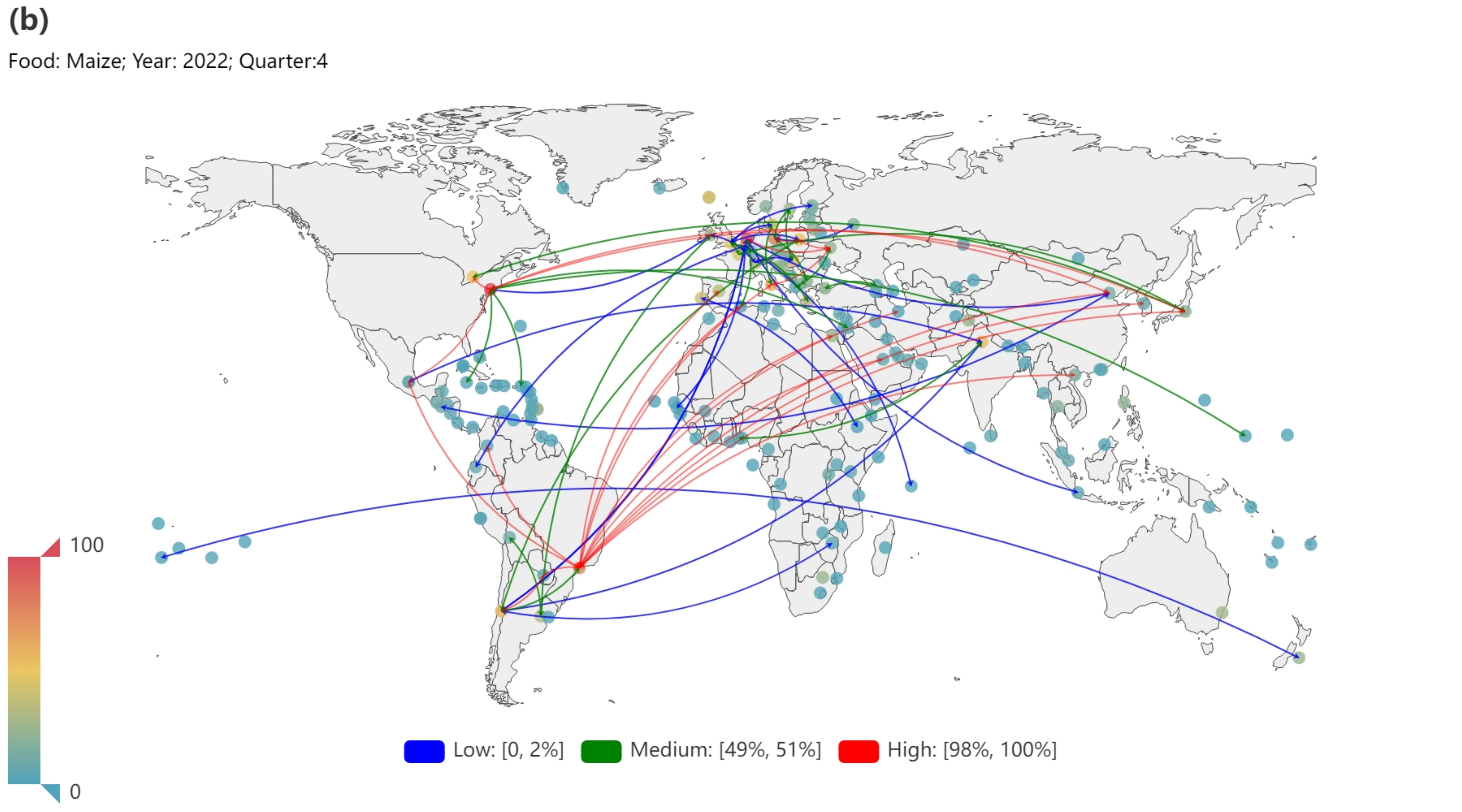}\\    
    \includegraphics[width=0.4733\linewidth]{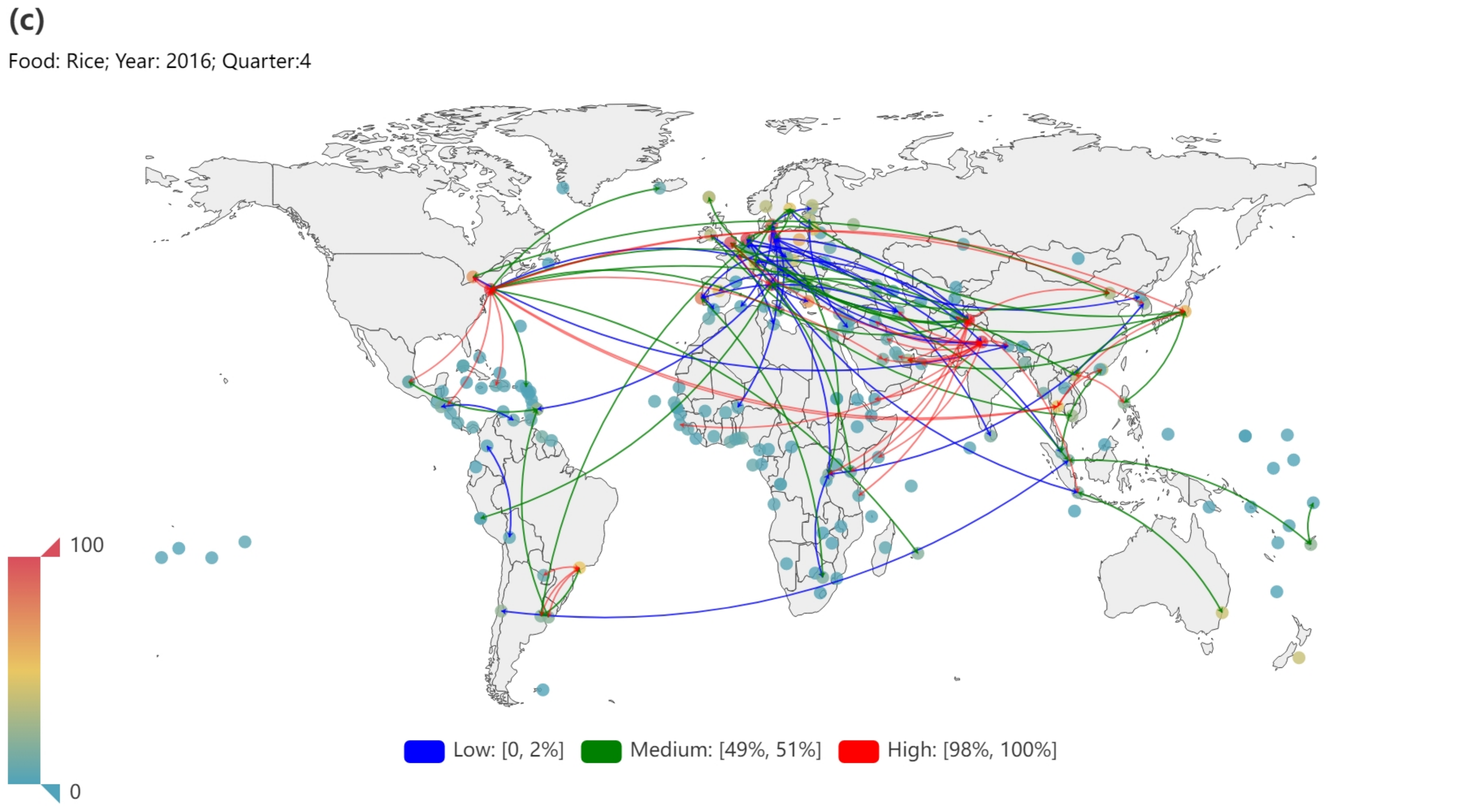}    
    \includegraphics[width=0.4733\linewidth]{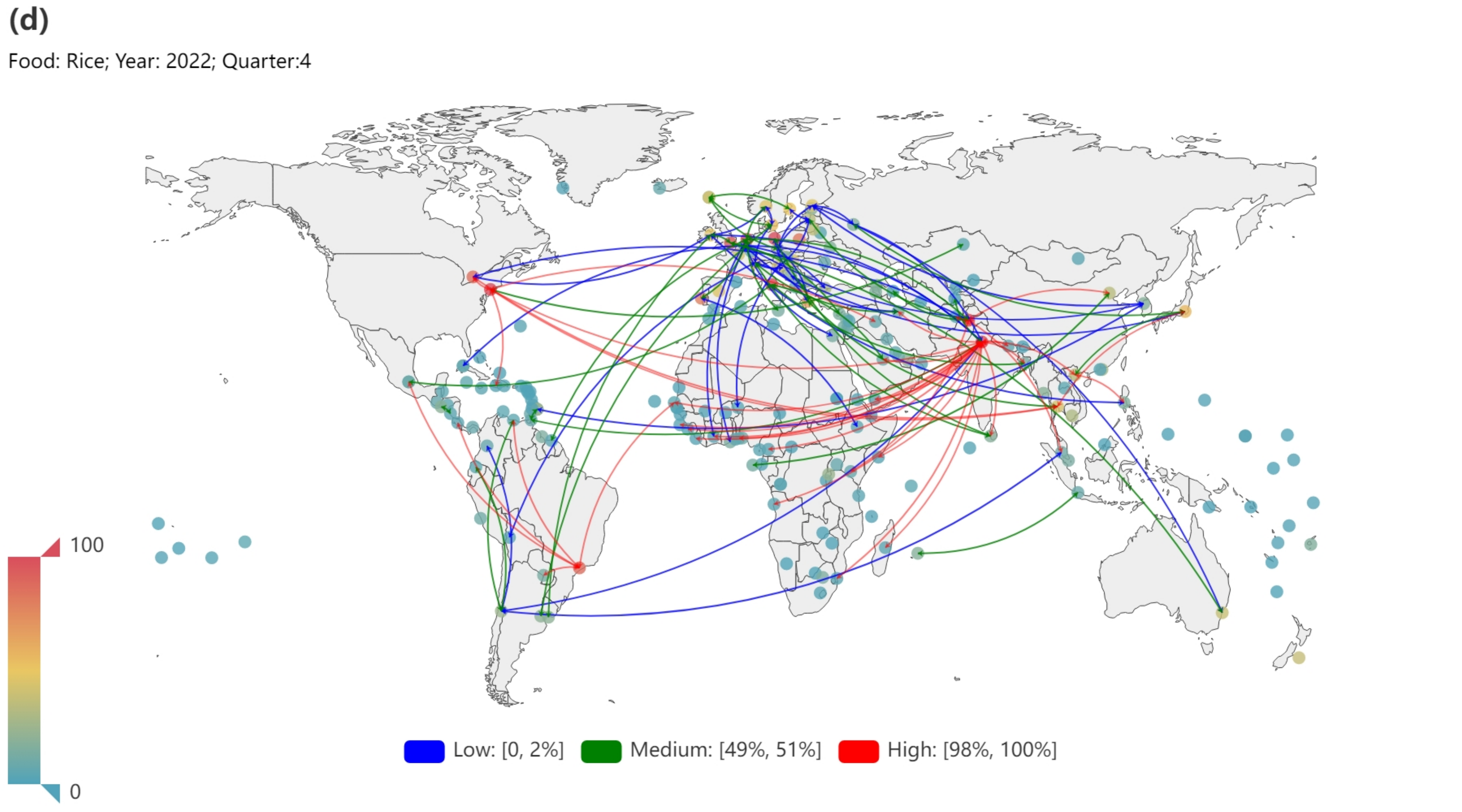}\\   
  \includegraphics[width=0.4733\linewidth]{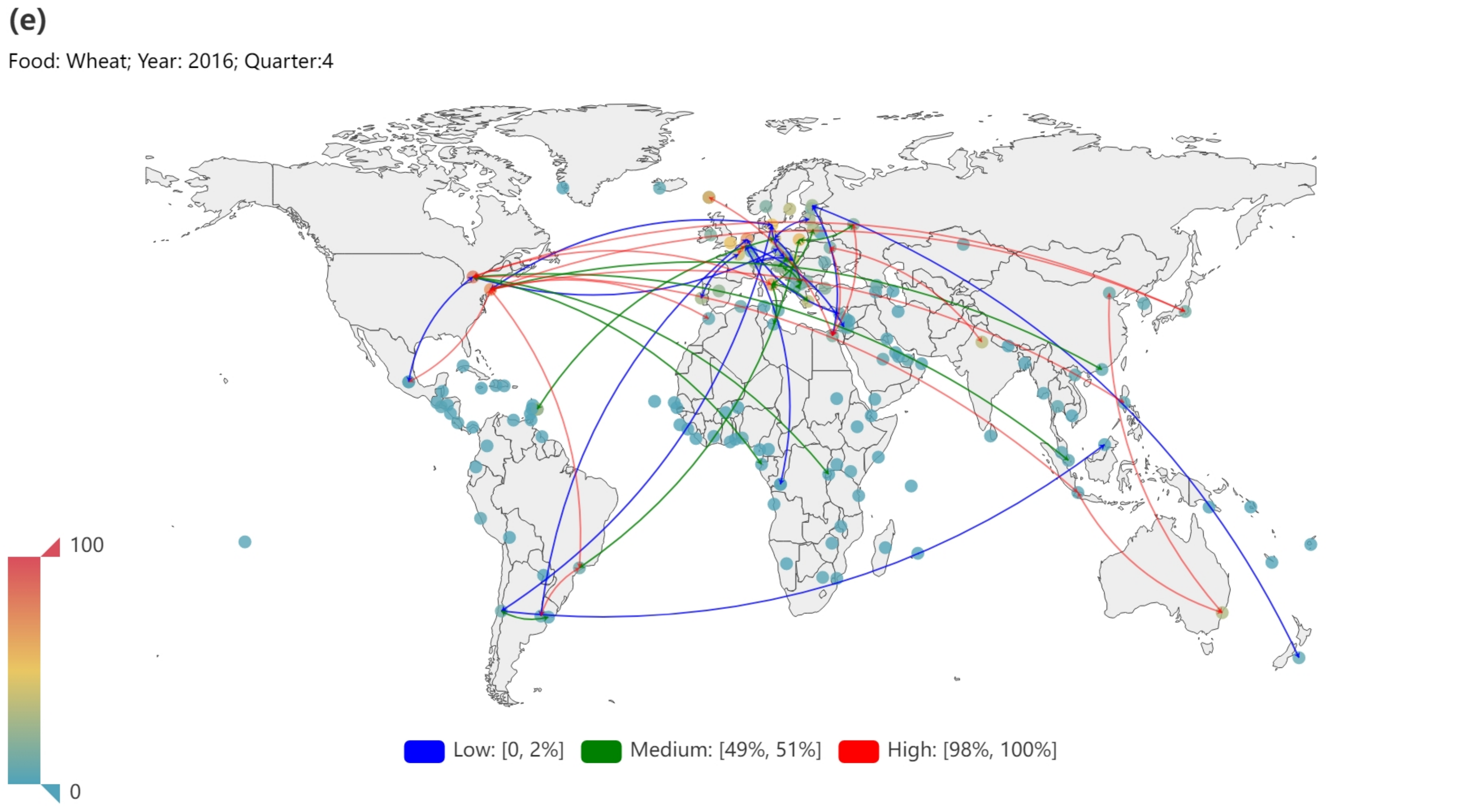}
    \includegraphics[width=0.4733\linewidth]{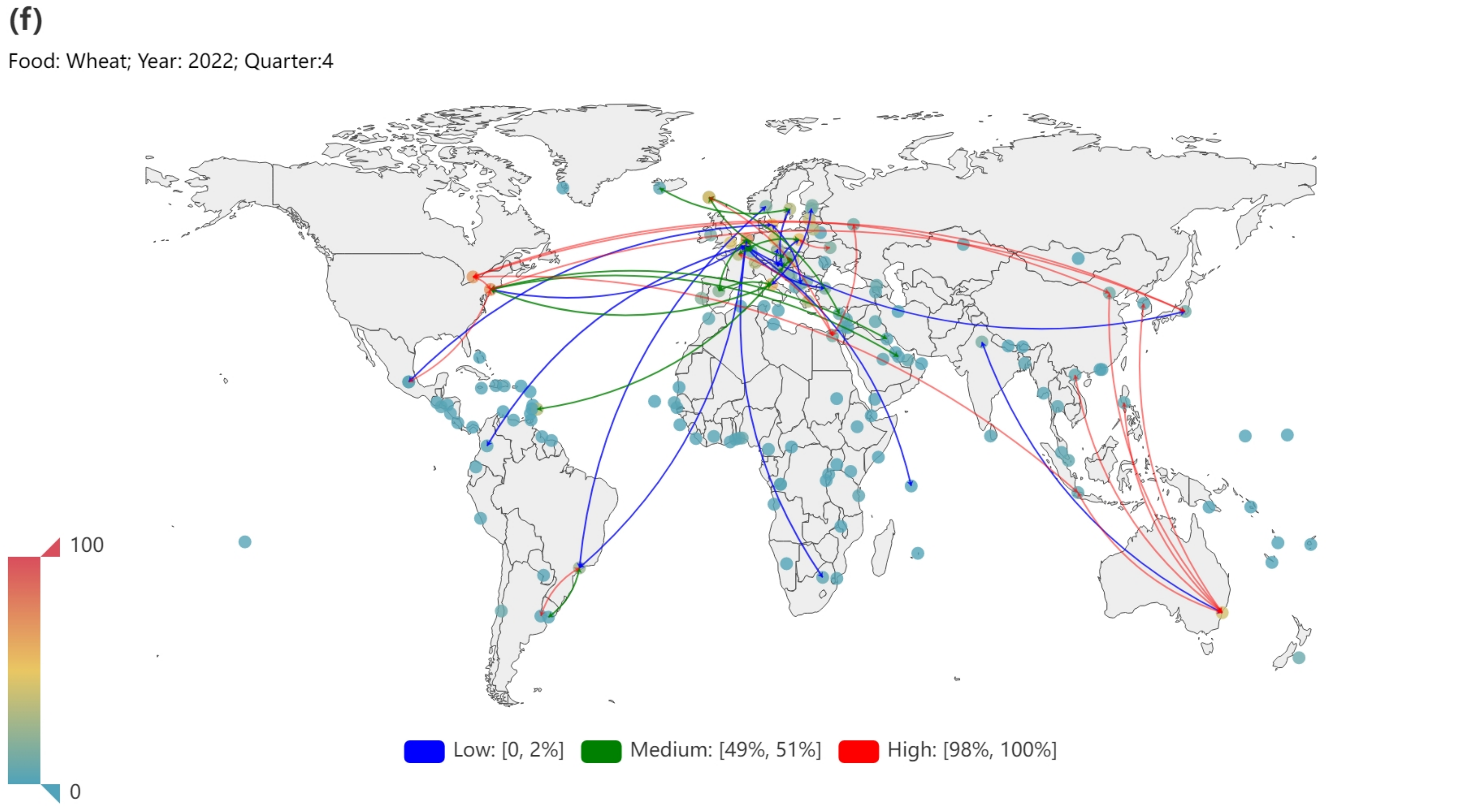}\\
  \includegraphics[width=0.4733\linewidth]{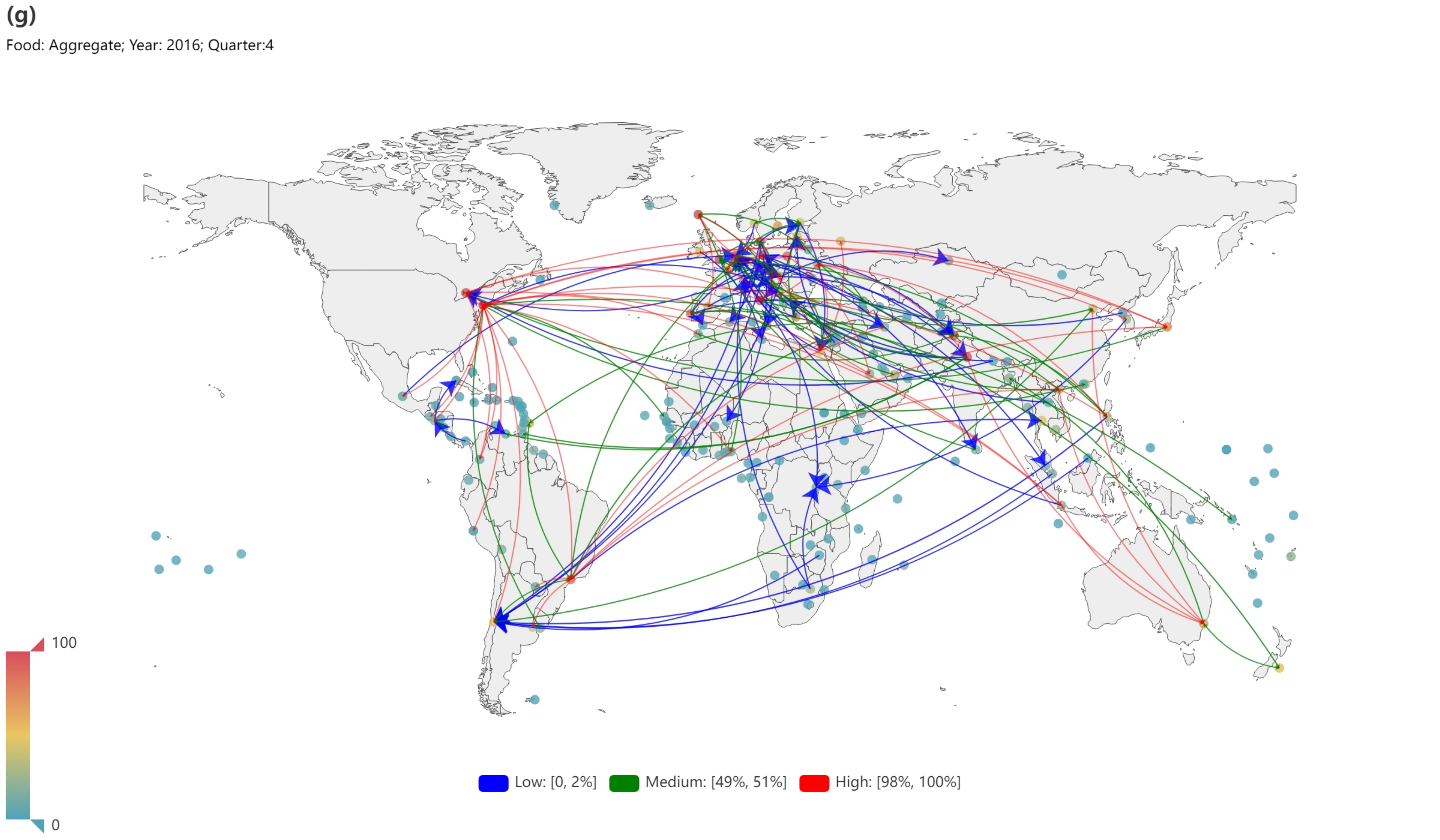}
    \includegraphics[width=0.4733\linewidth]{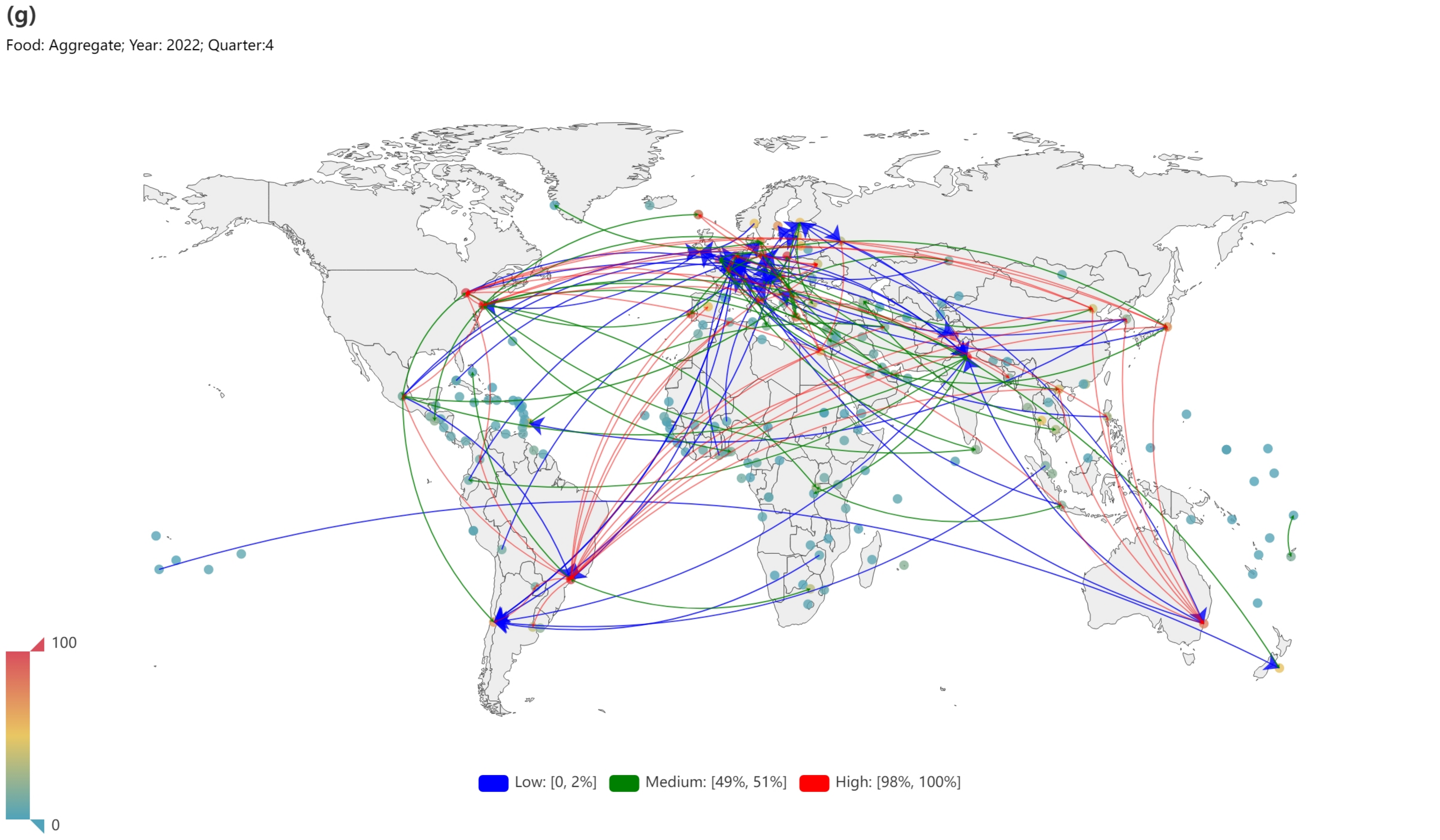}\\
    \caption{Three international crop trade networks (iCTNs) and the international food trade network (iFTN) in 2016/Q4 (left column) and 2022/Q4 (left column). The rows from top to bottom respectively describe maize, rice, wheat, and aggregate crops. For each graph, nodes stand for economies participating in crop trade. The colors of nodes correspond to degrees of economies. To enhance clarity, we show only links with trade volumes ranking in the top 2\%, medium 1\%, and bottom 2\% for each crop.}
    \label{Fig:iCTN:Geomap}
\end{figure}

Since monthly trade data possesses significant fluctuations, we restrict ourselves to the case of quarterly trade evolution. To eliminate the effects of crop prices and quality, we use food conversion data provided by the Food and Agriculture Organization of the United Nations (FAO) to convert the trade volumes into trade calories. Based on monthly trade data provided by the 24 economies selected, we derive the quarterly trade matrix $W^{crop}(t)$ for each $crop$ in each quarter $t$ ($crop=$ maize, rice, and wheat, $t=$ 2016/Q1, 2016/Q2, $\cdots$, 2022/Q4), where $w_{ij}^{crop}(t)$ denotes the caloric trade volume of $crop$ exported from economy $i$ to economy $j$. For each crop at each quarter from January 2016 to December 2022, we construct a time-varying international crop trade network (iCTN), $G^{crop}(t) = \left(\mathscr{V}^{crop}(t), W^{crop}(t) \right)$, where ${\mathscr{V}}^{crop}(t)$ denotes the set of network nodes, which includes reporting economies and their trading partners. Thus, we get three iCTNs: the international maize trade network (iMTN), the international rice trade network (iRTN), and the international wheat trade network (iWTN). At the same time, we construct the total international food trade network (iFTN), $G(t) = \left(\mathscr{V}(t), W(t) \right)$ by aggregating the three crops.

We compare the iCTNs and iFTN in different years but in the same quarter in order to exclude the seasonal impacts. Figure~\ref{Fig:iCTN:Geomap} displays the three iCTNs for maize, rice, and wheat as well as the iFTN in the fourth quarter of 2016 and 2022. The rows show trade flows for each crop and food (aggregate crops) in 2016/Q4 and 2022/Q4, respectively. To enhance clarity, we show only links with trade volumes ranking in the top 2\%, medium 1\%, and bottom 2\% for each crop. We find that the iFTN in the fourth quarter of 2022 (2022/Q4) has more links with high trade volumes with respect to the fourth quarter of 2016 (2016/Q4) (see Fig.~\ref{Fig:iCTN:Geomap}(g-h)). It indicates that new trade relationships are formed \cite{Zhang-Zhou-2022-FrontPhysics}. However, the iWTN has fewer links in 2022/Q4 compared to 2016/Q4, mirroring to a large extent the Russia-Ukraine conflict \cite{LiverpoolTasie-Reardon-Parkhi-Dolislager-2023-NatFood}. Figures.~\ref{Fig:iCTN:Geomap}(a-d) show that the the structure of the iMTN and iRTN covering selected economies has remained relatively stable. Further analysis based on topological indicators is required. 

\subsection{Node attributes}
\label{S2:NodeAttributes}

\subsubsection{Node degree}

The node degrees present the number of trade partners of economies. In a directed network, we define both in-degree and out-degree of a node to count incoming links and outgoing links respectively. The in-degree of node is defined as follows
\begin{equation}
    k_i^{\mathrm{in}} = \sum_{j\in\mathscr{V}-\{i\}} I_{\mathscr{E}}({e_{ji}})
    = \sum_{j=1}^{N_\mathscr{V}} I_{\mathscr{E}}({e_{ji}}),
\label{Eq:DiGraph:kin:j}
\end{equation}
where $I_{\mathscr{E}}({e_{ji}})$ is the indicator function:
\begin{equation}
I_{\mathscr{E}}({e_{ji}}) = 
\begin{cases}
      1, ~ \mathrm{if} ~~ e_{ji} \in \mathscr{E}\\
      0, ~ \mathrm{if} ~~ e_{ji} \notin \mathscr{E}
    \end{cases}
\end{equation}
The out-degree of node is defined as follows
\begin{equation}
    k_i^{\mathrm{out}} = \sum_{j\in\mathscr{V}-\{i\}} I_{\mathscr{E}}({e_{ij}})
    = \sum_{j=1}^{N_\mathscr{V}} I_{\mathscr{E}}({e_{ij}}).
    \label{Eq:DiGraph:kout:i}
\end{equation}

\subsubsection{Node strength}
Since the networks are weighted, we quantity node strengths, including in-strength $s_i^{\mathrm{in}}$ and out-strength $ s_i^{\mathrm{out}}$, which are defined as follows
\begin{equation}
    s_i^{\mathrm{in}} = \sum_{j\in\mathscr{V}-\{i\}} w_{ji}
    = \sum_{j=1}^{N_\mathscr{V}} w_{ji},
    \label{Eq:DiGraph:sin:j}
\end{equation}
\begin{equation}
    s_i^{\mathrm{out}} = \sum_{j\in\mathscr{V}-\{i\}} w_{ij}
    = \sum_{i=1}^{N_\mathscr{V}} w_{ij},
    \label{Eq:DiGraph:sout:i}
\end{equation} 
where $w_{jj}=0$ by definition.

\subsubsection{PageRank}
The PageRank algorithm was devised in 1997 to rank web pages in the Google search engine, and then it was used to measure the importance of nodes in a directed network in many fields. The PageRank computation proceeds iteratively over and over again to estimate the significance of a node \cite{Brin-Page-1998-ComputNetwISDNSyst}. Here, we omit the calculation process and apply PageRank to measure the influence of an economy in the iCTNs.

\subsection{Network metrics}

\subsubsection{Average node degree}

To show the average trade partners of all economies in the iCTNs and the iFTN, we calculate the average node degree,
\begin{equation}
    \left\langle{k^{\mathrm{in}}}\right\rangle
    = \left\langle{k^{\mathrm{out}}}\right\rangle
    = \frac{N_\mathscr{E}}{N_\mathscr{V}}
    \label{Eq:DiGraph:kin:ave:kout:ave}
\end{equation}
and
\begin{equation}
    \left\langle{k}\right\rangle_\mathscr{V} 
    = \left\langle{k^{\mathrm{in}}+k^{\mathrm{out}}}\right\rangle
    = \frac{2N_\mathscr{E}}{N_\mathscr{V}}.
    \label{Eq:DiGraph:k:ave}
\end{equation}

\subsubsection{Average node strength}

The average in-strength of nodes ($i\in\mathscr{V}$) is expressed as follows
\begin{equation}
    \left\langle{s^{\mathrm{in}}}\right\rangle_{\mathscr{V}}
    = \frac{1}{N_\mathscr{V}}\sum_{j=1}^{N_\mathscr{V}} s_j^{\mathrm{in}} 
    = \frac{1}{N_\mathscr{V}}\sum_{i=1}^{N_\mathscr{V}}\sum_{j=1}^{N_\mathscr{V}}\left(w_{ij}\right)^1
    = \frac{W}{N_{\mathscr{V}}},
    \label{Eq:DiGraph:sin:ave}
\end{equation}
Similarly, the average out-strength of nodes is expressed as follows
\begin{equation}
    \left\langle{s^{\mathrm{out}}}\right\rangle_{\mathscr{V}}
    = \frac{1}{N_\mathscr{V}}\sum_{i=1}^{N_\mathscr{V}} s_i^{\mathrm{out}} 
    = \frac{1}{N_\mathscr{V}}\sum_{i=1}^{N_\mathscr{V}}\sum_{j=1}^{N_\mathscr{V}}\left(w_{ij}\right)^1
    = \frac{W}{N_{\mathscr{V}}}.
    \label{Eq:DiGraph:sout:ave}
\end{equation}
Therefore, we have
\begin{equation}
    \left\langle{s^{\mathrm{in}}}\right\rangle_\mathscr{V} 
    = \left\langle{s^{\mathrm{out}}}\right\rangle_\mathscr{V} 
    = \frac{W}{N_\mathscr{V}}
    \label{Eq:DiGraph:sin:ave:sout:ave}
\end{equation}
and
\begin{equation}
    \left\langle{s}\right\rangle_\mathscr{V} 
    = \left\langle{s^{\mathrm{in}}+s^{\mathrm{out}}}\right\rangle_\mathscr{V} 
    = \frac{2W}{N_\mathscr{V}}.
    \label{Eq:DiGraph:s:ave}
\end{equation}

\subsubsection{Network density}

We use density to describe how connected nodes are in the network, which is defined as the portion of the potential links that are actual links:
\begin{equation}
    \rho = \frac{N_\mathscr{E}}{N_\mathscr{V}(N_\mathscr{V}-1)},
    \label{Eq:DiGraph:rho:t}
\end{equation}

\subsubsection{Link reciprocity}

Link reciprocity plays a pivotal role in shaping directed networks and is crucial for comprehending the observed network topology \cite{Garlaschelli-Loffredo-2004-PhysRevLett}. In conventional terms, the reciprocity of a node $i$ is defined as the ratio of the number of reciprocal links $k_i^R$ to the total number of links $k_i$, associated with node $i$ \cite{Newman-Forrest-Balthrop-2002-PhysRevE}:
\begin{equation}
    R_i = \frac{\sharp\left(\left\{j: e_{ij}\in{\mathscr{E}} ~\&~ e_{ji}\in{\mathscr{E}}\right\}\right)}
    {\sharp\left(\left\{j: e_{ij}\in{\mathscr{E}} ~{\mathrm{or}}~ e_{ji}\in{\mathscr{E}} \right\}\right)}
    = \frac{k_i^R}
    {k_i},
    \label{Eq:Net:Reciprocity:Node}
\end{equation}
where  
\begin{equation}
    k_i^R = \sharp\left(\left\{j: e_{ij}\in{\mathscr{E}} ~\&~ e_{ji}\in{\mathscr{E}}\right\}\right)
    = \sum_{j\neq{i}} \left(w_{ij} w_{ji}\right)^0
    \label{Eq:Net:Reciprocity:kiR}    
\end{equation}
is the number of reciprocal links node $i$ has. In Equation~(\ref{Eq:Net:Reciprocity:kiR}), we pose $0^0=0$. The investigation of link reciprocity provides valuable insights into the underlying dynamics and structure of complex networks, contributing to a more profound understanding of their behavior.

\subsubsection{Network efficiency}

Network efficiency, which is calculated as the average reciprocal of the shortest path lengths between pairs of nodes within a network \cite{Latora-Marchiori-2001-PhysRevLett,Zhou-Wang-Hang-2019-TRPE}. Here, we calculate the efficiency as follows:
\begin{equation}
  E =\frac1{N_{\mathscr{V}}(N_{\mathscr{V}}-1)}\sum_{i\neq j}\frac1{d_{ij}},
\label{Eq:efficiency}
\end{equation} 
where $d_{ij}$ is the shortest path between node $i$ and node $j$. In the iCTNs, efficiency serves as a measure of how effectively crops are transported.

\subsubsection{Natural connectivity}

Natural connectivity offers a valuable approach for studying network resilience, avoiding complex computations and instead relying on the inherent structure of the network. Its primary application lies in quantifying the redundancy of alternative pathways within the network. This is achieved by computing a weighted sum of closed walks of different lengths \cite{Wu-Barahona-Tan-Deng-2011-IEEETransSystManCybernPaartA-SystHum}. Mathematically, the initial natural connectivity can be defined as follows:
\begin{equation}
\begin{aligned}
  \label{Eq:connectivity:natural}
NC=\ln \left(\frac{SC}{N_{\mathscr{V}}}\right)=\ln \left(\frac{\sum_{i=1}^{N_{\mathscr{V}}} e^{\lambda_{i}}}{N_{\mathscr{V}}}\right), 
\end{aligned}
\end{equation}
where
\begin{equation}
\begin{aligned}
  \label{Eq:centrality:subgraph }
SC=\sum_{l=0}^{\infty} \frac{\mu_{l}}{l !}=\sum_{{i}=0}^{N_{\mathscr{V}}}\sum_{l=0}^{\infty} \frac{\lambda_{i}^{l}}{l !}=\sum_{i=1}^{N_{\mathscr{V}}} {e}^{\lambda_{i}},
\end{aligned}
\end{equation}
where $\mu_{l}$ is the number of closed walks of length $l$, $SC$ is the initial weighted sum of the numbers of closed walks \cite{Estrada-2000-ChemPhysLett}, and $\lambda$ is the eigenvalue of the adjacency matrix $A$.

\section{Empirical results}
\label{sec3:Results}
Russia and Ukraine are major global producers and exporters of crops \cite{Lin-Li-Jia-Feng-Huang-Huang-Fan-Ciais-Song-2023-GlobFoodSecur-AgricPolicy}, with their crops such as maize and wheat holding significant positions in the global food trade \cite{Bentley-Donovan-Sonder-Baudron-Lewis-Voss-Rutsaert-Poole-Kamoun-Saunders-Hodson-Hughes-Negra-Ibba-Snapp-Sida-Jaleta-Tesfaye-BeckerReshef-Govaerts-2022-NatFood}. The ongoing conflict between Russia and Ukraine is likely to exacerbate global food supply shortages, leading to a rise in food prices and endangering the food security. In this study, we analyze the evolution of the structural characteristics of three iCTNs and the iFTN from January 2016 to December 2022. To better understand the impact of the Russia-Ukraine conflict, We focus on two time points before and after the outbreak of the Russia-Ukraine conflict (the second, third and forth quarters of 2021 and 2022, respectively).

\subsection{Impact on the iCTNs}
\label{subsec3:iCTNs}

\begin{figure}[h!]
    \centering
    \includegraphics[width=0.321\linewidth]{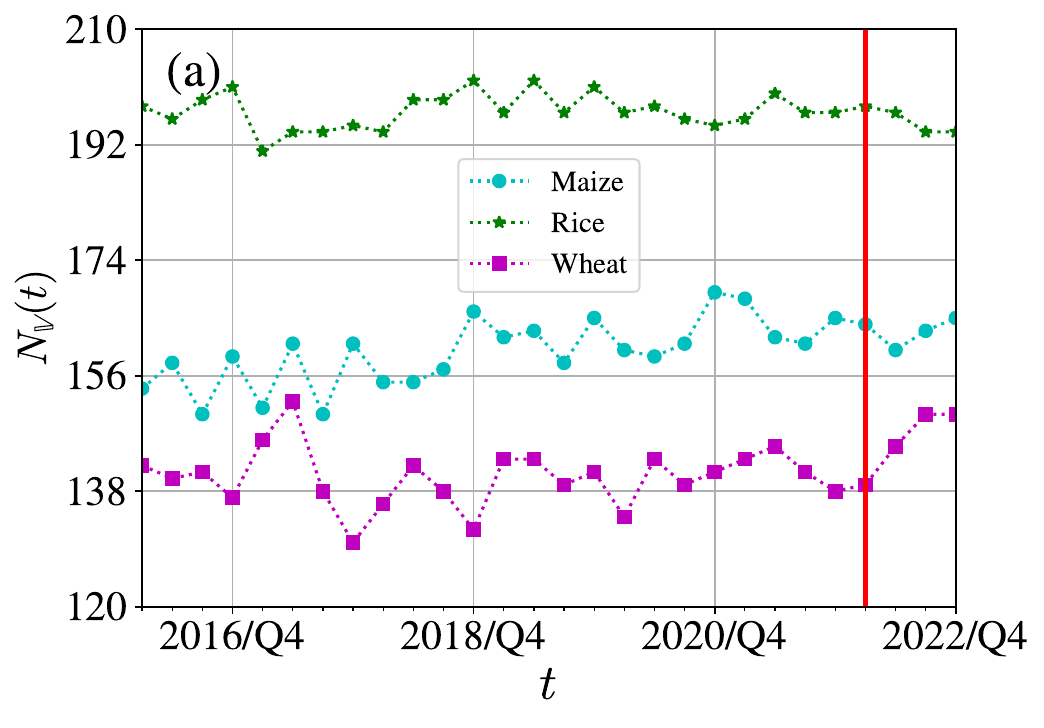}   
    \includegraphics[width=0.321\linewidth]{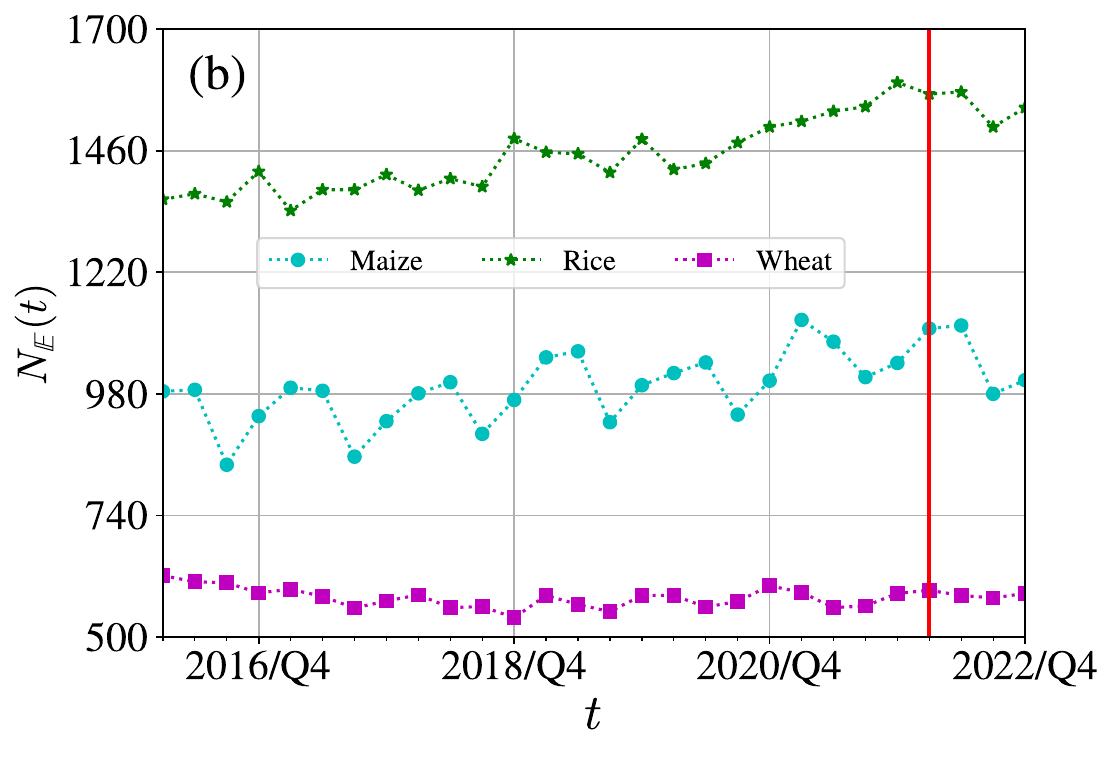}
    \includegraphics[width=0.321\linewidth]{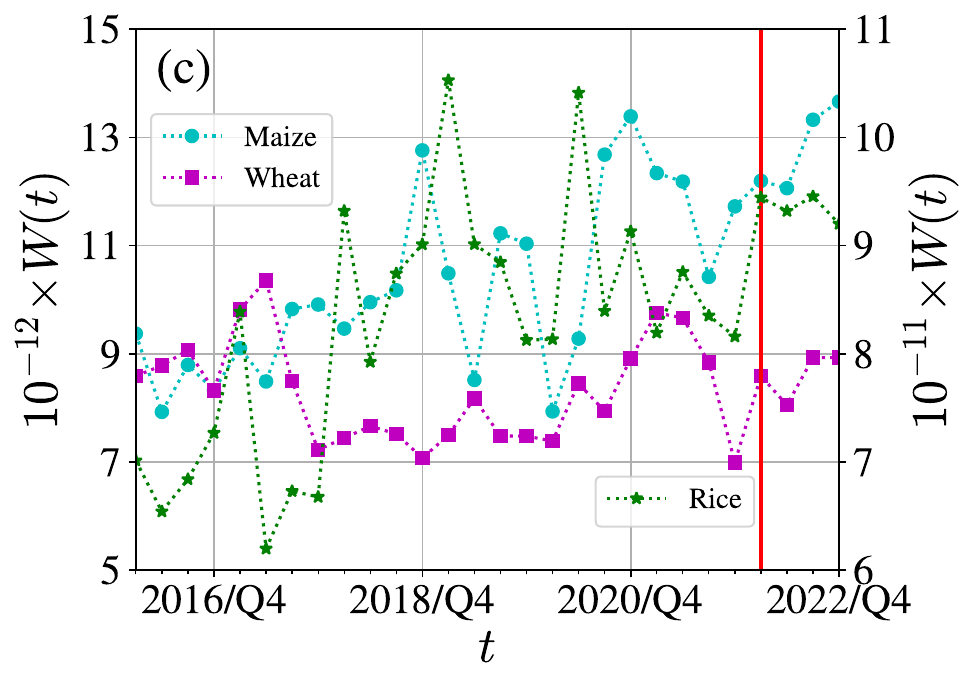}   
    \includegraphics[width=0.321\linewidth]{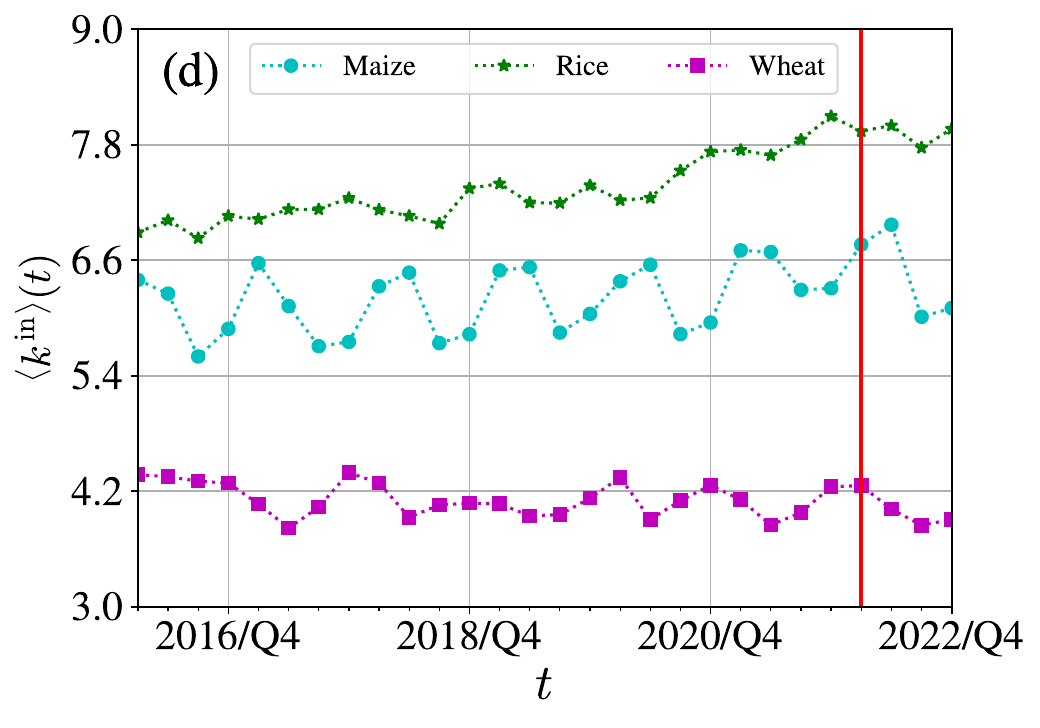}   
    \includegraphics[width=0.321\linewidth]{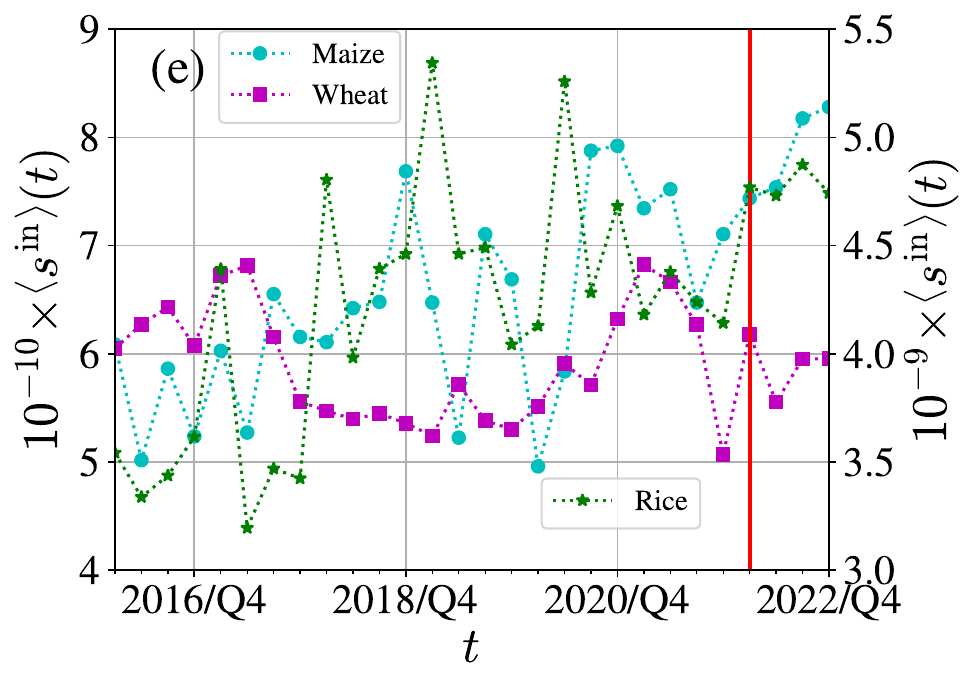}   
    \includegraphics[width=0.321\linewidth]{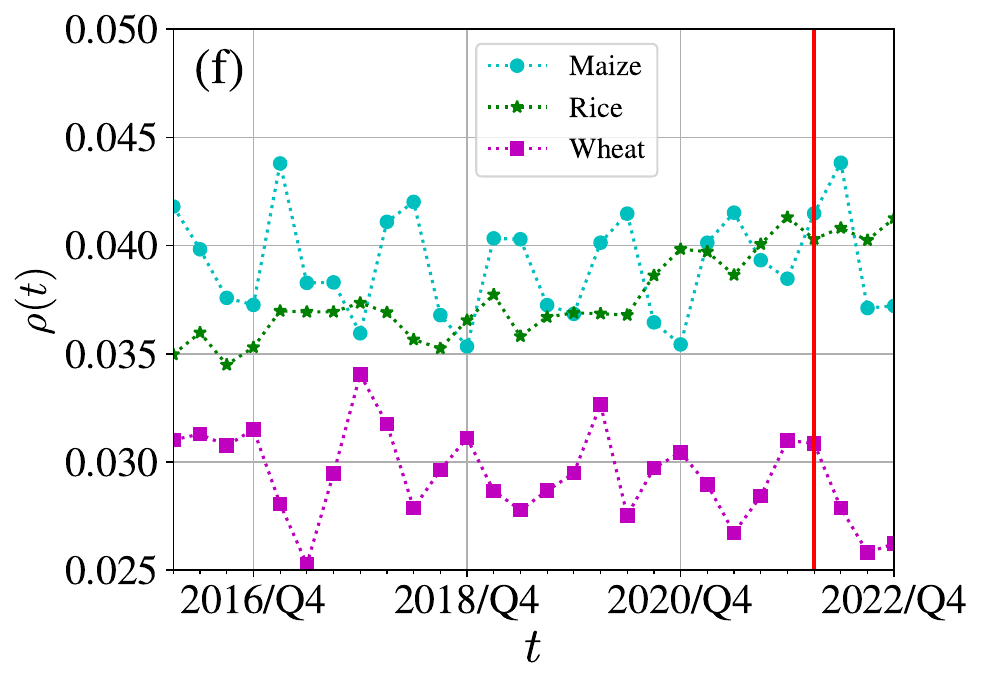}
    \includegraphics[width=0.321\linewidth]{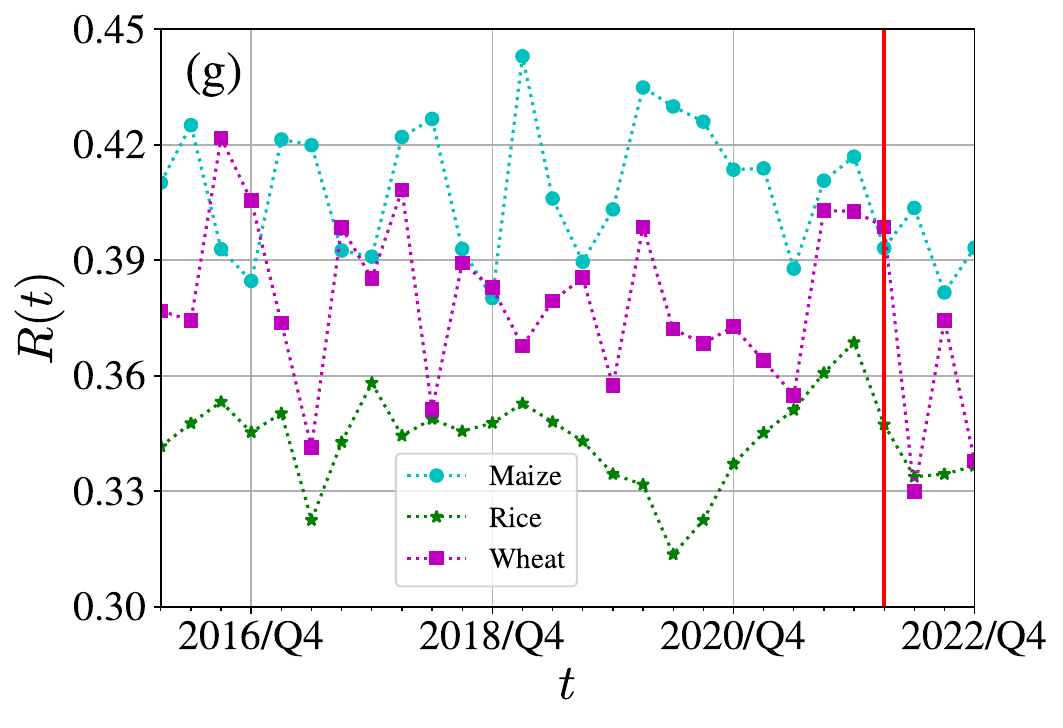}  
    \includegraphics[width=0.321\linewidth]{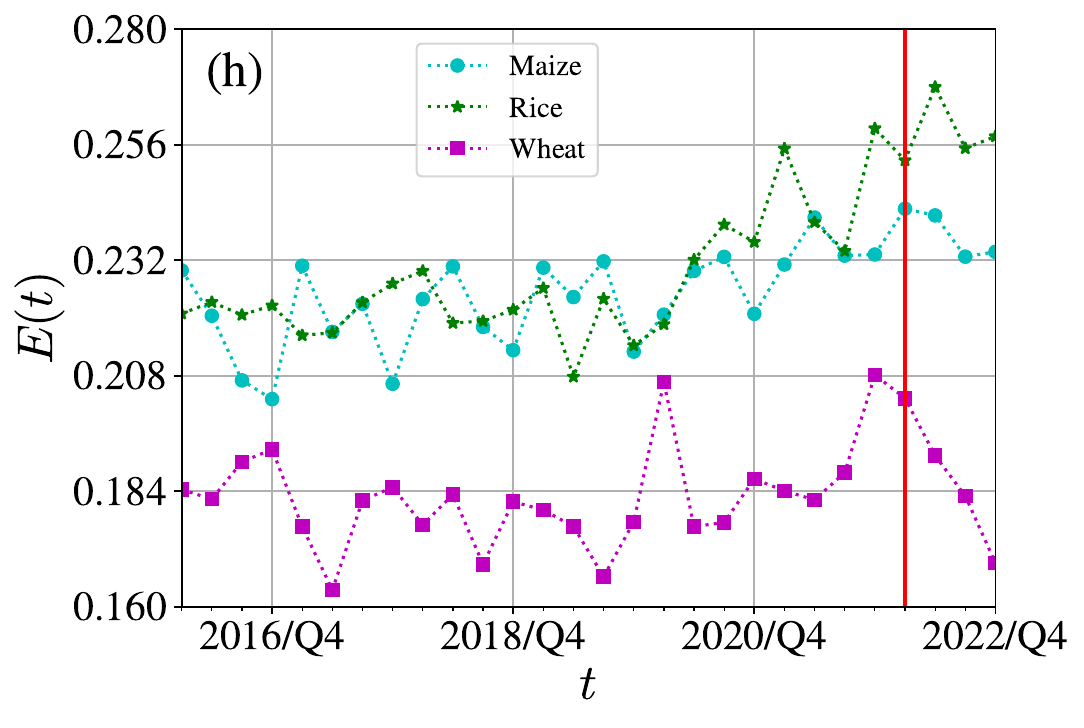}
    \includegraphics[width=0.321\linewidth]{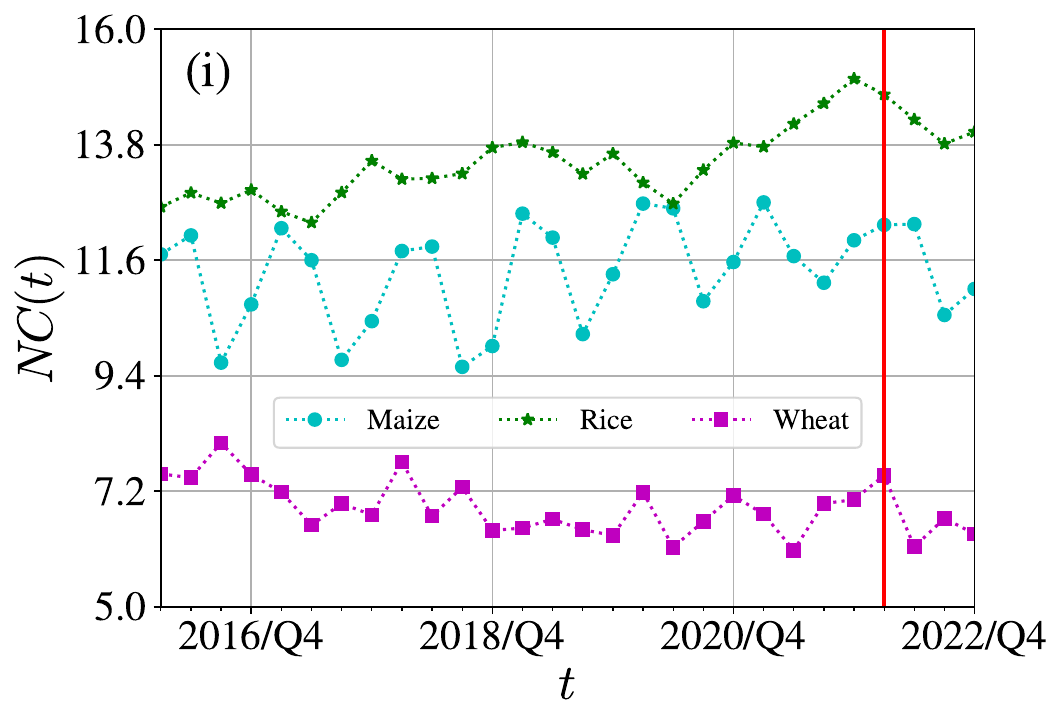}
    \caption{Quarterly evolution of the structure of the three iCTNs from 2016/Q1 to 2022/Q4. The structural metrics include: (a) number of nodes $N_{\mathscr{V}}$, (b) number of links $N_{\mathscr{E}}$, (c) total link weight $W$ in units of calories, (d) average in-degree $\left\langle{k^{\mathrm{in}}}\right\rangle$, (e) average in-strength $\left\langle{k^{\mathrm{in}}}\right\rangle$, (f) network density $\rho$, (g) link reciprocity $R$, (h) network efficiency $E$, and (i) natural connectivity $NC$. Curves with different colored markers correspond to different crops. The vertical red line corresponds to 24 February 2022 when the Russia-Ukraine conflict began.}
    \label{Fig:iCTN:NetStructure:t}
\end{figure}

We present the evolution of three iCTNs over the period from 2016/Q1  to 2022/Q4, highlighting various network properties such as the number of nodes, edges, total edge weight, average degree, average strength, density, clustering coefficient, efficiency, and natural connectivity, as shown in Fig.~\ref{Fig:iCTN:NetStructure:t}. Figures.~\ref{Fig:iCTN:NetStructure:t}(a-b) and (d) show that the numbers of nodes and edges do not have a significant change. It indicates that the trade relationships between these 24 economies in the iCTNs are stable. It is consistent with the overall trend of the density (see Fig.~\ref{Fig:iCTN:NetStructure:t}(f)). However, the density of the iWTN decreased significantly after the first quarter of 2022. It would reflect the impact of the Russia-Ukraine conflict on the international wheat trade. From Figs.~\ref{Fig:iCTN:NetStructure:t}(c) and (e), we find that the structure of the three iCTNs has significant seasonal fluctuations from 2016 to 2022 but has an increasing trend. Figure.~\ref{Fig:iCTN:NetStructure:t}(g) presents that the properties of bidirectional trading relationships between economies in the iCTNs are stable. From Fig.~\ref{Fig:iCTN:NetStructure:t}(h), we can see that the efficiency of the iCTNs increased. It implies that there are multiple efficient pathways for propagating crop trade flows. What’s more, natural connectivity is stable in general.

To better understand the structural changes in the iCTNs before and after the Russia-Ukraine conflict, we calculate percentage change $r(t)$ to compare the network's structure between two consecutive quarters spanning the period from 2016 to 2022,
\begin{equation}
    r(t)=\frac{x(t)-x(t-1)}{x(t-1)}  
    \label{Eq:PercentageChange }
\end{equation}
where $x(t)$ means the value of the topological metric at time $t$. We focus on topological metrics that have undergone notable changes following the Russia-Ukraine conflict.

\begin{figure}[h!]
    \centering
    \includegraphics[width=0.431\linewidth]{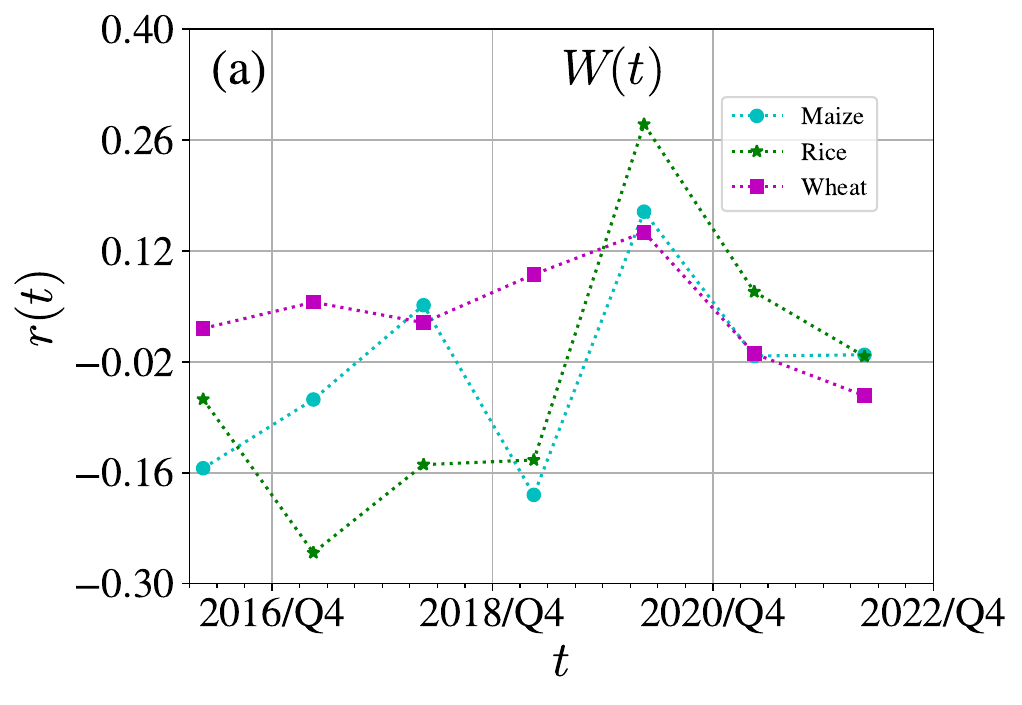}   
    \includegraphics[width=0.431\linewidth]{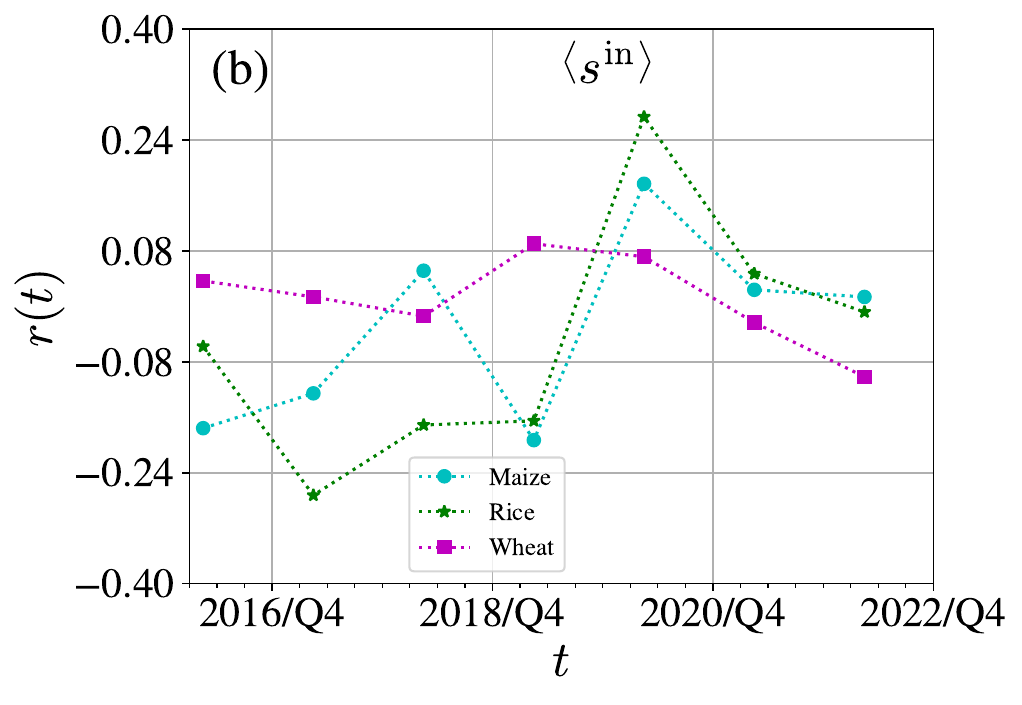}   
    \includegraphics[width=0.431\linewidth]{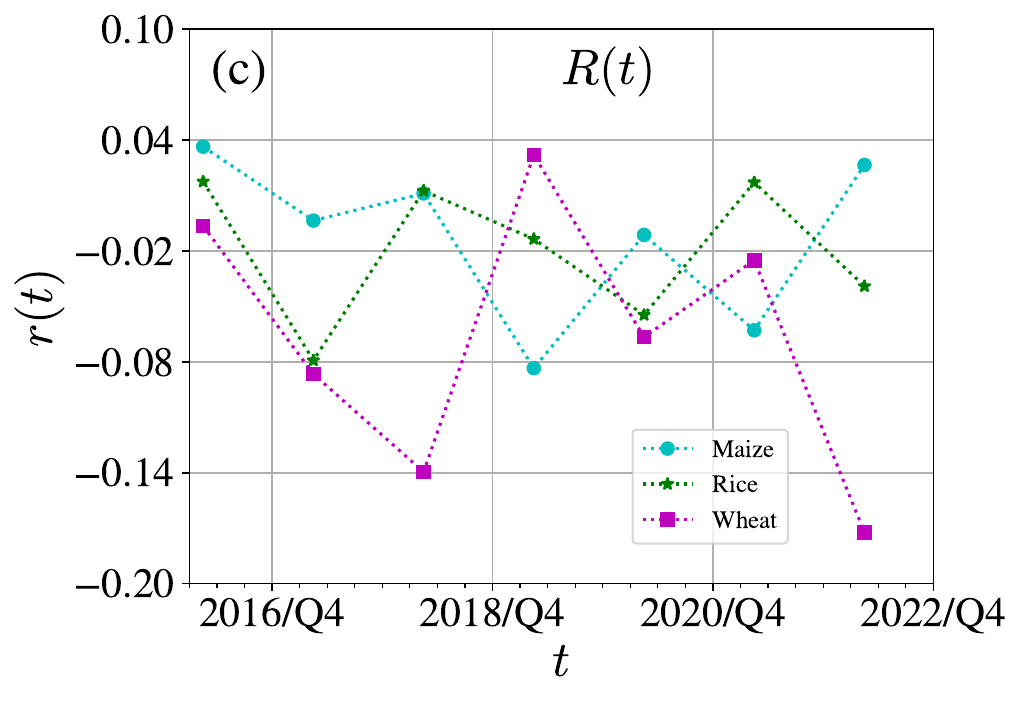}   
    \includegraphics[width=0.431\linewidth]{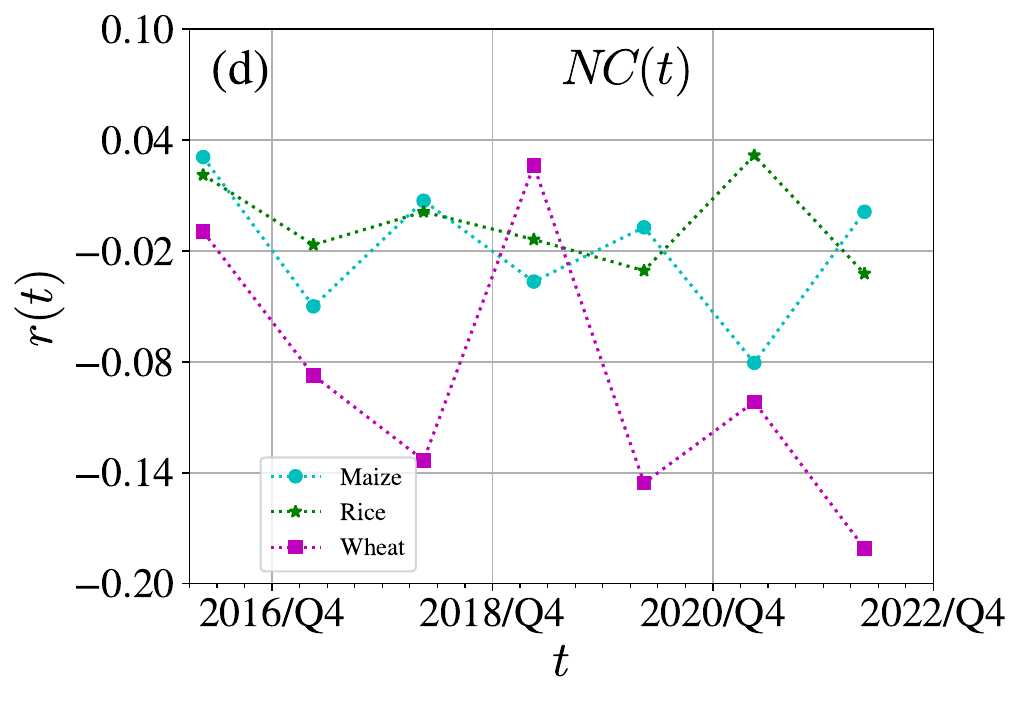}
    \caption{Percentage change in the structure of the iCTNs comparing two adjacent quarters (the first and second quarter) from 2016 to 2022. $r(t)=(x(t)-x(t-1))/x(t-1)$, where $x(t)$ means the value of the topological metric at time $t$ and $t=$2016/Q2, 2017/Q2, $\cdots$,  2022/Q2. We select topological metrics that have changed significantly after the Russia-Ukraine conflict. Topological metrics including: (a) percentage change in total link weight $W$ in units of calories, (b) percentage change in average in-strength $\left\langle{s^{\mathrm{in}}}\right\rangle$ comparing the first and the second quarter, (c) percentage change in link reciprocity $R$ comparing the first and the second quarter, (d) percentage change in natural connectivity $NC$ comparing the first and the second quarter, and (d) percentage change in average in-degree $\left\langle{k^{\mathrm{in}}}\right\rangle$.}
    \label{Fig:iCTN:NetStructure:D12}
\end{figure}

\begin{figure}[h!]
    \centering
    \includegraphics[width=0.321\linewidth]{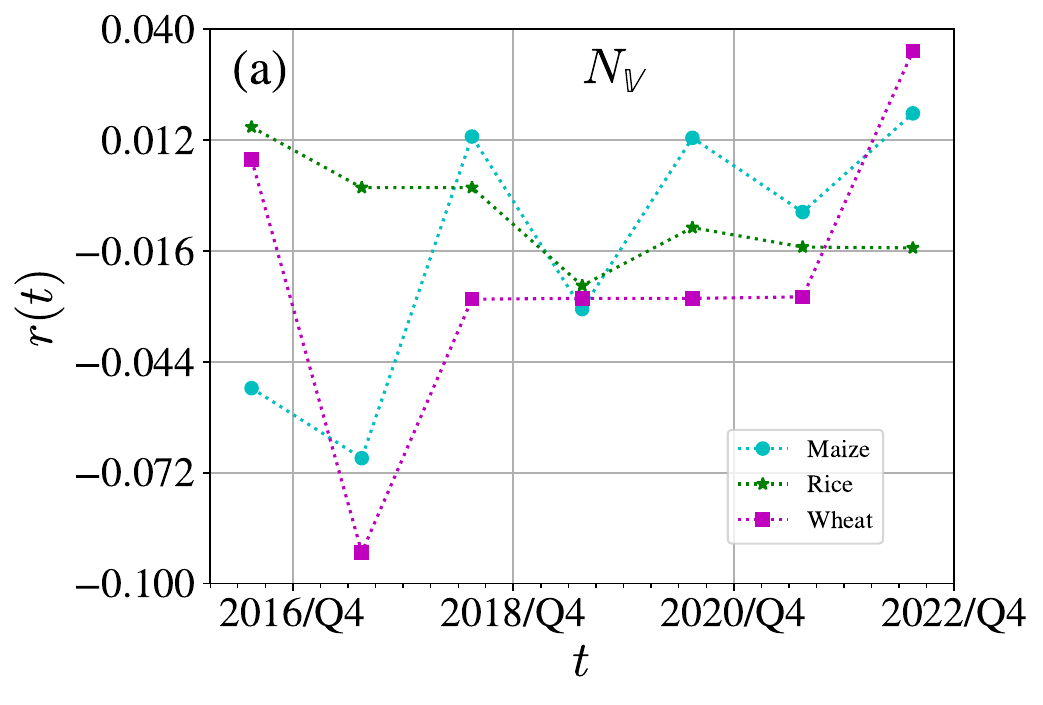}   
    \includegraphics[width=0.321\linewidth]{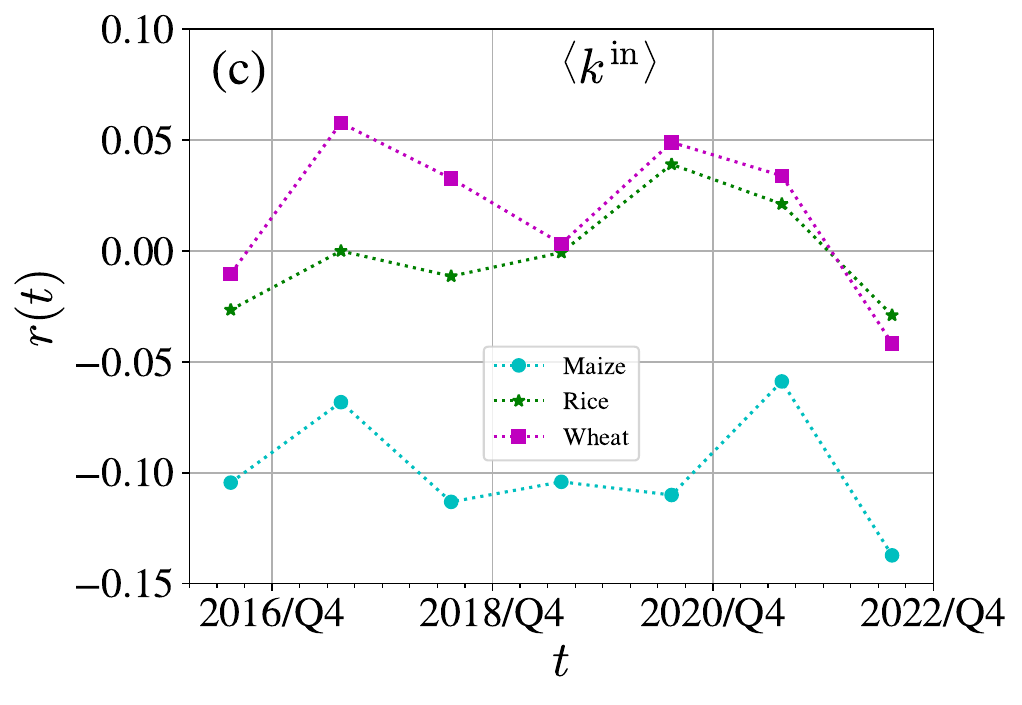}
    \includegraphics[width=0.321\linewidth]{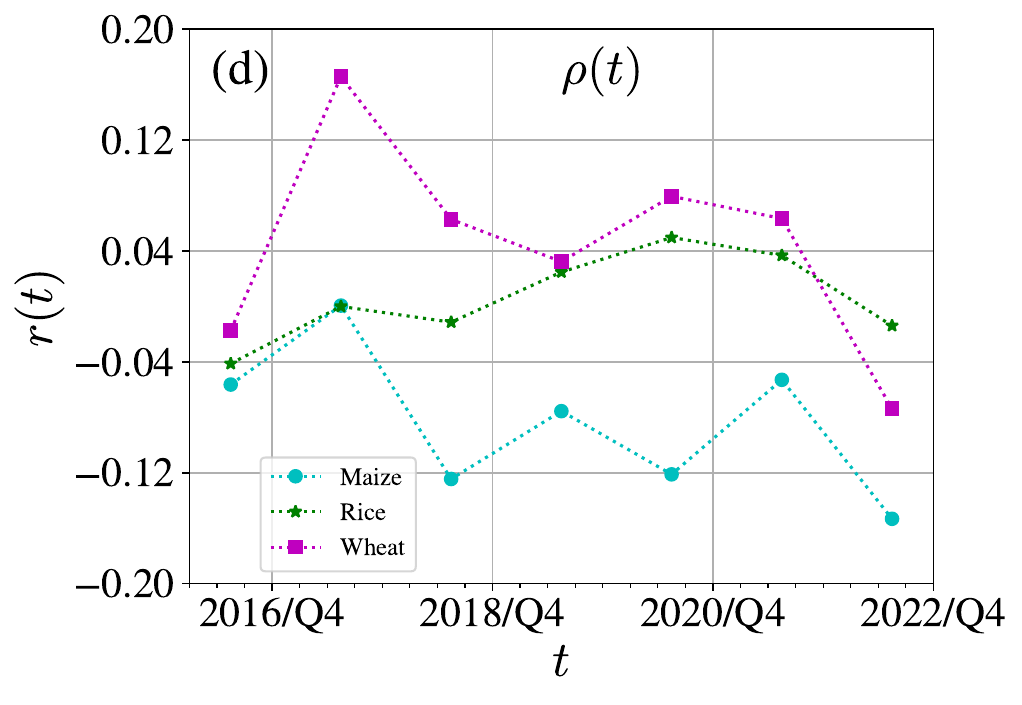}
    \caption{Percentage change in the structure of the international food trade network comparing two adjacent quarters (the second and third quarter) from 2016 to 2022. $r(t)=(x(t)-x(t-1))/x(t-1)$, where $x(t)$ means the value of the topological metric at time $t$. The first to third rows correspond to $t=$2017/Q3, $\cdots$, 2022/Q3. We select 
   topological metrics that have changed significantly after the Russia-Ukraine conflict. Topological metrics including: (a) percentage change in number of nodes $N_{\mathscr{V}}$, (b) percentage change in average in-degree $\left\langle{k^{\mathrm{in}}}\right\rangle$, and (c) percentage change in network density $\rho$.}
    \label{Fig:iCTN:NetStructure:D23}
\end{figure}

The Russia-Ukraine conflict has had a more significant impact on the structure of the iWTN than the iMTN and iRTN in a short time. 
Figure~\ref{Fig:iCTN:NetStructure:D12} provides an overview of the percentage change in link weights, average in-strength, link reciprocity, and natural connectivity between the first and second quarters from 2016 to 2022. In 2022/Q2, particularly for the iWTN, both link weights and average in-strength experienced a significant decrease—marking the first quarter after the Russia-Ukraine conflict. Although the iMTN and iRTN witnessed more substantial decreases in these two metrics before 2022, it remains inconclusive whether these changes were directly influenced by the Russia-Ukraine conflict. However, the iWTN exhibited a noteworthy decline in link weights and average in-strength specifically during 2022/Q2 compared to 2022/Q1, suggesting a negative impact on international wheat trade stemming from the conflict. Furthermore, both link reciprocity and natural connectivity of the iWTN experienced significant decreases in 2018/Q2 and 2022/Q2. It indicates that the connectivity of the iWTN decreased. The reason would be that Northern and Eastern Europe experienced wheat yield losses in 2018 due to extreme weather \cite{Clarke-Hess-HaroMonteagudo-Semenov-Knox-2021-AgricForMeteorol}. Meanwhile, the conflict has also disrupted the structure of the iWTN.

All three iCTNs experienced disturbances within six months following the conflict. Figure~\ref{Fig:iCTN:NetStructure:D23} shows the evolution of percentage change in number of nodes, average in-degree, and density. It is interesting that the number of nodes increased instead of decreasing in 2022/Q3. Notably, the iWTN experienced a significant rise in the number of nodes during this quarter. Conversely, both the average in-degree and density of the iCTNs exhibited a decrease. These findings indicate that the conflict influenced the structure and connectivity of the iCTNs, but it did not impede the participation of economies in international crop trade.

\begin{figure}[h!]
    \centering
    \includegraphics[width=0.231\linewidth]{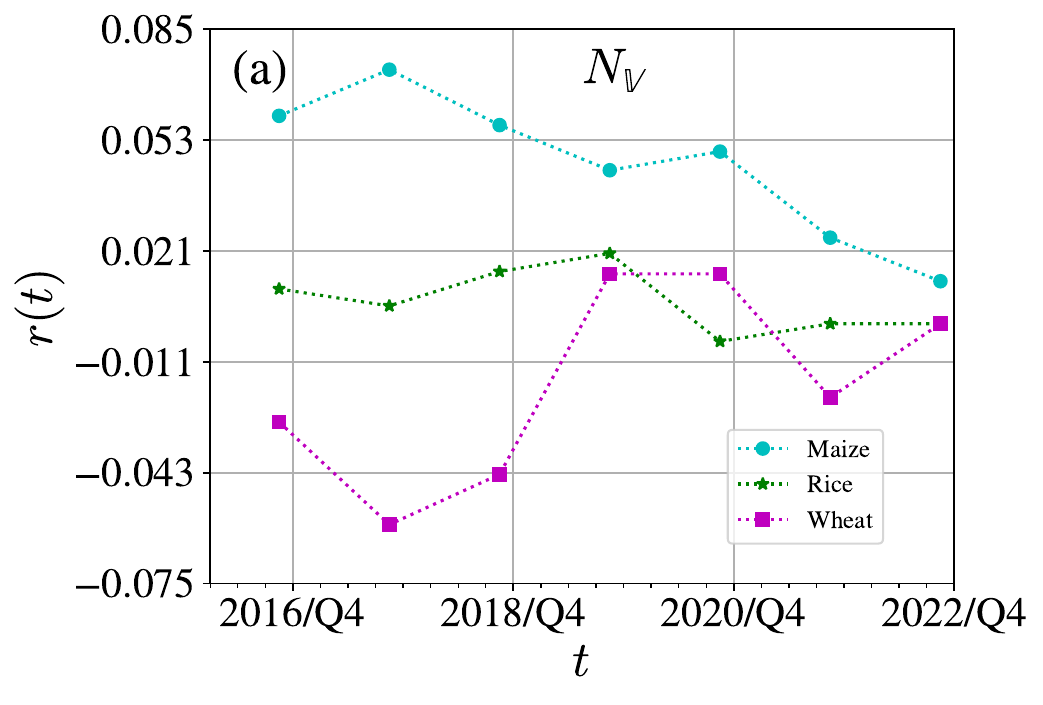}   
    \includegraphics[width=0.231\linewidth]{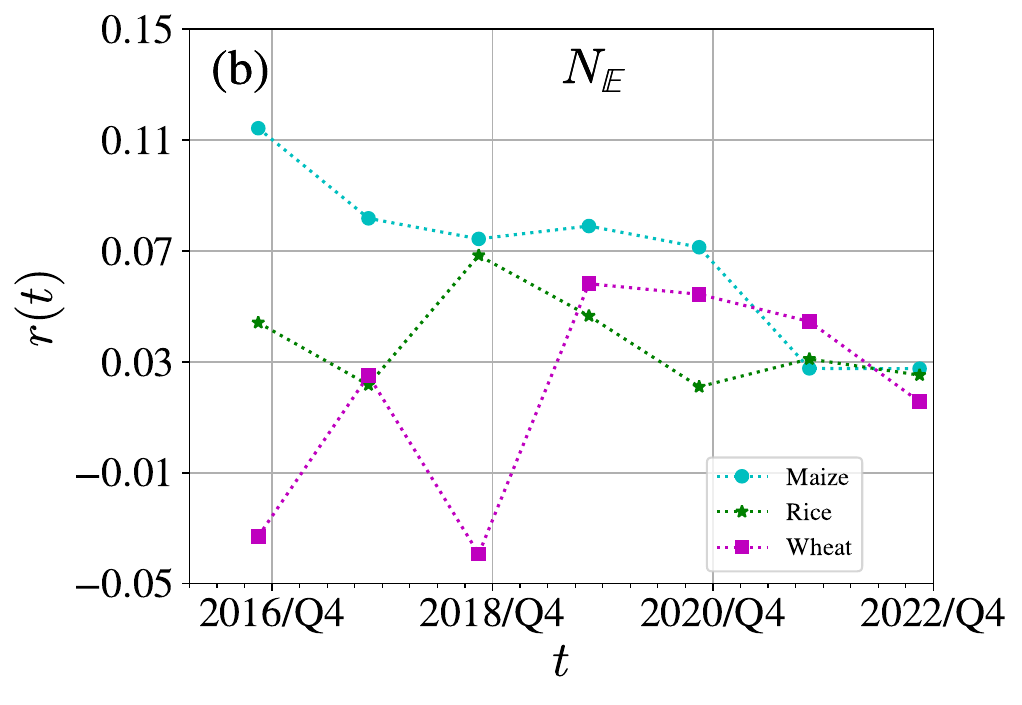}
    \includegraphics[width=0.231\linewidth]{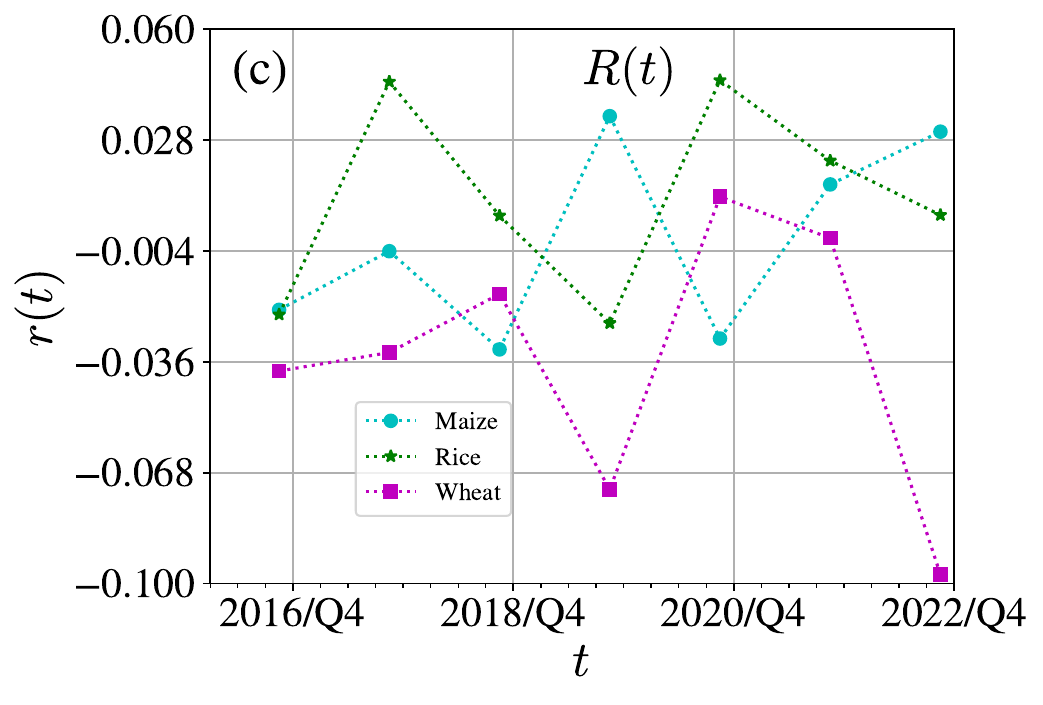}   
    \includegraphics[width=0.231\linewidth]{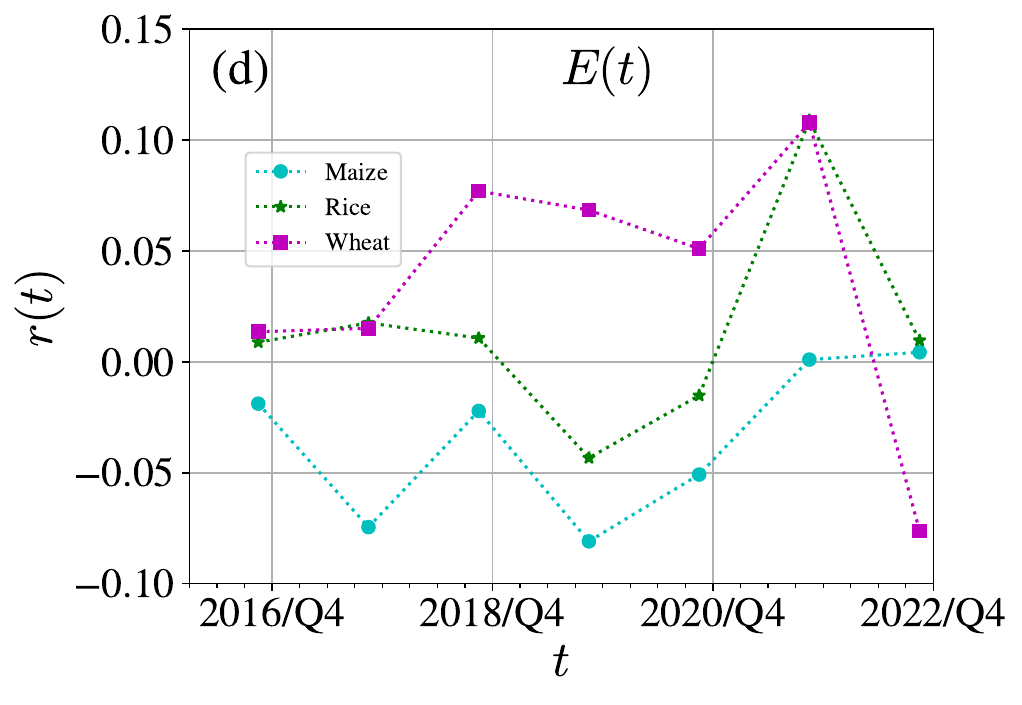}
    \caption{Percentage change in the structure of the international food trade network comparing two adjacent quarters  (the third and fourth quarter) from 2016 to 2022. $r(t)=(x(t)-x(t-1))/x(t-1)$, where $x(t)$ means the value of the topological metric at time $t$ and $t=2017/Q4, \cdots, 2022/Q4$. We select topological metrics that have changed significantly after the Russia-Ukraine conflict. Topological metrics including: (a) percentage change in number of nodes $N_{\mathscr{V}}$, (b) percentage change in number of links $N_{\mathscr{E}}$, (c) percentage change in link reciprocity $R$, and (d) efficiency $E$.}
    \label{Fig:iCTN:NetStructure:D34}
\end{figure}

The ongoing Russia-Ukraine conflict has consistently exerted a negative impact on the iCTNs, particularly on the iWTN. As depicted in Figs.~\ref{Fig:iCTN:NetStructure:D34}(a-b), both the number of nodes and edges of the iCTNs decreased in 2022/Q4. This is in part because the escalating conflict has hindered the development of free trade. From Figs.~\ref{Fig:iCTN:NetStructure:D34}(c-d), the link reciprocity and efficiency of the iWTN showed a more substantial reduction in 2022/Q4. 

Overall, the impact of the Russia-Ukraine conflict on the iCTNs differs across different crop sectors. The iWTN is shown to face the most severe disturbance. Wheat yield losses in Ukraine and wheat export restrictions would explain the shifts in the structure of the iWTN. Russia and Ukraine are major wheat producers and traders, especially for European economies. Both Russia and Ukraine have imposed sanctions on each other, including trade restrictions. These measures have hindered the flow of wheat and other agricultural products between the economies and affected the availability of wheat in European markets.

\subsection{Impact on the iFTN}
\label{subsec3:iFTN}

\begin{figure}[h!]
    \centering
    \includegraphics[width=0.321\linewidth]{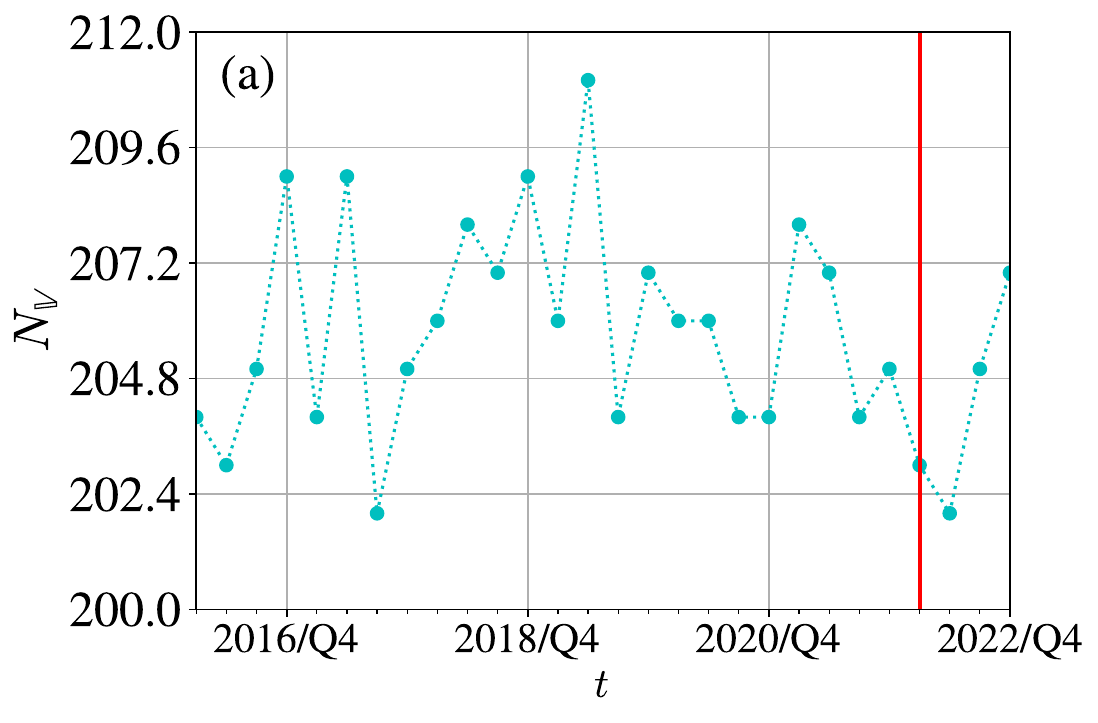}   
    \includegraphics[width=0.321\linewidth]{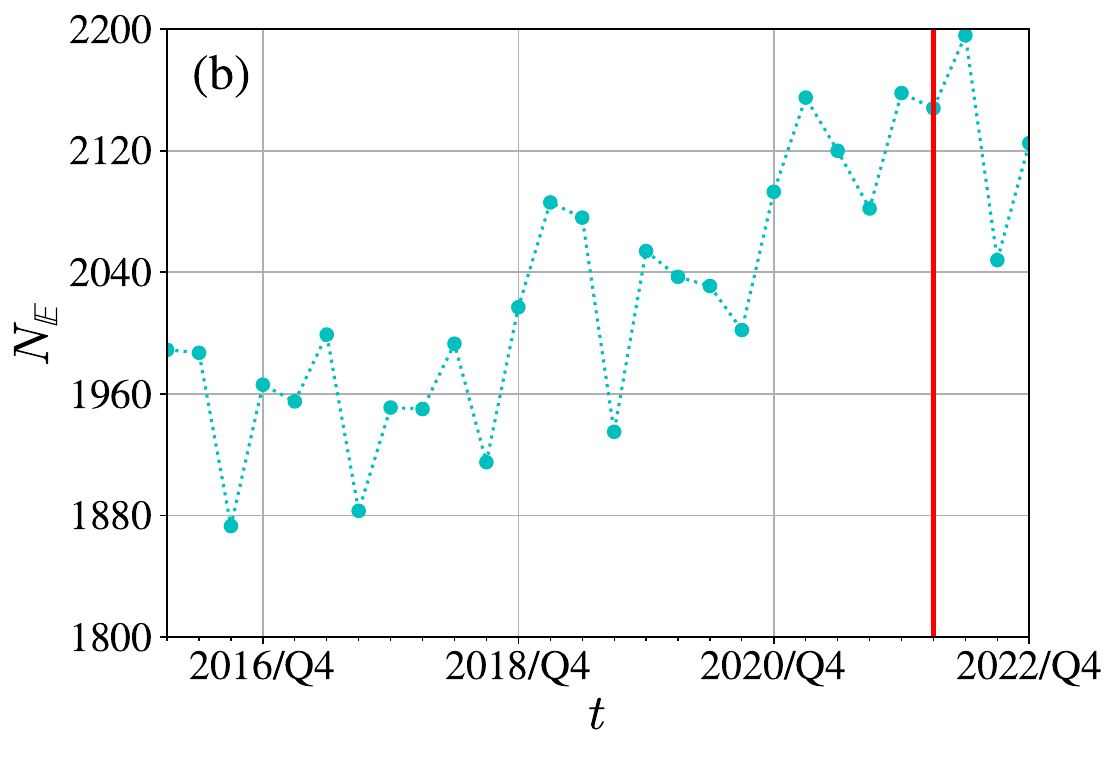}
    \includegraphics[width=0.321\linewidth]{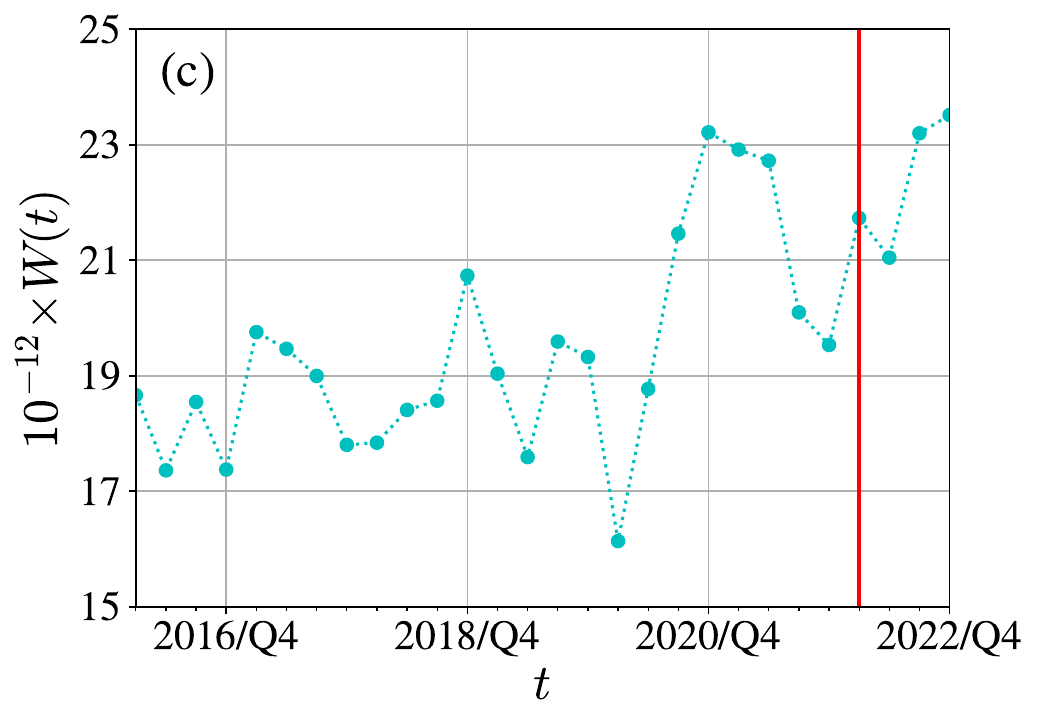}   
    \includegraphics[width=0.321\linewidth]{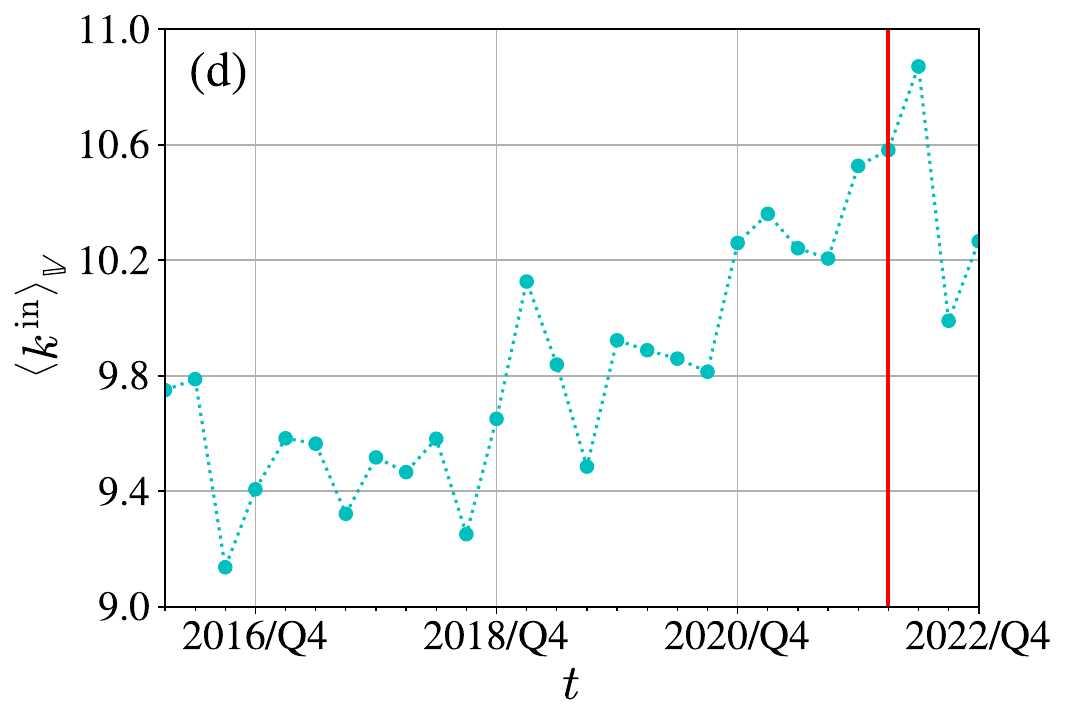}   
    \includegraphics[width=0.321\linewidth]{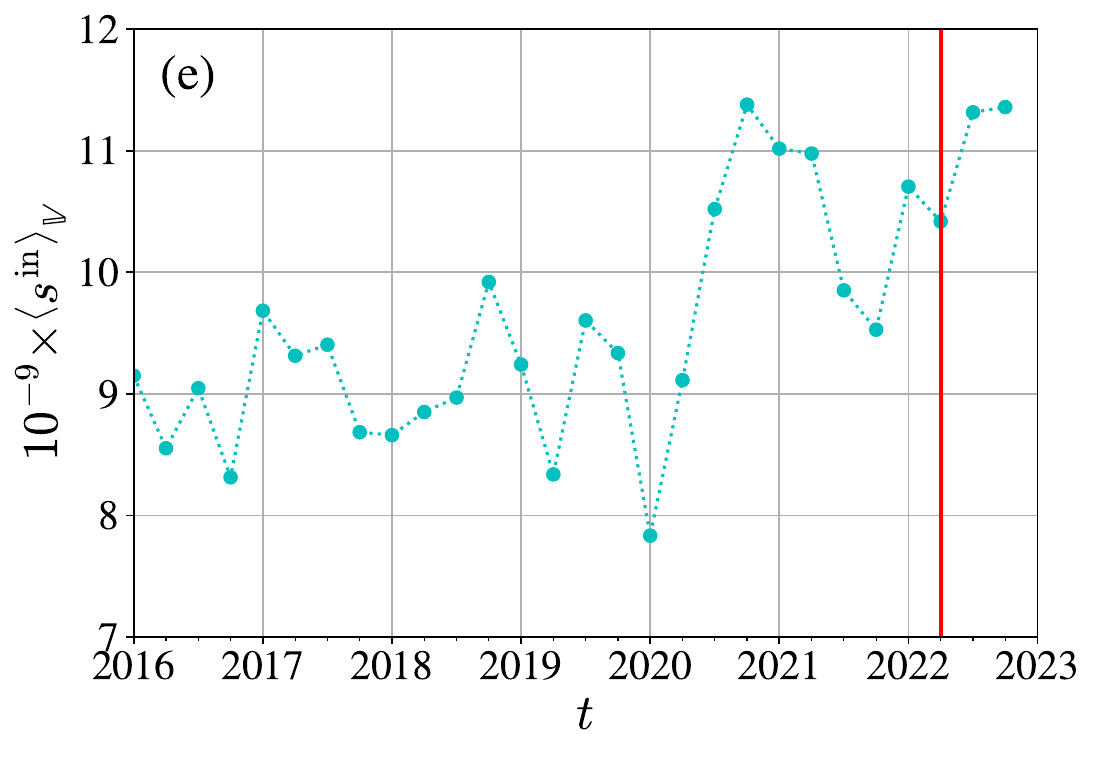}   
    \includegraphics[width=0.321\linewidth]{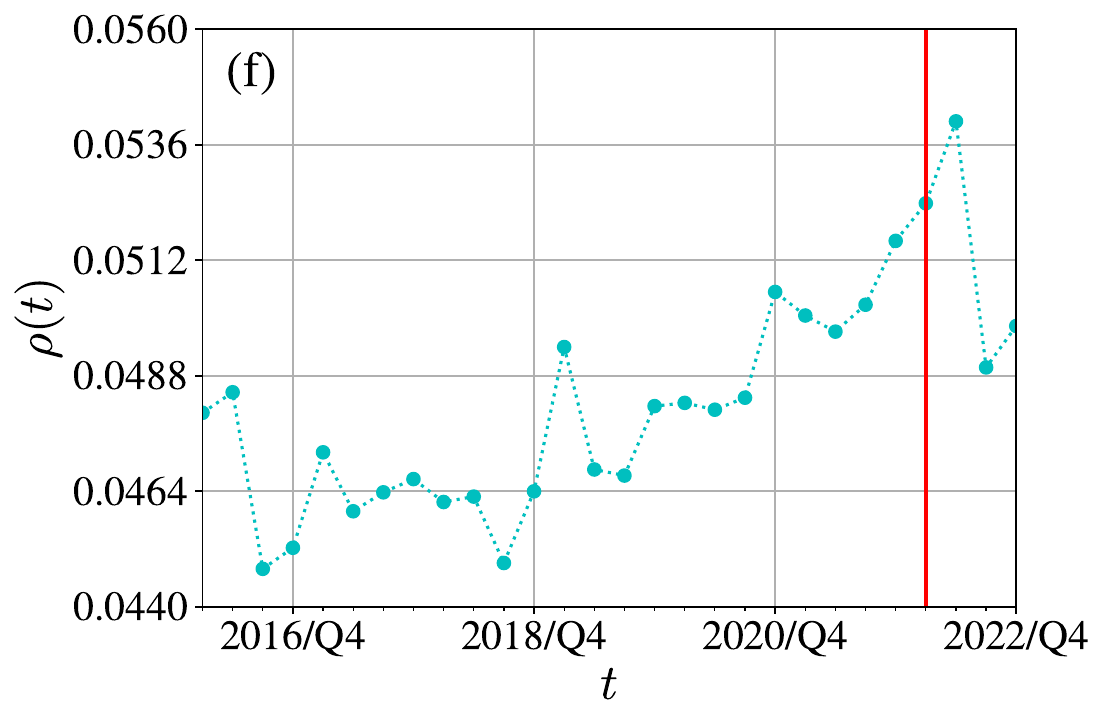}
    \includegraphics[width=0.321\linewidth]{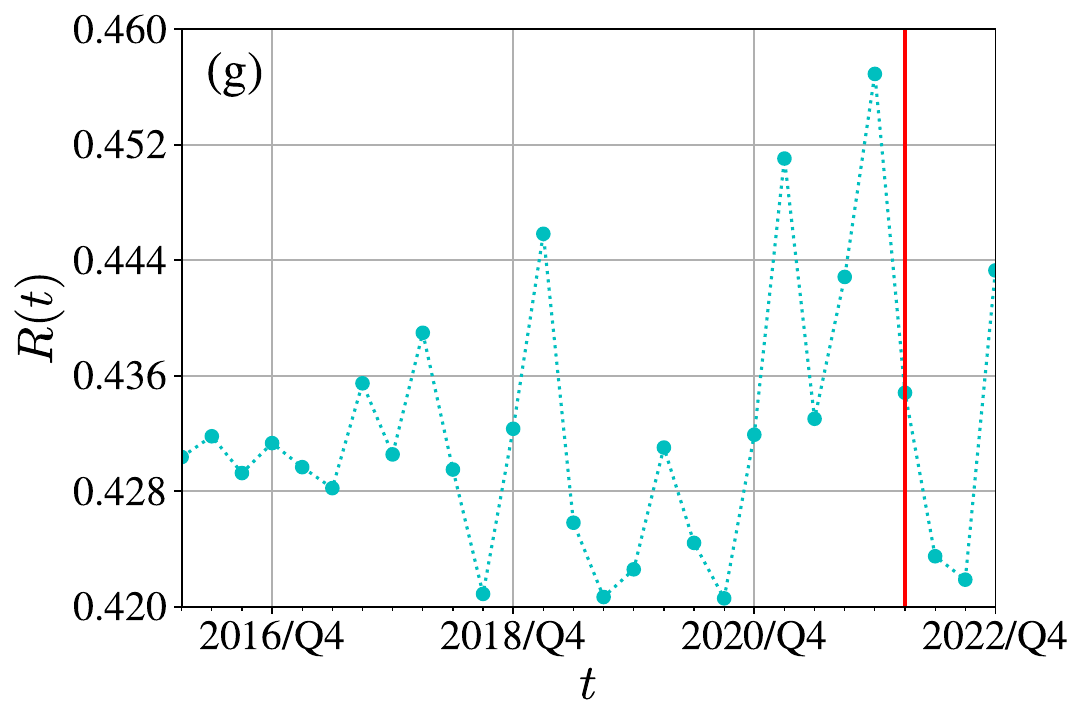}   
    \includegraphics[width=0.321\linewidth]{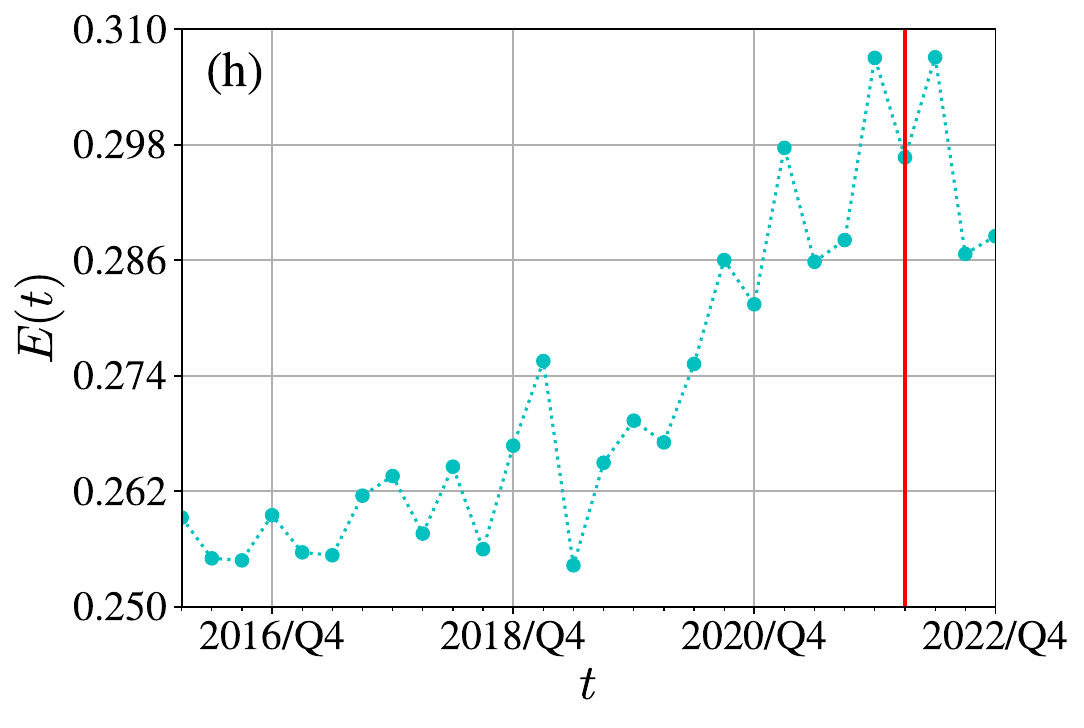}
    \includegraphics[width=0.321\linewidth]{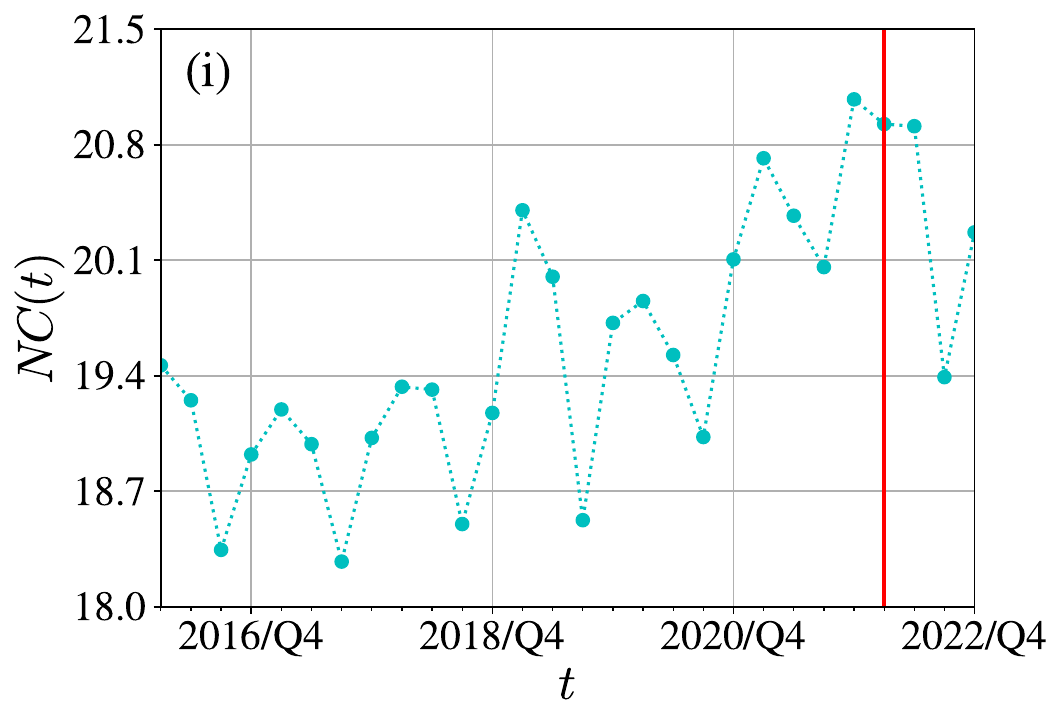}
    \caption{Quarterly evolution of the structure of the iFTN from 2016/Q1 to 2022/Q4. (a) number of nodes $N_{\mathscr{V}}$, (b) number of links $N_{\mathscr{E}}$, (c) total link weight $W$ in units of calories, (d) average in-degree $\left\langle{k^{\mathrm{in}}}\right\rangle$, (e) average in-strength $\left\langle{k^{\mathrm{in}}}\right\rangle$, (f) network density $\rho$, (g) link reciprocity $R$, (h) network efficiency $E$, and (i) natural connectivity $NC$. The purple line shows the month when Russia-Ukraine conflict began.}
    \label{Fig:iFTN:NetStructure:t}
\end{figure}

We construct the iFTN by aggregating maize, rice, and wheat to gain some insight into the impact of the Russia-Ukraine conflict on the total crop trade. 

The number of economies included in the iFTN was not affected by the conflict. We show the quarterly evolution of the structure of the iFTN from 2016/Q1 to 2022/Q4 in Fig.~\ref{Fig:iFTN:NetStructure:t}. We find that all topological metrics display an overall upward trend, which differs from the behavior observed in the individual iCTNs. Notably, the number of nodes, the link weights, the average in-strength, and the link reciprocity maintained a consistent trend in 2022. Nevertheless, changes were observed in the number of edges, average in-degree, density, efficiency, and natural connectivity following the onset of the conflict.

\begin{figure}[h!]
    \centering
    \includegraphics[width=0.475\linewidth]{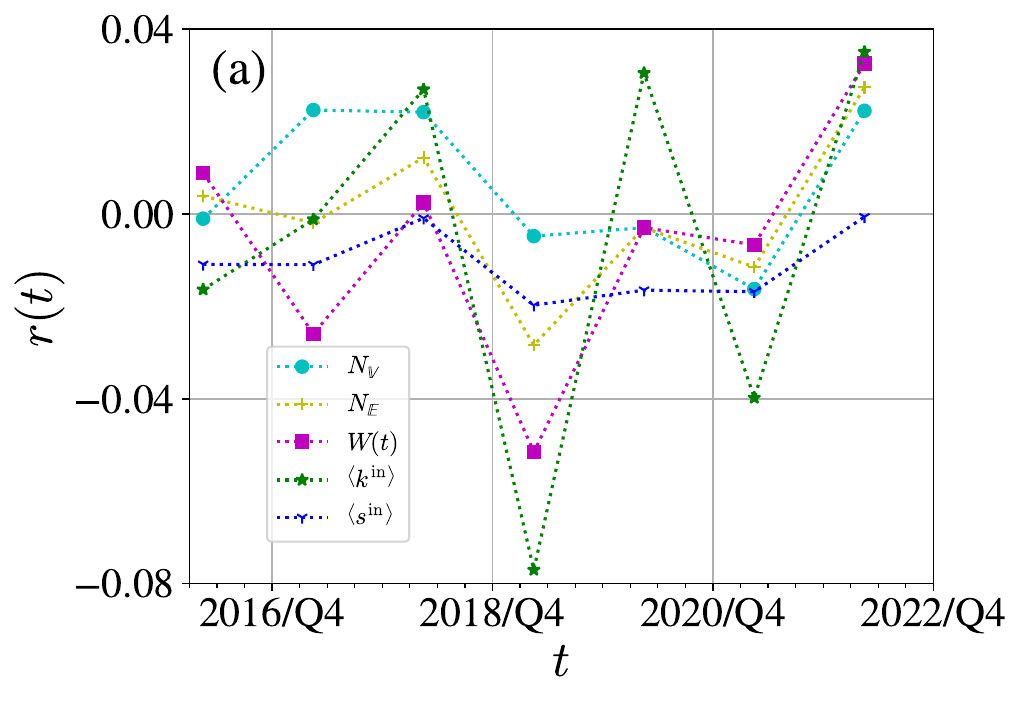} 
    \includegraphics[width=0.475\linewidth]{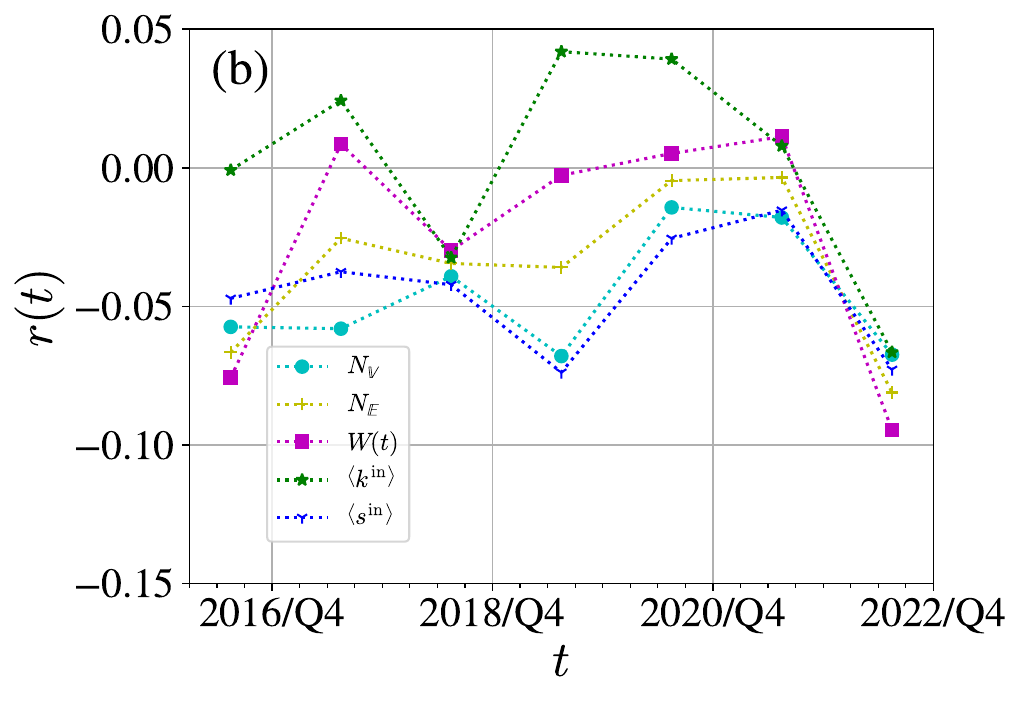}   
    \caption{Percentage change $r(t)$ in the structure of the iFTN comparing the first quarter with the second quarter (a) and the second quarter with the third quarter (b) from 2016 to 2022.}
    \label{Fig:iFTN:NetStructure:D:12_34}
\end{figure}

Our focus is solely on analyzing the structural changes in the iFTN during the ongoing conflict. Figures~\ref{Fig:iFTN:NetStructure:D:12_34}(a) and (b) illustrate the percentage change ($r(t)$) in the iFTN's structure, comparing the first quarter to the second quarter and the second quarter to the third quarter, respectively, from 2016 to 2022. Due to the absence of significant structural changes between the third and fourth quarters, we omit these results. It is evident that in 2019/Q2, there was a significant decrease in the number of edges, average in-degree, density, efficiency, and natural connectivity. Another notable finding is that these network metrics increased in 2022/Q2 but decreased in 2022/Q3. This suggests that the conflict initially had a minimal impact on the iFTN within a short timeframe but eventually had a negative effect as the conflict endured. This finding contrasts somewhat with the results of the iWTN, which show a decrease in connectivity. One possible explanation for this disparity lies in the substitution effect among maize, rice, and wheat \cite{LiverpoolTasie-Reardon-Parkhi-Dolislager-2023-NatFood}. Since these staple crops can often be used as substitutes for one another in various food products, when the production or exports of wheat decline, consumers and businesses tend to shift their preferences to alternative crops like rice and maize. This increased demand for rice and maize can lead to higher exports of these crops to meet international market demand. Consequently, the iFTN was not severely affected in the short term.

\subsection{Impact on economies}    
\label{subsec3:economies}

To gain a deeper understanding of the Russia-Ukraine conflict’s impact across economies, we conduct economy-scale analyses from two aspects, including the topological properties of Russia and Ukraine and the trade relationships between NATO economies and both Russia and Ukraine. 
Russia and Ukraine present different responses to the conflict.

\begin{figure}[h!]
    \subfigbottomskip=-1pt
    \subfigcapskip=-5pt
    \centering
    \subfigure[]{\label{level.sub.9}\includegraphics[width=0.237\linewidth]{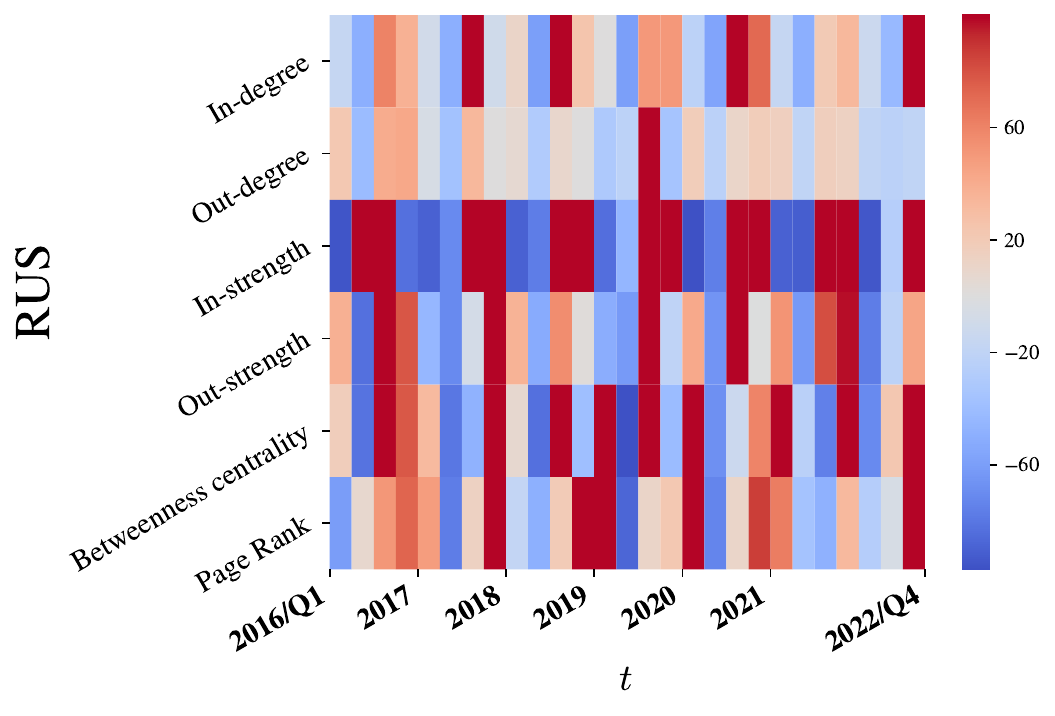} }     \subfigure[]{\label{level.sub.10}\includegraphics[width=0.237\linewidth]{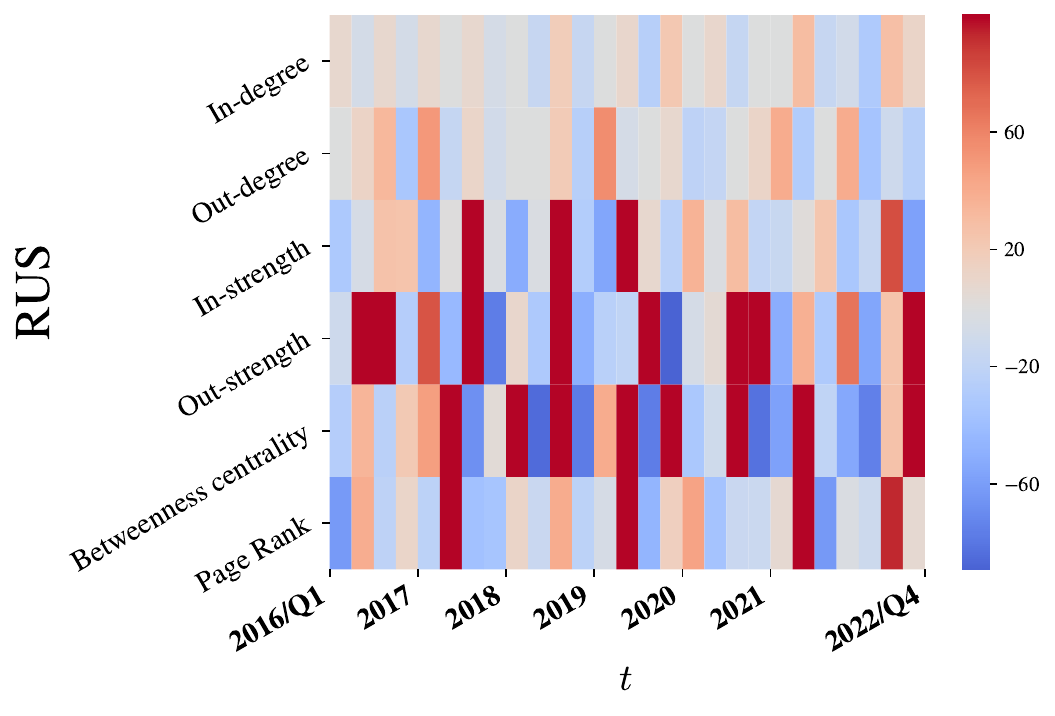}}  
    \subfigure[]{\label{level.sub.11}\includegraphics[width=0.237\linewidth]{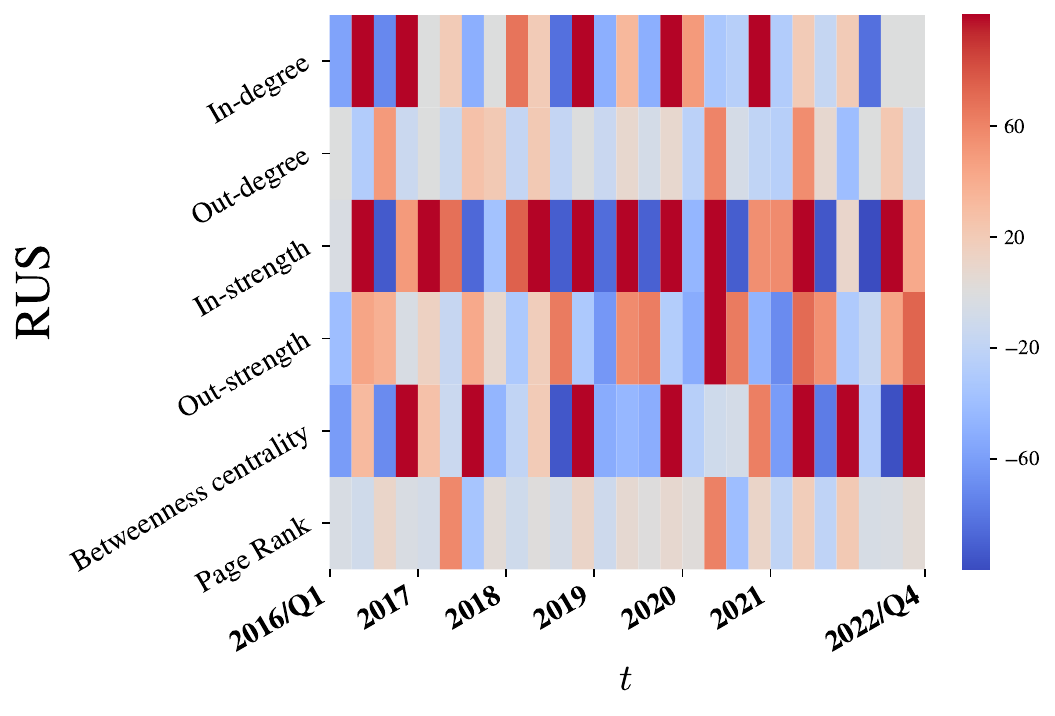}}   
    \subfigure[]{\label{level.sub.12}\includegraphics[width=0.237\linewidth]{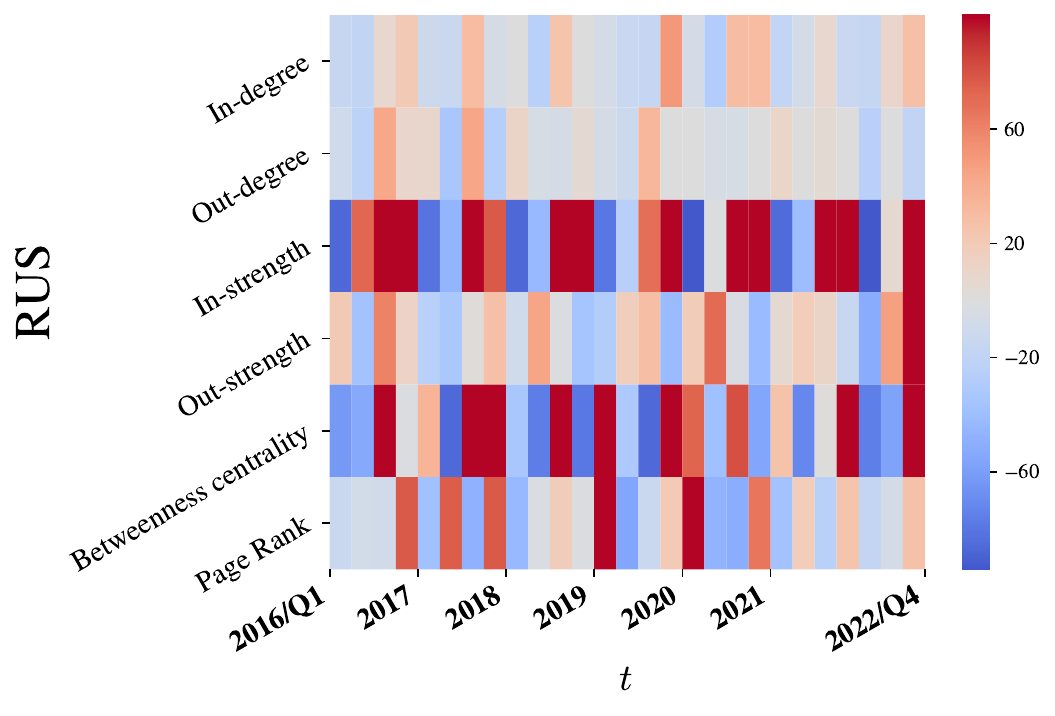}}  
    \subfigure[]{\label{level.sub.13}\includegraphics[width=0.237\linewidth]{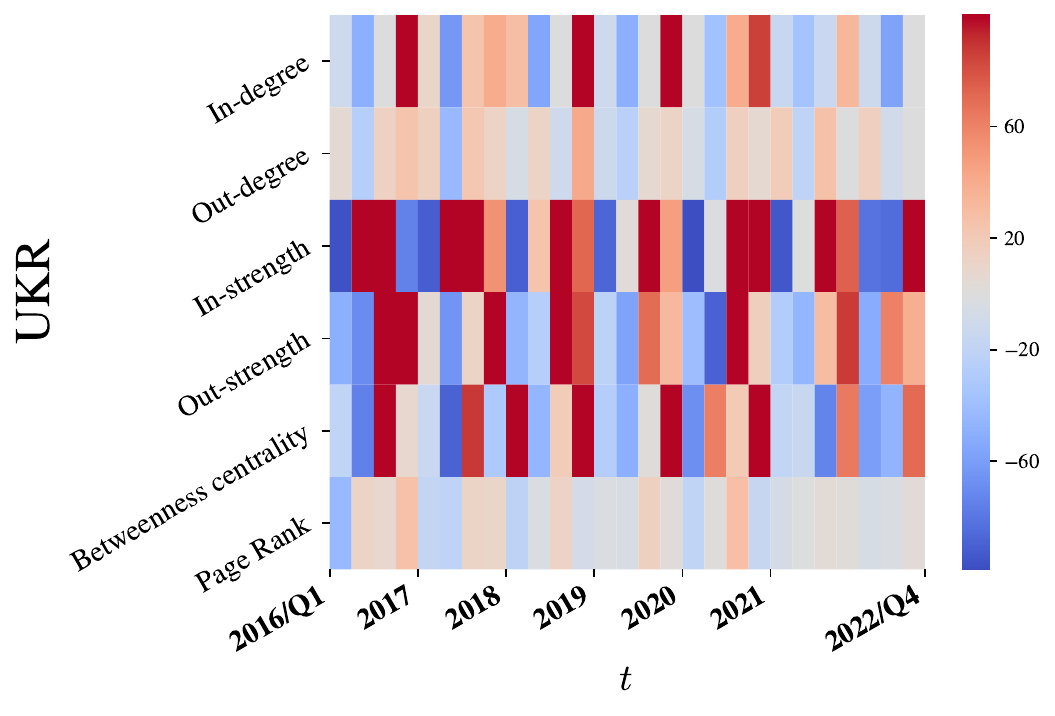}}    \subfigure[]{\label{level.sub.14}\includegraphics[width=0.237\linewidth]{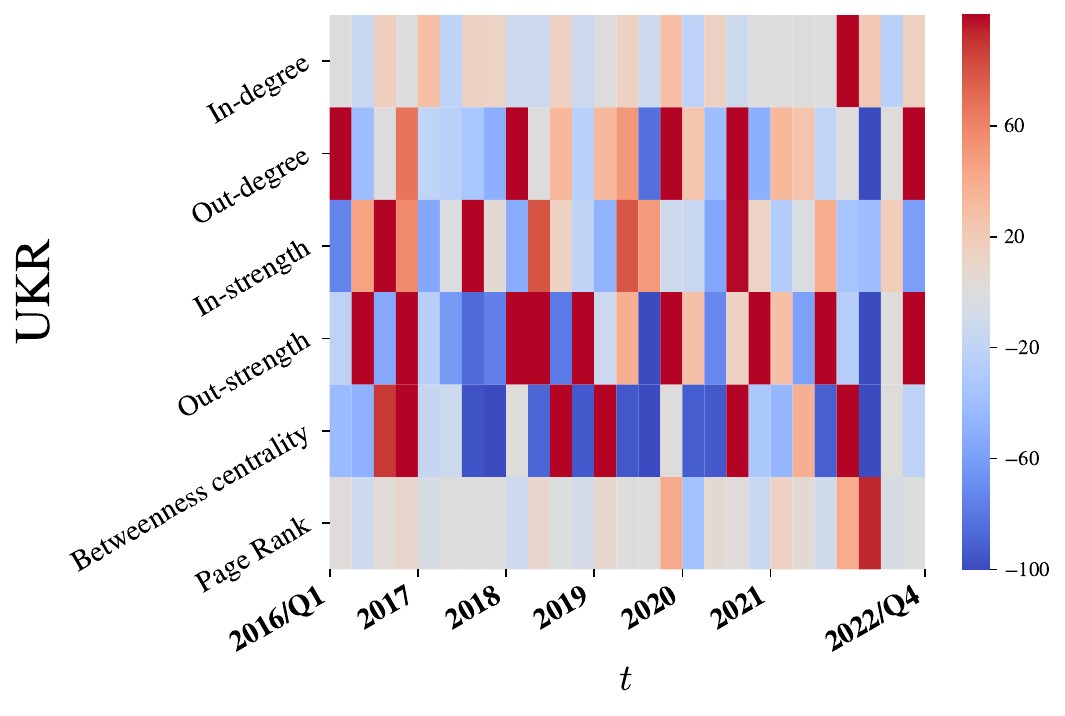}}    
    \subfigure[]{\label{level.sub.15}\includegraphics[width=0.237\linewidth]{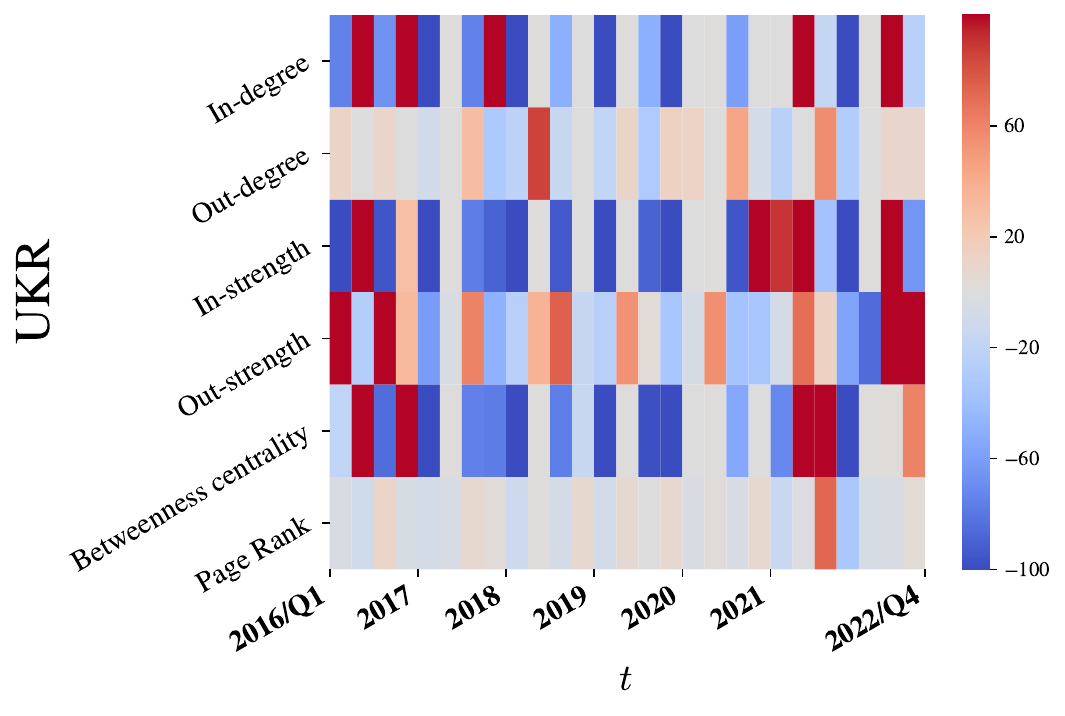}}    
    \subfigure[]{\label{level.sub.16}\includegraphics[width=0.237\linewidth]{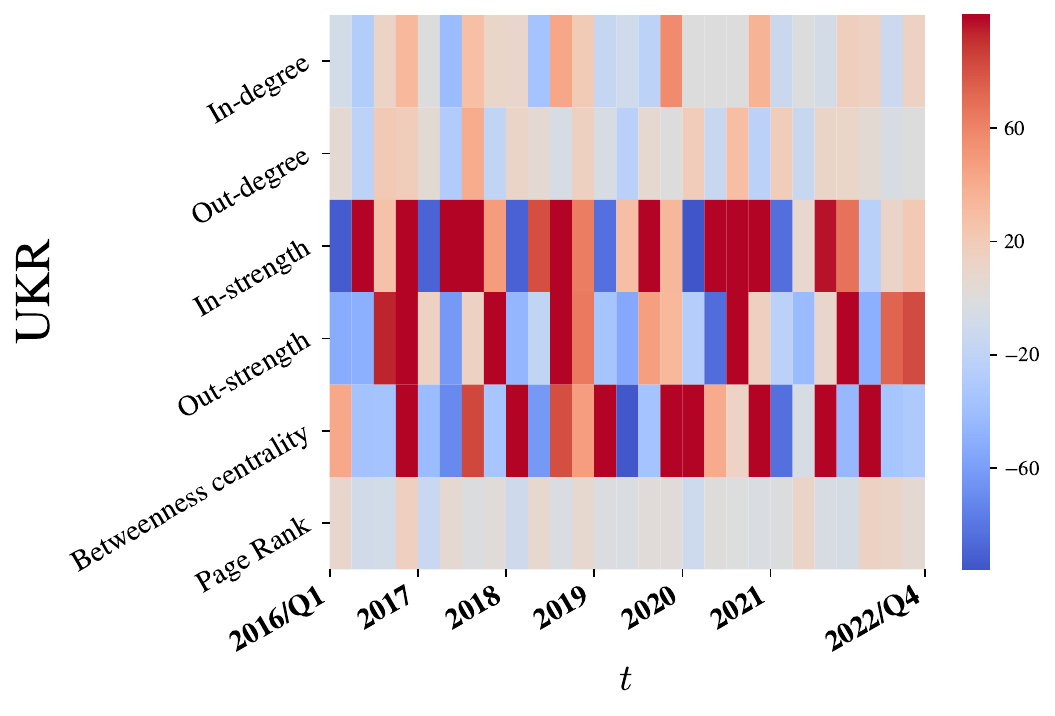}} 
    \caption{Quarterly percentage change in in/out-degrees, in/out-strengths, betweenness centrality and PageRank for Russia and Ukraine. The X-axis refers to quarters (2016/Q2 means the second quarter of 2016), and the y-axis is the topological indexes, including in-degree, out-degree, in-strength, out-strength, imp-trade ratio and exp-trade ratio. The columns from left to right, respectively, describe maize, rice, soybean and wheat. The top row corresponds to Russia, and the bottom row corresponds to Ukraine. The color scheme from blue to red indicates a decline in the growth of topological indexes.}
    \label{Fig:iCTN:RUS:UKR:k:s:t}
    \end{figure}

Figure~\ref{Fig:iCTN:RUS:UKR:k:s:t} presents the quarterly percentage change in in/out-degrees, in/out-strengths, betweenness centrality, and PageRank for Russia and Ukraine. As shown in Figs.~\ref{Fig:iCTN:RUS:UKR:k:s:t}(a-d), all the topological properties of Russia decreased in 2022/Q2. However, a partial recovery is observed in 2022/Q3 and 2022/Q4, where some of these properties show an increase. These findings suggest that the Russia-Ukraine conflict has had a negative short-term impact on Russia's crop trade. Compared to Russia, Ukraine is less affected by the Russia-Ukraine conflict, as shown in Figs.~\ref{Fig:iCTN:RUS:UKR:k:s:t}(e-h).

The relationships between NATO economies and both Russia and Ukraine are complicated. As a political and military alliance, NATO has consistently shown support for Ukraine and expressed criticism of Russia's actions. NATO has strongly condemned Russia's annexation of Crimea and its involvement in the conflict in Eastern Ukraine. Consequently, this has resulted in a deterioration of relations between NATO and Russia, with significant implications for trade and cooperation. Thus, our analysis also reveals heterogeneity in trade structure between Russia and Ukraine.

\begin{figure}[h!]
    \centering
    \includegraphics[width=0.475\linewidth]{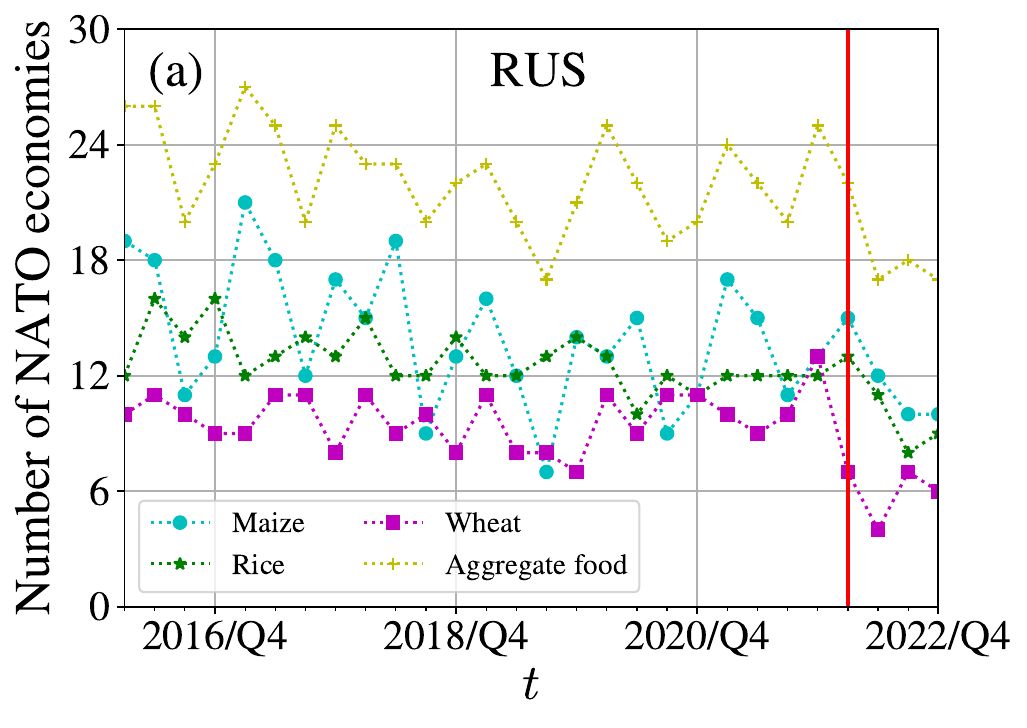}
    \includegraphics[width=0.475\linewidth]{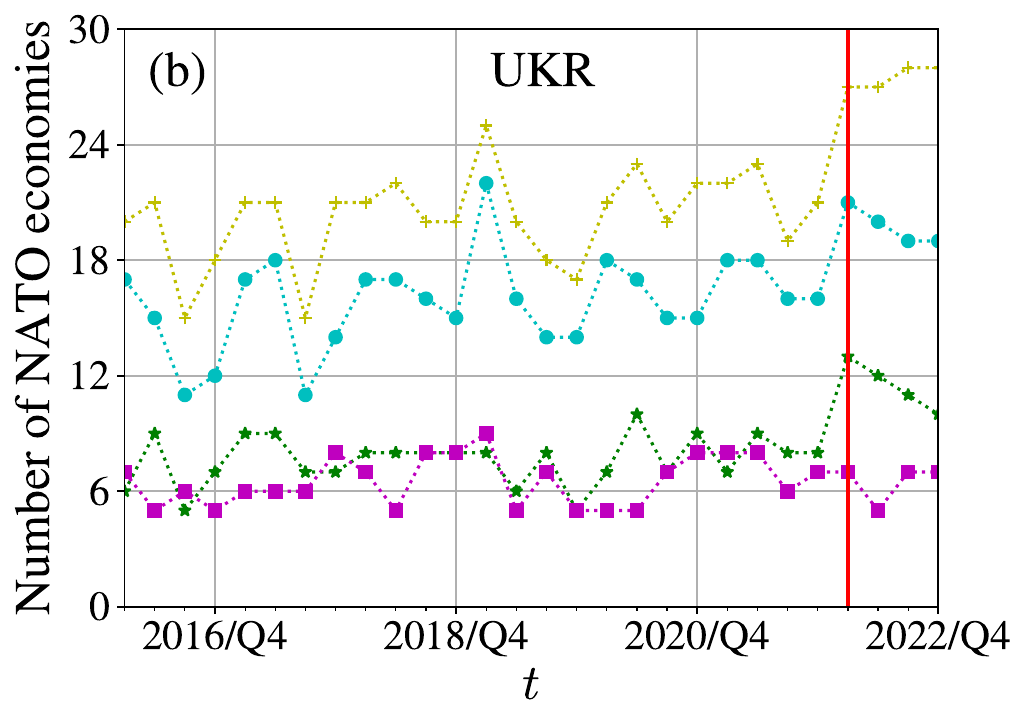} 
    \caption{The number of NATO economies importing or exporting crops from or to Russia (a) and Ukraine (b) from 2016/Q1 to 2022/Q4.}
    \label{Fig:iCTN:NATONum}
\end{figure}

Figure~\ref{Fig:iCTN:NATONum} presents the number of NATO economies that have trade links with Russia and Ukraine from 2016/Q1 to 2022/Q4. Notably, in 2022/Q2, there was a significant decline in trade links between Russia and NATO economies for all crops and aggregate food, as depicted in Fig.~\ref{Fig:iCTN:NATONum}(a). It might be explained by the conflict containment in trade relationships between Russia and NATO economies. However, the impact of the conflict eased as time progressed. What's more, Figure~\ref{Fig:iCTN:NATONum}(b) shows that Ukraine decreased the trade links with the NATO economies for all crops following the occurrence of the conflict and alleviated the trade relationship for wheat. But the trade links for aggregate food show a slight increase during 2022/Q2 and 2022/Q4. It suggests that the conflict had a more severe impact on the crop trade relationships between Russia and NATO economies compared to Ukraine and NATO economies.

\begin{figure}[h!]
    \subfigbottomskip=-1pt
    \subfigcapskip=-5pt
    \centering
     \subfigure[]{\label{level.sub.1}{\includegraphics[width=0.475\linewidth]{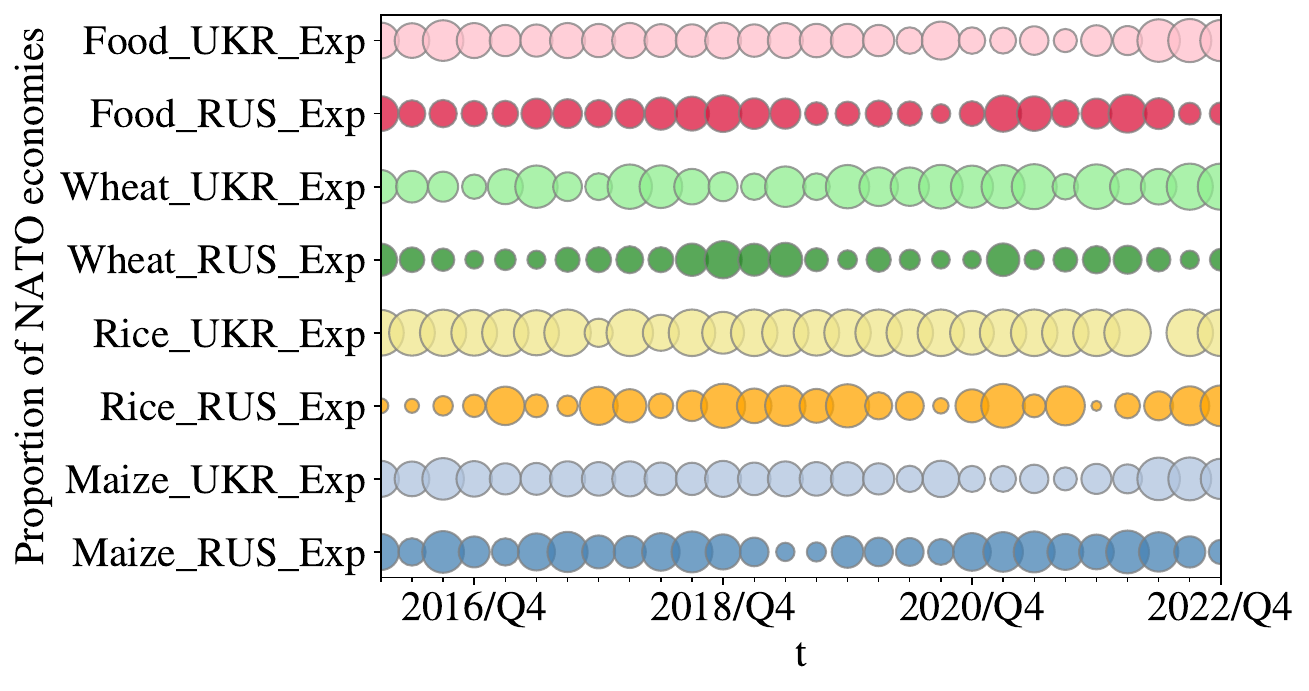}}}
     \subfigure[]{\label{level.sub.2}{\includegraphics[width=0.475\linewidth]{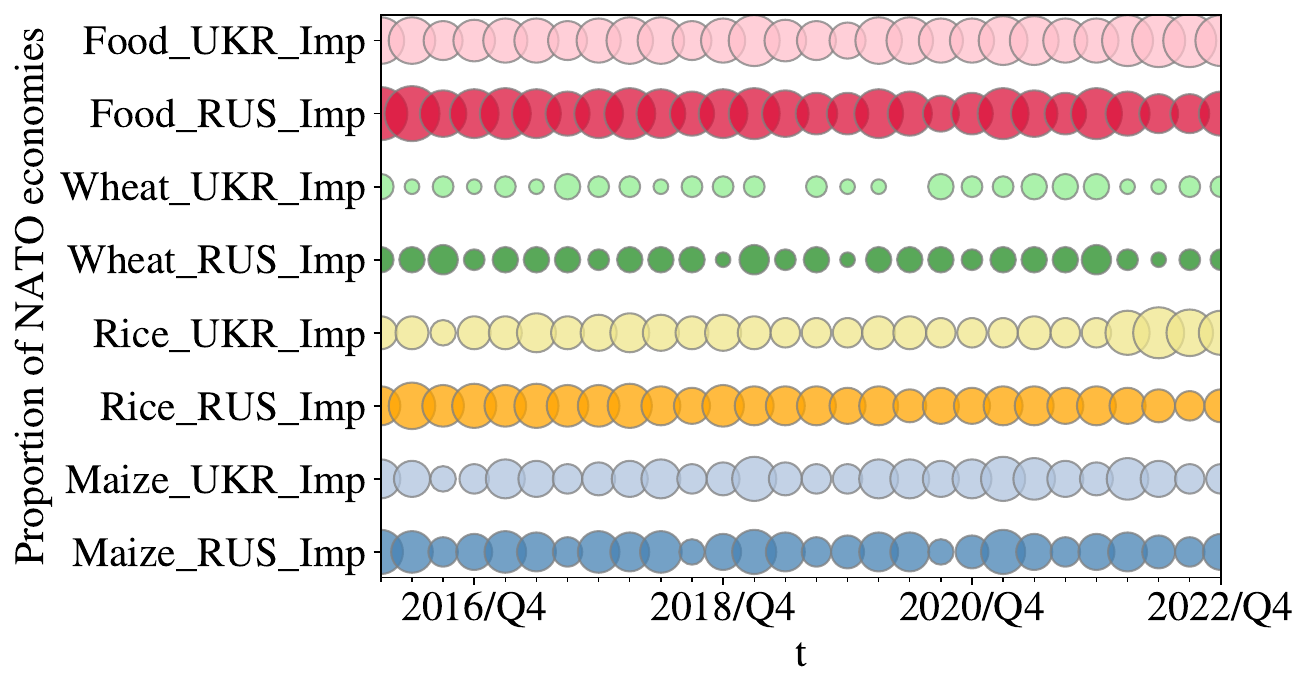}}}    
    \caption{Evolution of the proportion of food caloric trade volumes between Russia and NATO economies to all 24 economies, and between Ukraine and NATO economies to all 24 economies from 2016/Q1 to 2022/Q4. (a) Evolution of the proportion of NATO economies' exports from Russia and Ukraine for three crops and food from 2016/Q1 to 2022/Q4. (b) Evolution of the proportion of NATO economies' imports from Russia and Ukraine for three crops and food from 2016/Q1 to 2022/Q4. The size of the bubbles in a subgraph represents the size of the proportion.}
    \label{Fig:iCTN:NATO:Evolution}
\end{figure}

We display the evolution of the proportion of food caloric trade volumes between Russia and NATO economies and between Ukraine and NATO economies to all 24 economies from 2016/Q1 to 2022/Q4 in Fig.~\ref{Fig:iCTN:NATO:Evolution}. We find that following the occurrence of the conflict, the proportion of maize and rice exports between Russia and NATO economies, as well as between Ukraine and NATO economies, did not undergo significant changes. This indicates that these crop exports remained relatively stable amidst the conflict. However, a substantial decline in the proportion of wheat and aggregate food exports between Russia and NATO economies is evident. This decline suggests a noticeable impact of the conflict on these specific trade relationships, with a decrease in the relative volume of these exports. Furthermore, the proportion of NATO economies' imports from Russia and Ukraine for the three crops, as well as for overall food, from 2016/Q1 to 2022/Q4, remained unaffected by the conflict.

\begin{figure}[h!]
    \centering
    \includegraphics[width=0.237\linewidth]{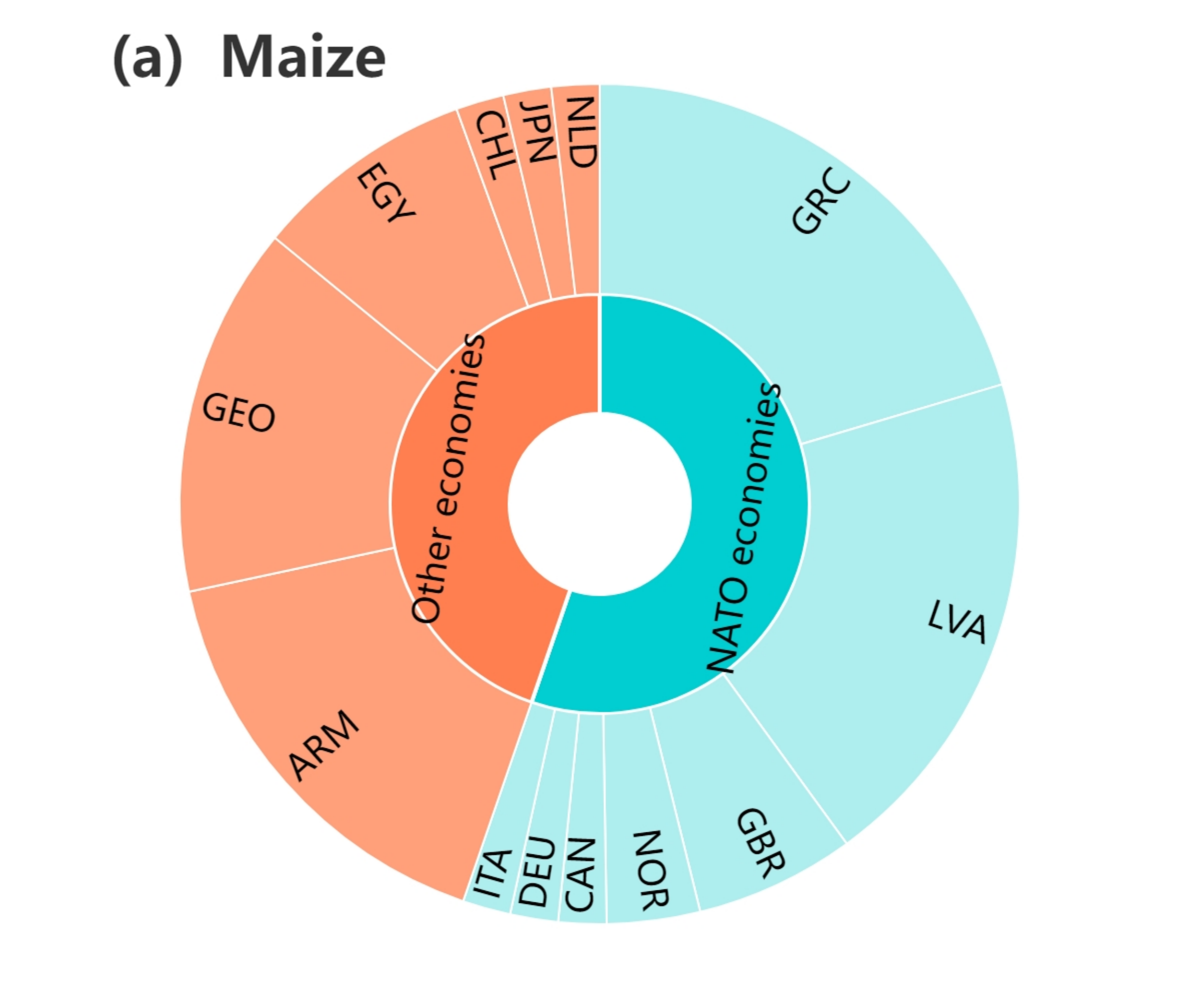}
    \includegraphics[width=0.237\linewidth]{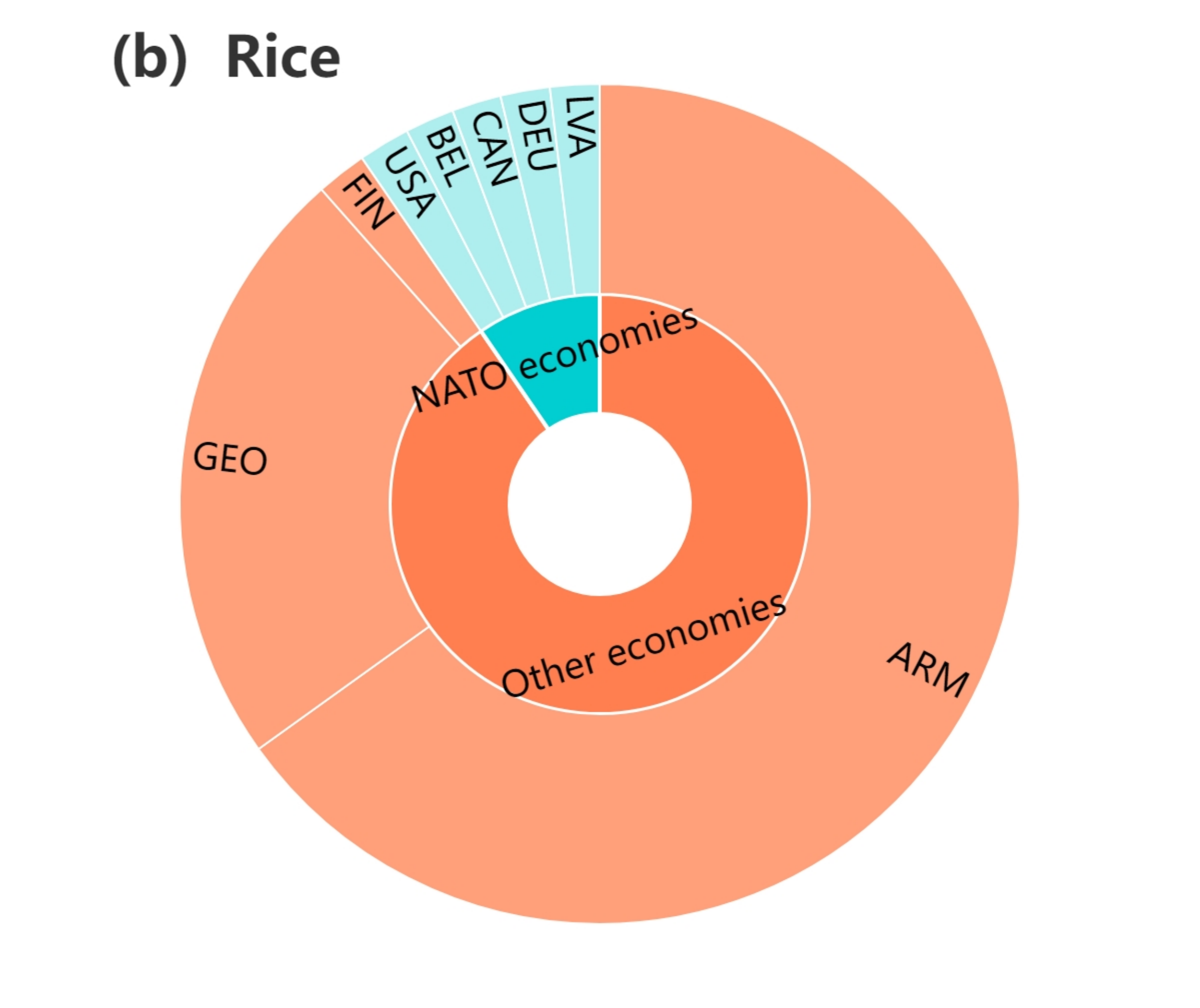}     
    \includegraphics[width=0.237\linewidth]{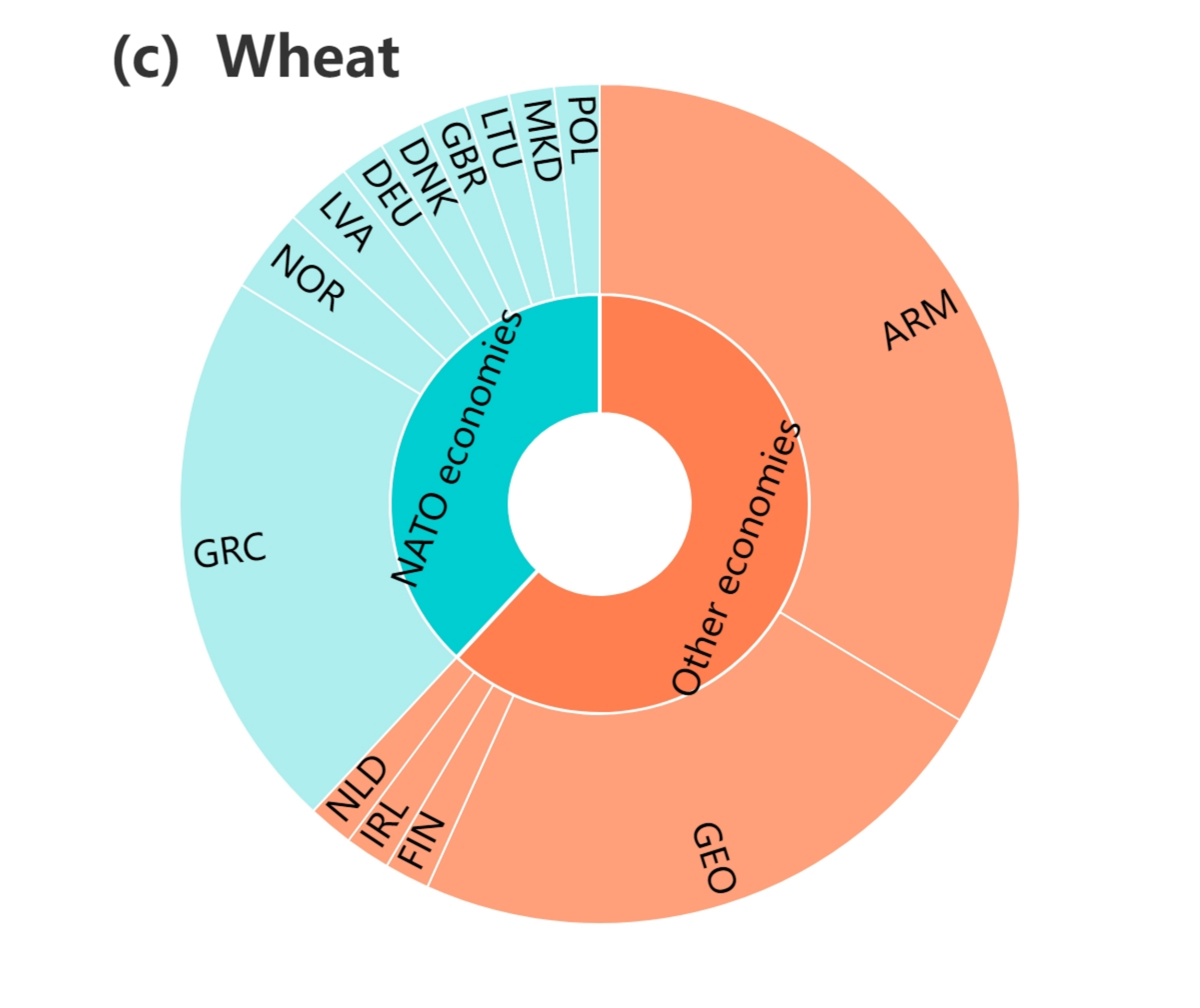} \includegraphics[width=0.237\linewidth]{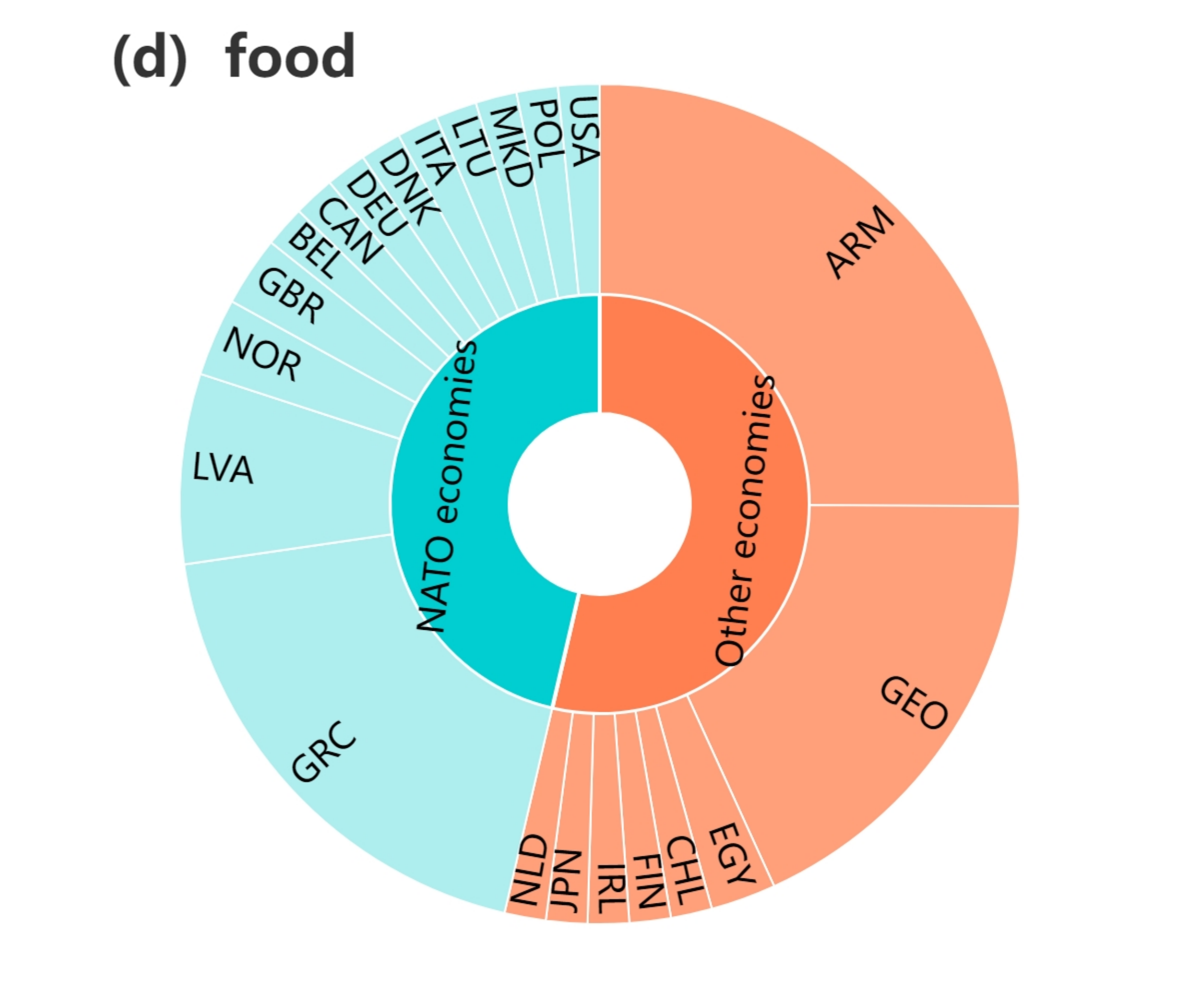}   
    \includegraphics[width=0.237\linewidth]{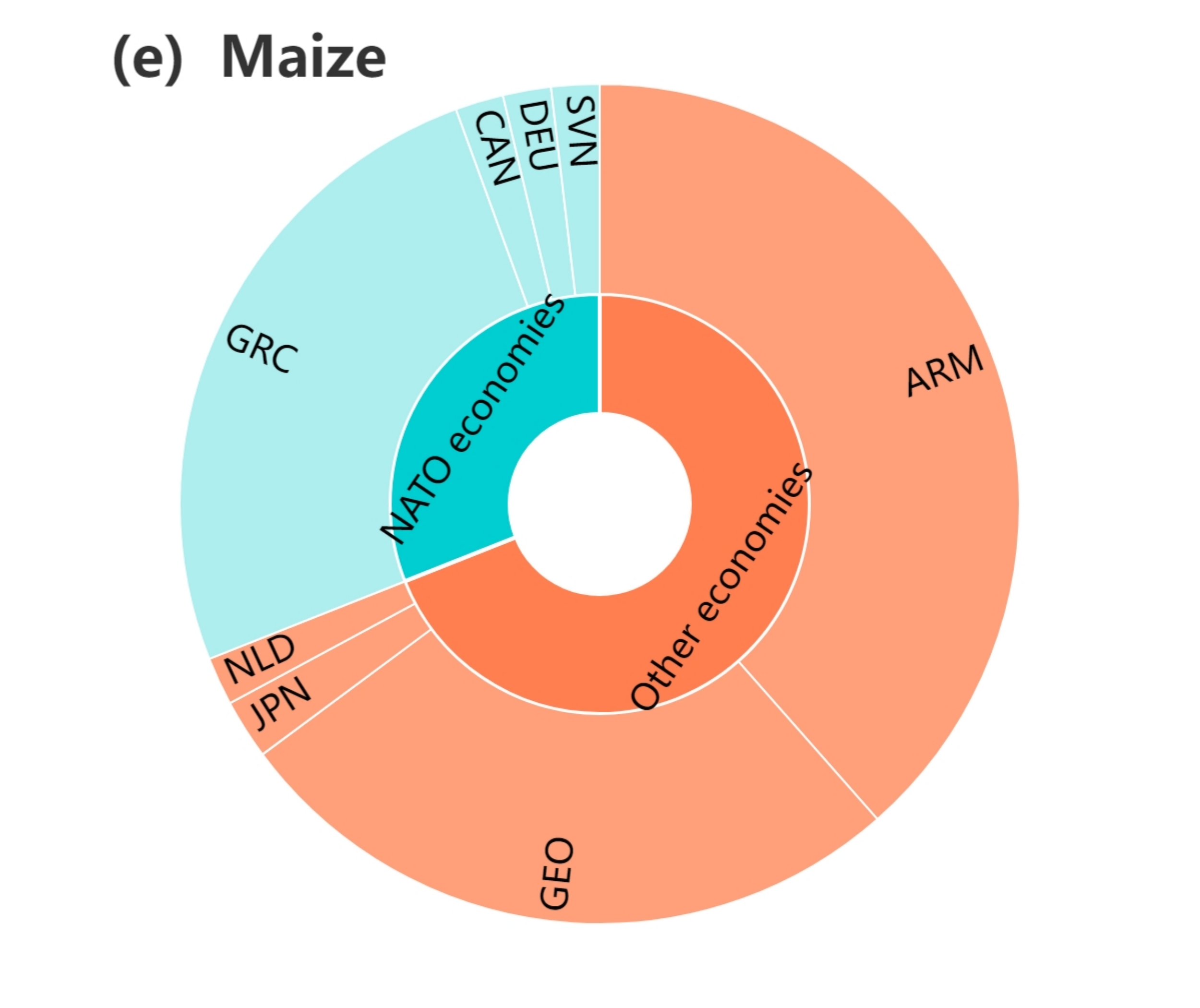}    \includegraphics[width=0.237\linewidth]{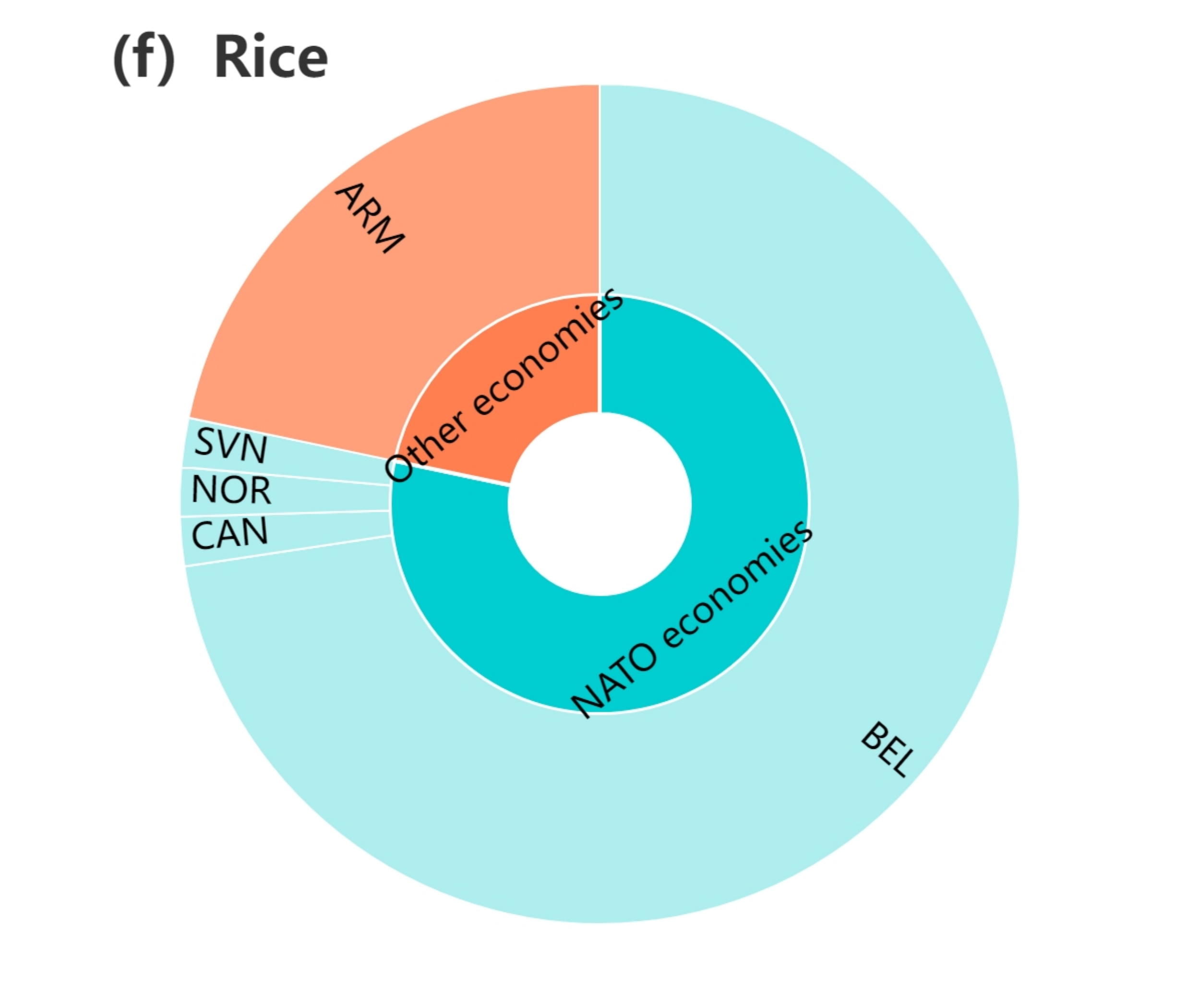}
    \includegraphics[width=0.237\linewidth]{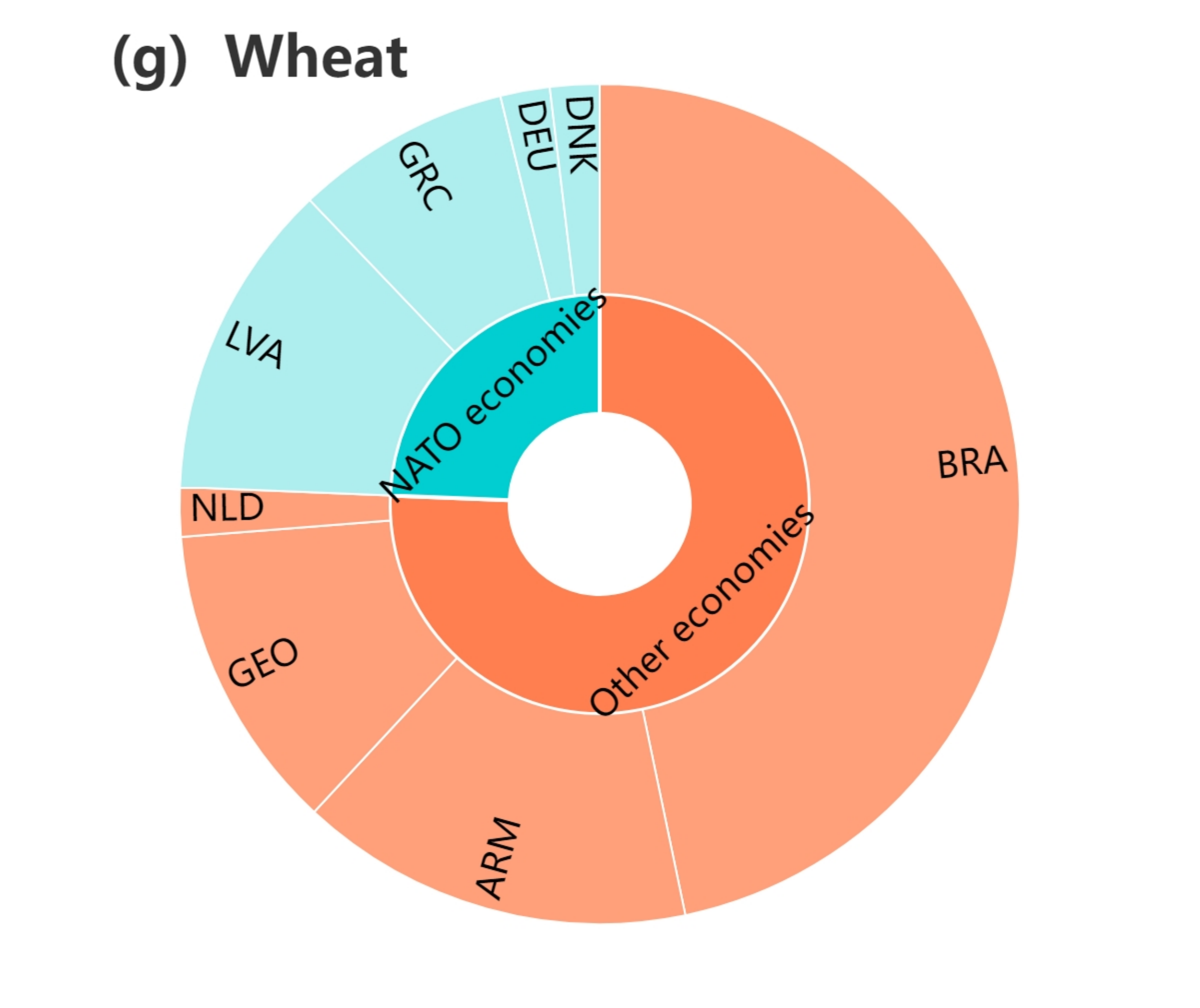}   
    \includegraphics[width=0.237\linewidth]{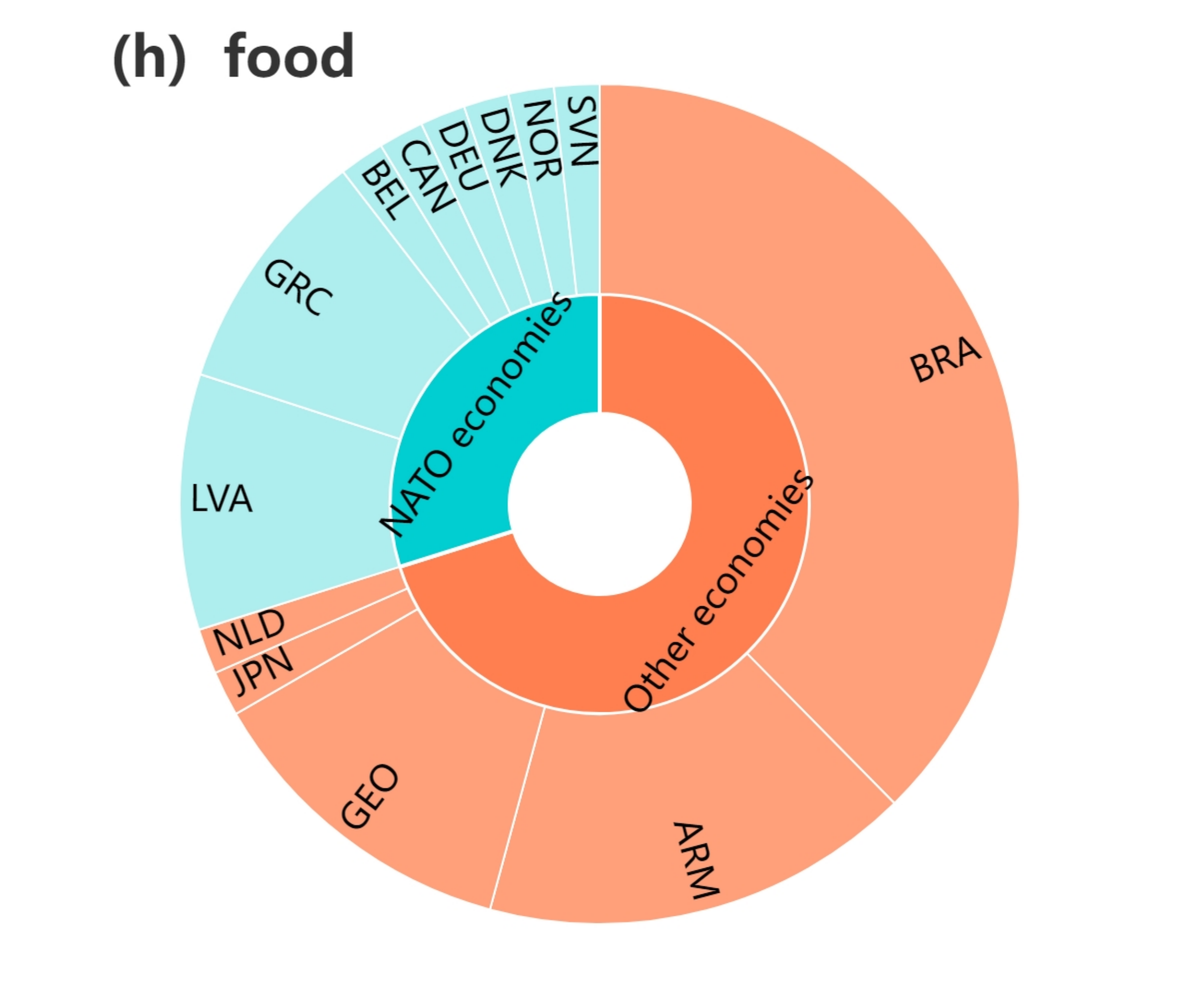}  
    \caption{Russia's crop export structure in 2021/Q4 (a-d) and 2022/Q4 (e-h). The inner circle shows the ratio of crop calories imported from Russia by NATO economies and other economies, with blue indicating NATO economies and orange indicating other economies. The outer circle displays the ratio of crop calories imported from Russia by different economies. Since Russia has not provided crop trade data since March 2022, the economies involved here are only those that continuously report crop trade data from 2016/1 to 2022/12.}
    \label{Fig:iCTN:tradeflow:RUS}
\end{figure}

\begin{figure}[h!]
    \centering
    \includegraphics[width=0.237\linewidth]{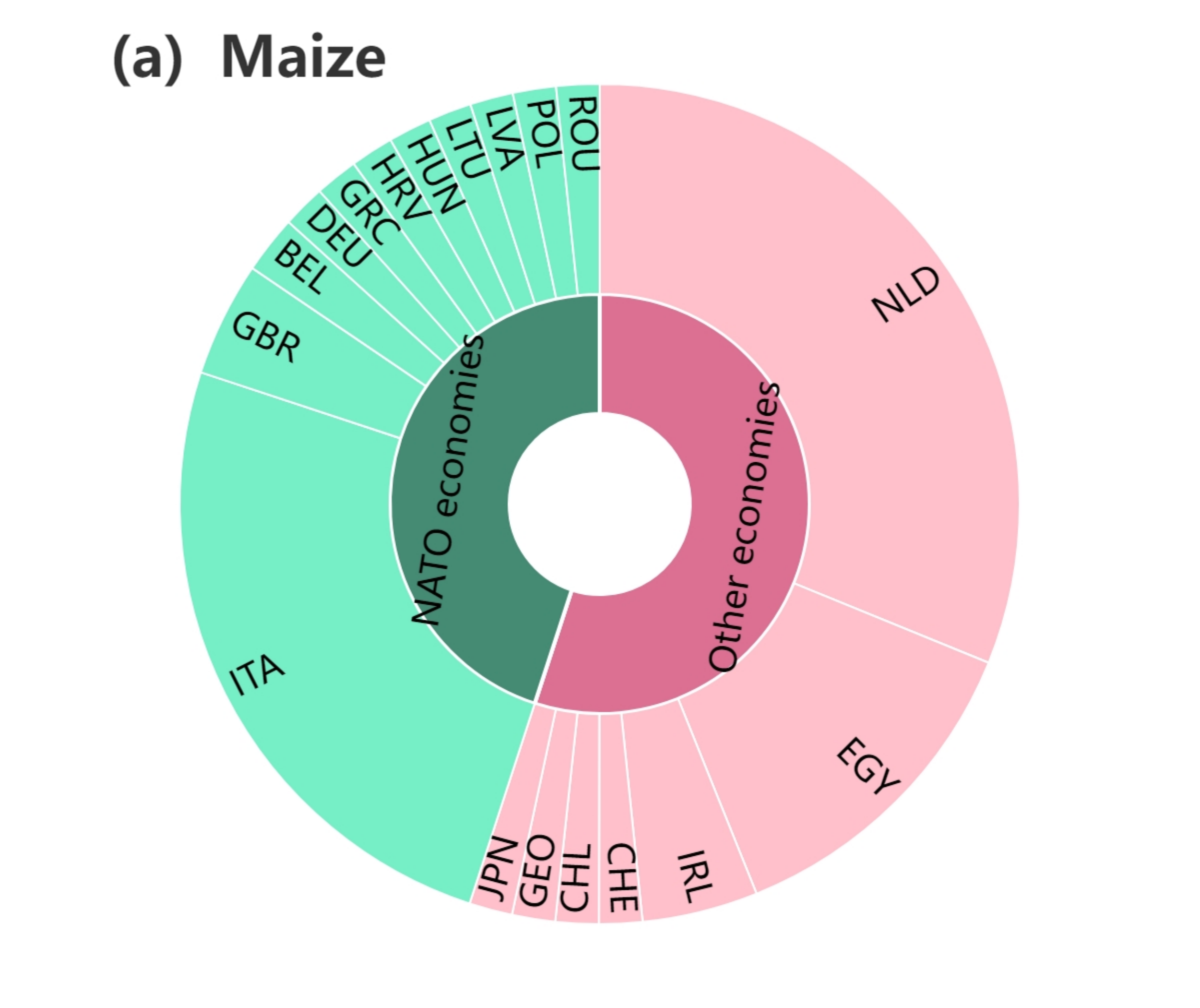}
    \includegraphics[width=0.237\linewidth]{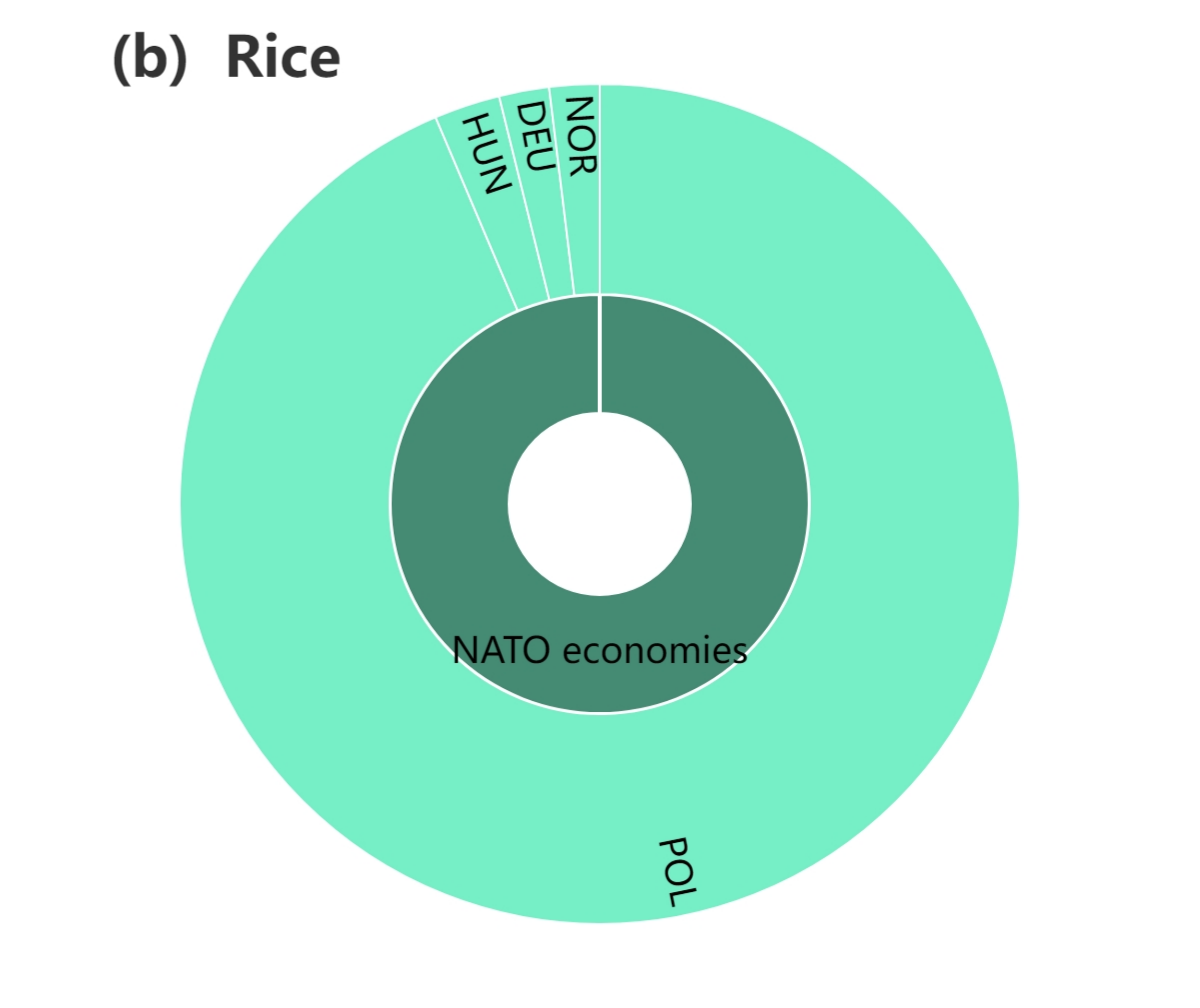}     
    \includegraphics[width=0.237\linewidth]{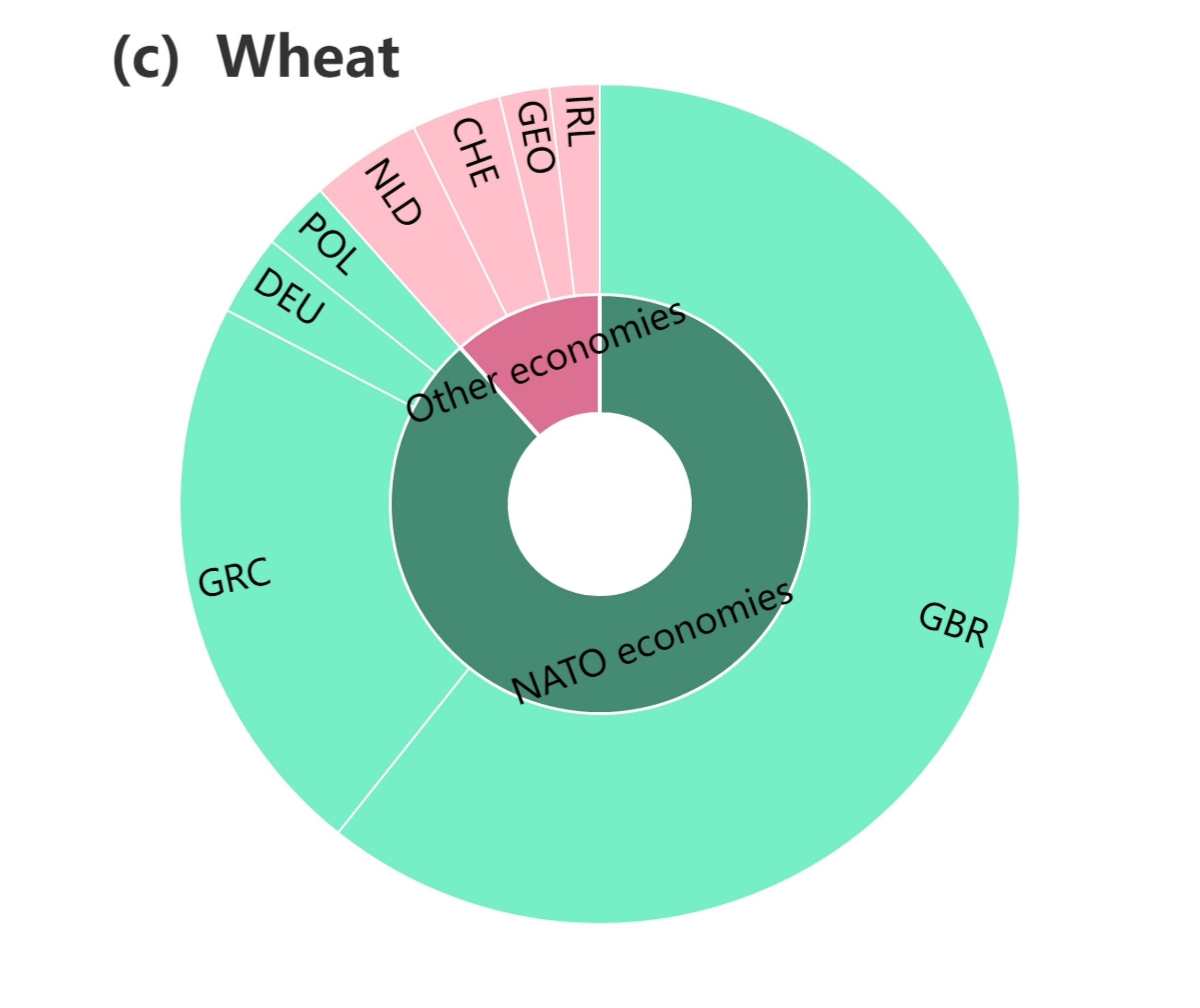} \includegraphics[width=0.237\linewidth]{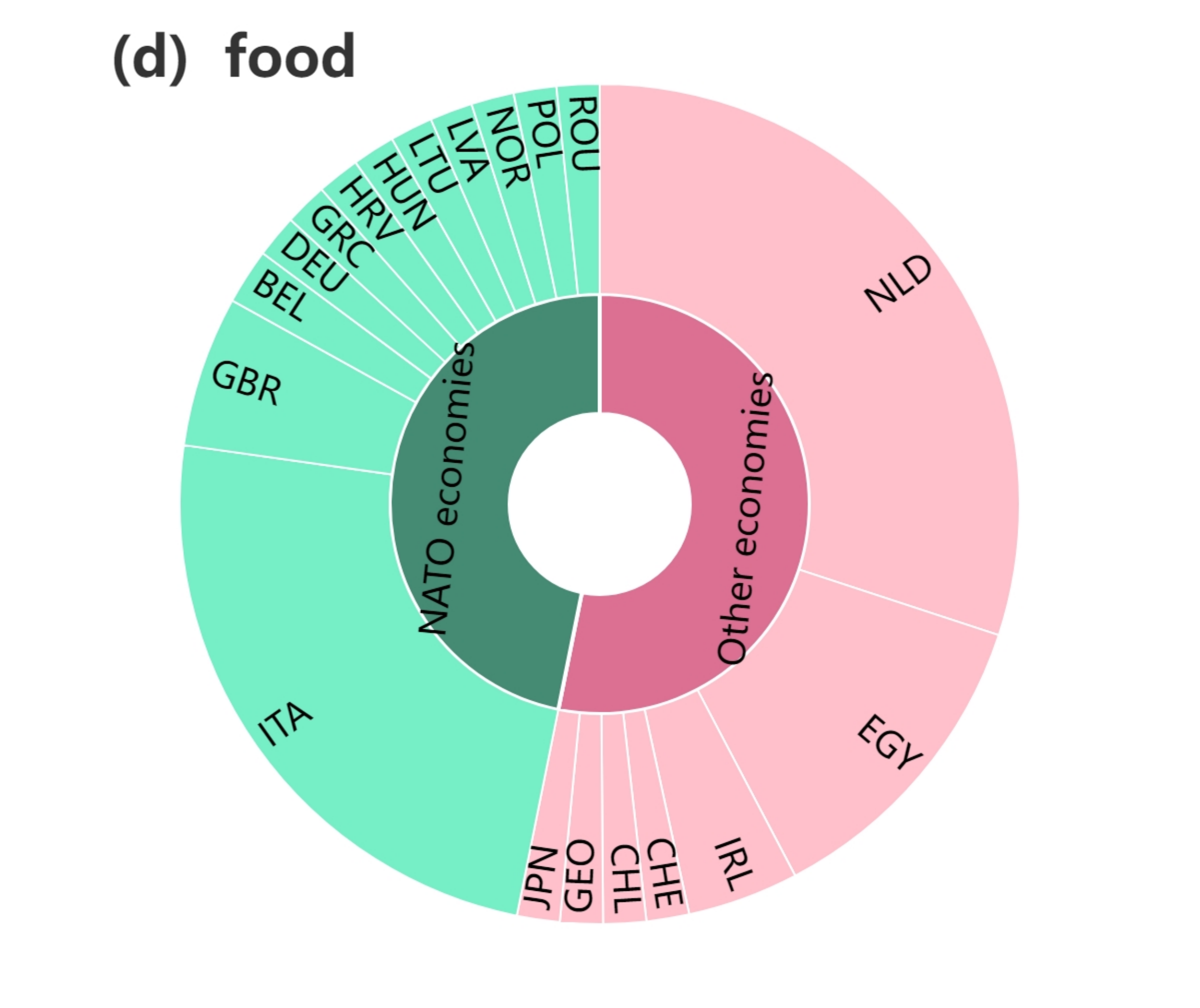}   
    \includegraphics[width=0.237\linewidth]{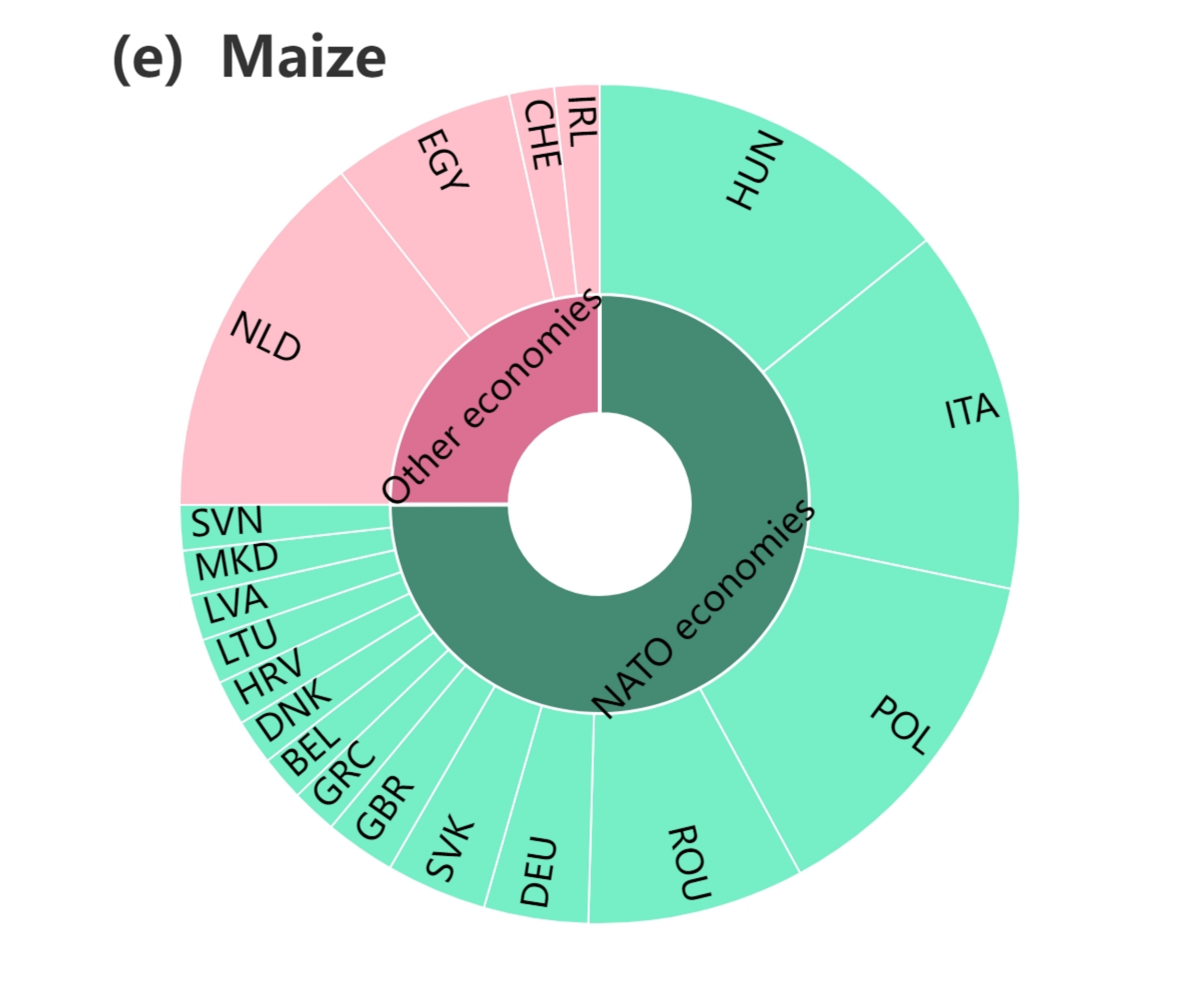}    \includegraphics[width=0.237\linewidth]{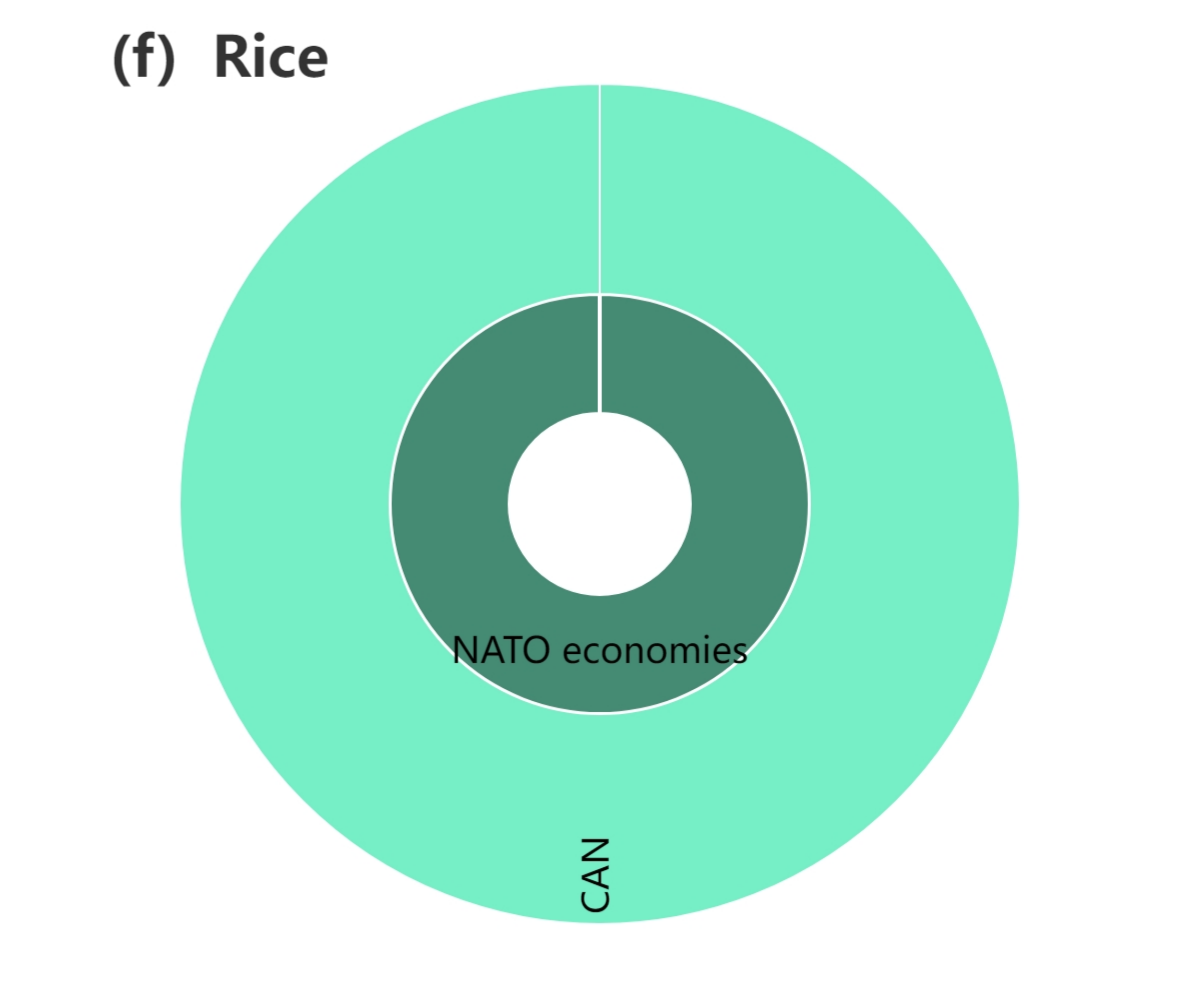}
    \includegraphics[width=0.237\linewidth]{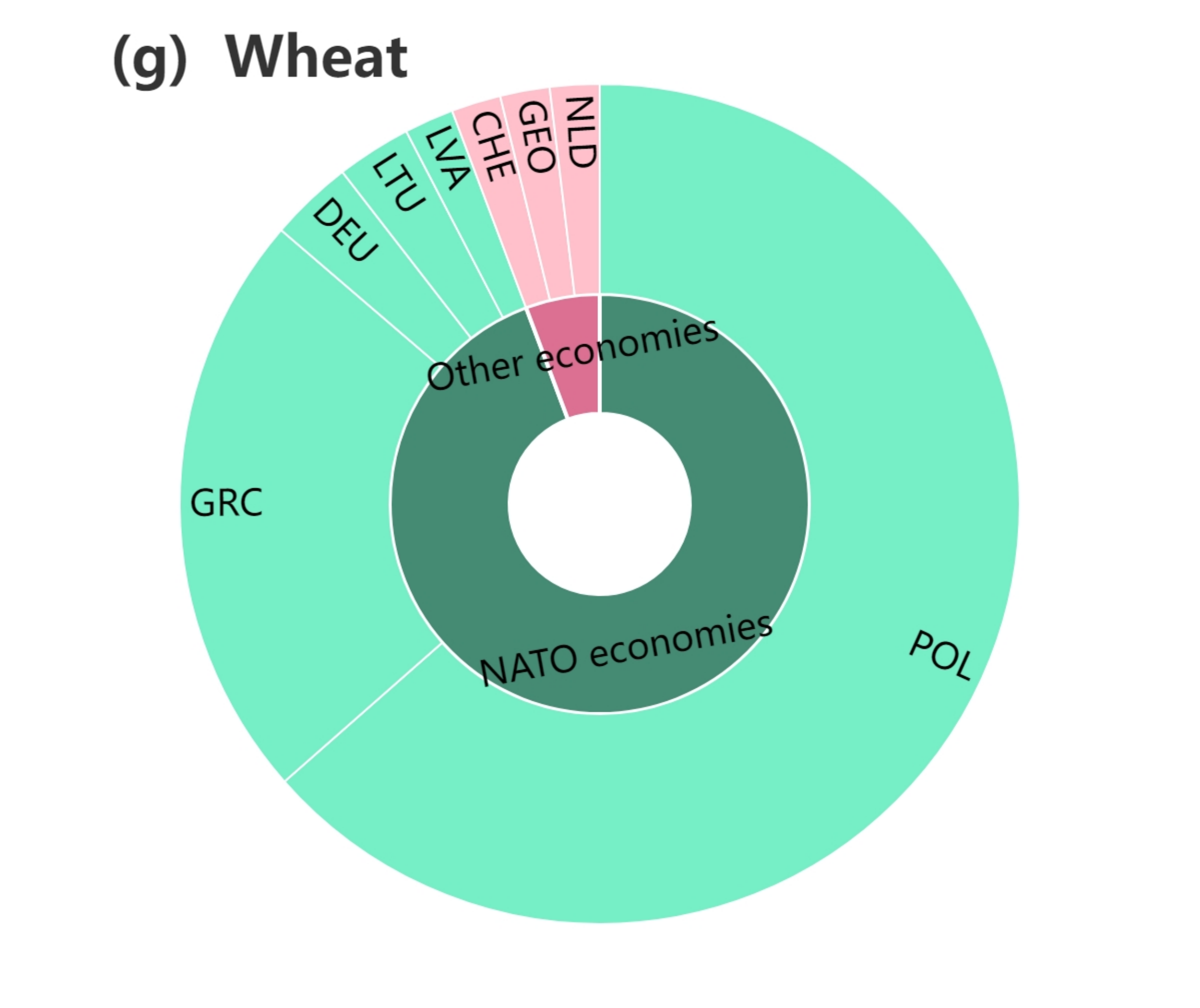}   
    \includegraphics[width=0.237\linewidth]{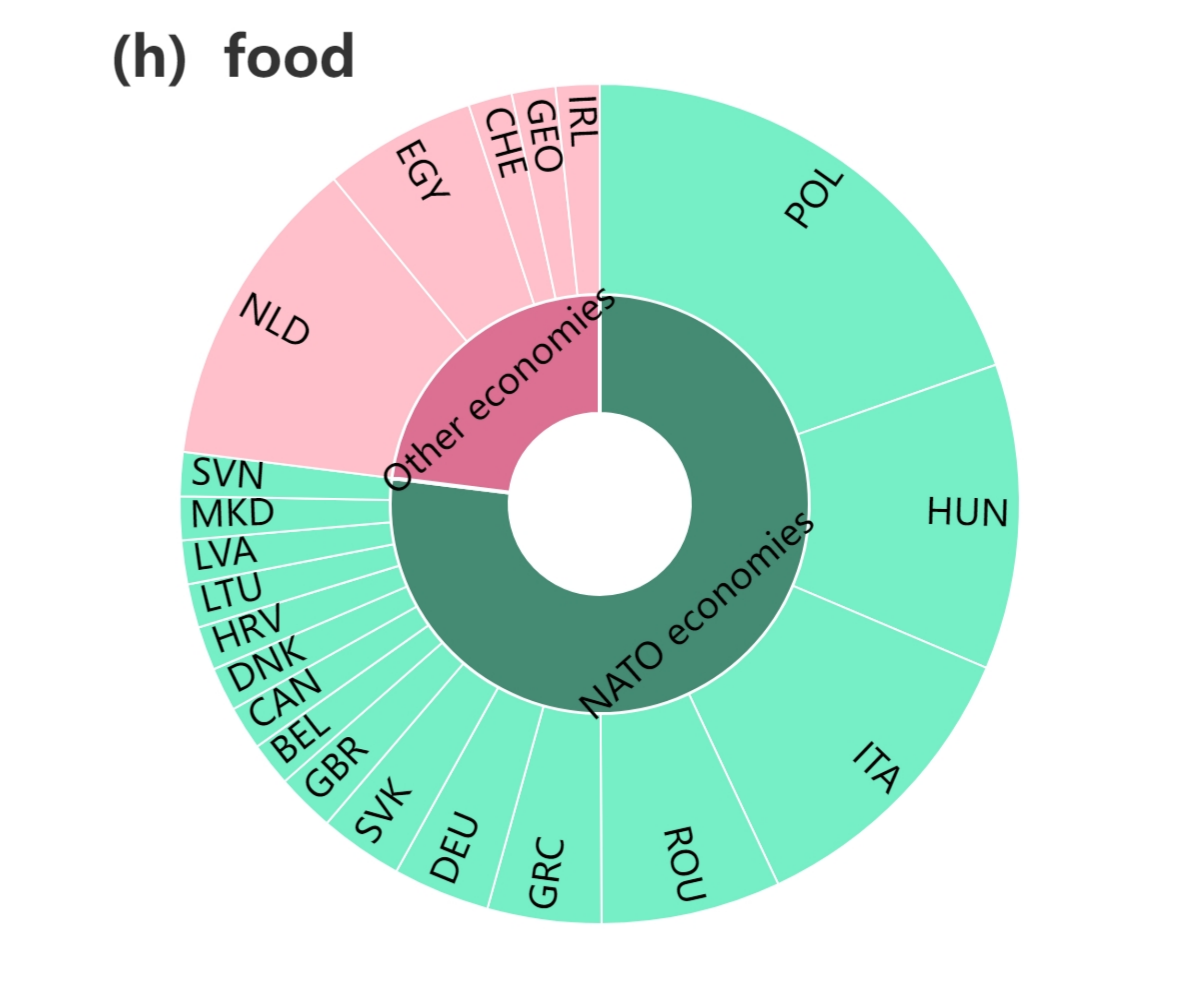}  
    \caption{ Ukraine's crop export structure in 2021/Q4 (a-d) and 2022/Q4 (e-h). The inner circle shows the ratio of crop calories imported from Ukraine by NATO economies and other economies, with blue indicating NATO economies and orange indicating other economies. The outer circle displays the ratio of crop calories imported from Ukraine by different economies. Since Ukraine has not provided crop trade data since March 2022, the economies involved here are only those that continuously report crop trade data from 2016/1 to 2022/12.}
    \label{Fig:iCTN:tradeflow:UKR}
\end{figure}

Since Russia and Ukraine are major crop exporters, we compare Russia's and Ukraine's crop export structures in 2021/Q4 and 2022/Q4 in Fig.~\ref{Fig:iCTN:tradeflow:RUS} and Fig.~\ref{Fig:iCTN:tradeflow:UKR}. It is worth noting that the cessation of export trade between Russia and some NATO economies resulted in a marked decline in the proportion of maize, wheat, and aggregate food exports to NATO economies to Russia's total maize, wheat, and aggregate food exports after the Russo-Ukrainian conflict, as shown in Figs.~\ref{Fig:iCTN:tradeflow:RUS}(a), (c), (d), (e), (g), and (h). Conversely, as illustrated in Fig.~\ref{Fig:iCTN:tradeflow:RUS}(b) and Fig.~\ref{Fig:iCTN:tradeflow:RUS}(c), there was a discernible increase in the proportion of rice exports to NATO economies to Russia's total rice exports.

In contrast to Russia, Ukraine experienced a different scenario, as demonstrated by Fig.~\ref{Fig:iCTN:tradeflow:UKR}, which showcases an escalation in maize, wheat, and aggregate food export trade with NATO economies, leading to an increase in the proportion of exports to NATO economies to Ukraine's total maize, rice, wheat, and aggregate food exports.

Overall, the impact of the Russia-Ukraine conflict differs across economies and crops. The conflict has different impacts on Russia and Ukraine. It significantly altered the topological properties of Russia and had a notable impact on the trade relationships between Russia and NATO economies, particularly regarding wheat. Although less so than in Russia, the conflict also had an impact on Ukraine. The findings indicate that the conflict had a more substantial impact on Russia's trade dynamics and network properties than on Ukraine.

\section{Conclusions}
\label{sec4:Conclusions}

The Russia-Ukraine conflict has sparked global apprehension regarding international trade and food security. As significant grain exporters, both Russia and Ukraine play a crucial role, accounting for approximately 40\% of global grain exports. Notably, these economies were responsible for about 30\% of the world's wheat exports in 2021. Consequently, the conflict has triggered declines in the global food supply and substantial increases in global food prices. In this paper, we consolidate three iCTNs and an iFTN and compare structural changes by using crop trade data from the UN Comtrade Database. Additionally, we analyze the topological attributes of Russia and Ukraine and the shifts in crop trade relationships between NATO economies and both Russia and Ukraine. Through this comprehensive analysis, we aim to provide valuable insights into the impact of the conflict on the network structures and trade dynamics within the global agrifood sector.

First, we present the structural evolution of three iCTNs and the iFTN over the period from 2016/Q1 to 2022/Q4, including the number of nodes, edges, total edge weight, average degree, average strength, density, clustering coefficient, efficiency, and natural connectivity. We found that the Russia-Ukraine conflict has affected the structure of the iCTNs and the iFTN. And the impact of the Russia-Ukraine conflict on the iCTNs differs across different crop sectors. It seems plausible to assume that wheat yield losses in Ukraine and wheat export restrictions make the iWTN face the most severe disturbance.

In terms of the iFTN, all topological metrics display an overall upward trend, which differs from the behavior observed in the individual iCTNs. Notably, the number of nodes, the link weights, the average in-strength, and the link reciprocity maintained a consistent trend in 2022. Nevertheless, changes were observed in the number of edges, average in-degree, density, efficiency, and natural connectivity following the onset of the conflict. Another notable finding is that these network metrics increased in 2022/Q2 but decreased in 2022/Q3. This suggests that the conflict initially had a minimal impact on the iFTN within a short time but eventually had a negative effect as the conflict endured. This finding contrasts somewhat with the results of the iWTN, which show a decrease in connectivity. One possible explanation for this disparity lies in the substitution effect among maize, rice, and wheat \cite{LiverpoolTasie-Reardon-Parkhi-Dolislager-2023-NatFood}.

We examine the effects of the Russia-Ukraine conflict specifically on Russia and Ukraine. Our study investigates the quarterly percentage changes in in/out-degrees, in/out-strengths, betweenness centrality, and PageRank for both countries. The findings indicate that the Russia-Ukraine conflict has had a negative short-term impact on Russia's crop trade. In comparison, Ukraine has experienced relatively lesser effects from the conflict. Additionally, we analyze the trade relationships between NATO economies and both Russia and Ukraine before and after the conflict. It is observed that the conflict has had a more significant impact on the crop trade relationships between Russia and NATO economies, compared to Ukraine and NATO economies. The discontinuation of export trade between Russia and certain NATO economies has resulted in a notable decline in the proportion of maize, wheat, and aggregate food exports from Russia to NATO economies in relation to its total maize, wheat, and aggregate food exports.

To conclude, this study investigates the changes in international crop trade networks following the occurrence of the Russia-Ukraine conflict, with a particular focus on European economies. By tracking the changes in the topological properties of the three iCTNs and the iFTN during 2022/Q2 and 2022/Q4 compared with 2021, we have identified significant shifts in the number of edges, average degree, density, efficiency, and natural connectivity in 2022/Q3, especially in the iWTN. We have also uncovered notable changes in the trade relationships between NATO economies and Russia. Although the networks analyzed in this paper are based on data from 24 economies and certain trade links are not included—such as those involving many Asian and African economies, which are major crop trade partners of Russia and Ukraine—our study still sheds light on the impact of the conflict on a segment of the global food trade network and some important economies. Our work provides insights into the impact of geopolitical conflicts on the global food system and encourages a series of effective strategies to mitigate the negative impact of the conflict on global food trade.

\section*{Data Availability}
\label{S1:Data}
Publicly available datasets were analyzed in this study. This data can be found here: \href{https://comtradeplus.un.org}{https://comtradeplus.un.org}.

\section*{CRediT authorship contribution statement}
\label{S2:CRediT}
Funding acquisition, Wei-Xing Zhou; Investigation, Yin-Ting Zhang and Mu-Yao Li; Methodology, Yin-Ting Zhang and Wei-Xing Zhou; Supervision, Wei-Xing Zhou; Writing – original draft, Yin-Ting Zhang and Mu-Yao Li; Writing – review \& editing, Yin-Ting Zhang and Wei-Xing Zhou.

\section*{Declaration of Competing Interest}
\label{S3:Declaration}

The authors declare that they have no known competing financial interests or personal relationships that could have appeared to influence the work reported in this paper.

\section*{Acknowledgements}
\label{S4:Acknowledgements}

This work was partly supported by the National Natural Science Foundation of China (72171083), the Shanghai Outstanding Academic Leaders Plan, and the Fundamental Research Funds for the Central Universities.


\end{document}